\documentclass[twocolumn,floatfix]{aastex631}
\usepackage{multirow, makecell}
\usepackage{microtype}

\usepackage{graphicx,subfigure}

\usepackage{textcomp}
\begin{document}

\title{UVIT Study of the MAgellanic Clouds (U-SMAC) II. 
A Far-UV catalog of the Small Magellanic Cloud: Morphology and Kinematics of young stellar population}
\author[0009-0000-3938-162X]{Sipra Hota}
\affiliation{Indian Institute of Astrophysics,
$2^{nd}$ Block, Koramangala,
Bangalore-560034, India}
\affiliation{Pondicherry University,
 R.V.Nagar, Kalapet,
 Puducherry-605014, India\\}
 
\author[0000-0003-4612-620X]{Annapurni Subramaniam}
\affiliation{Indian Institute of Astrophysics,
$2^{nd}$ Block, Koramangala,
Bangalore-560034, India}

\author[0000-0002-4638-1035]{Prasanta K. Nayak}
\affiliation{ Institute of Astrophysics, Pontificia Universidad Cat\'olica de Chile, Av. Vicu\~na MacKenna 4860, Santiago 7820436, Chile }

\author[0000-0002-5331-6098]{Smitha Subramanian}
\affiliation{Indian Institute of Astrophysics,
$2^{nd}$ Block, Koramangala,
Bangalore-560034, India}

\begin{abstract}

The Small Magellanic Cloud (SMC) is an irregular dwarf galaxy that has recently undergone an interaction with the Large Magellanic Cloud. The young massive stars in the SMC formed in the disturbed low-metallicity environment are important targets in astrophysics. We present a catalog of $\sim$ 76,800 far ultraviolet (FUV) sources towards the SMC detected using the Ultra Violet Imaging Telescope (UVIT) onboard AstroSat. We created an FUV catalog with $\sim$ 62900 probable SMC members which predominantly comprise main-sequence, giant, and subgiant stars. We selected 4 young populations (Young 1, Young 2, Young 3, and Blue Loop (BL) stars) identified from the Gaia optical color-magnitude diagram to study the morphology and kinematics of the young SMC using this catalog. We detect a clumpy morphology with a broken bar, a shell-like structure, and the inner SMC Wing for the 4 stellar populations. The eastern region and the northeastern regions are mainly populated by Young 1, 2, and 3. The central region predominantly has the Young 2 and 3 populations, whereas the SW has BL stars, Young 2 and 3.  The 2-D kinematic study using proper motion (PM) reveals that Young 2 and 3 populations show two kinematically distinct sub-populations with low and high PM dispersion, whereas the Young 1 and BL stars show two kinematically distinct populations with low dispersion. Our analysis points to a kinematic disturbance along the RA direction for stars younger than $\sim$ 150 Myr located in the eastern region, with no significant disturbance along the Declination. 

\end{abstract}


\section{Introduction}
\label{sec:Introduction}

The Large and the Small Magellanic Clouds (LMC and SMC) are the closest interacting galaxies to the Milky Way (MW), at a distance of $\sim$ 50 kpc \citep{2014..de..Grijs,2019Pietrzy..LMC..distance} and $\sim$ 60 kpc \citep{2015..de..Grijs,2020ApJ...904...13Graczyk..SMC..dist}, respectively. Both the Clouds, the Magellanic Bridge, the Magellanic Stream, and the Leading Arm are together known as the Magellanic System. The Magellanic Bridge, which connects the LMC and the SMC, is composed of gas and stars, and its origin is due to the recent collision between the two galaxies at $\sim$ 200 Myr \citep{1963..Hindman..MB,1985..Irwin..MB,1994...Gardiner,2007..Muller_MBR,2019Zevick..MB}. The Magellanic Stream \citep{1992..Liu..MS,1994...Gardiner} is a gaseous tail behind the Clouds, orbiting the MW, and the counterpart of the Stream, known as the Leading Arm \citep[first detected by][]{1998..mw-lmc-smc-inter} is approaching the Galactic disk \citep[][]{2020..Luchin..MS}. Both the Stream and the Leading Arm are the tidal features formed due to MW--LMC--SMC interactions \citep{1992..Liu..MS,1994...Gardiner,Nidever_2008..MS,2010..Nidever...MS,2016A..Onghia..MS,2020..Luchin..MS}.  The LMC and the SMC have undergone repeated interactions between each other \citep{2010..Glatt..MCs..Interaction, 2011..Indu..MCs..int,2018..Rubele..MCs..int,2018..PK..Nayak,2019..Joshi..MCs..int} and they are in their first infall toward the MW \citep{2006..Kallivayali..firts_passage_MCs, 2007..Besla..first..passage..MCs, 2023chandra..first..passage..MCs}. However, \citet{2023..Vasiliev} suggested the second passage of the LMC around the MW.

The SMC is a gas-rich irregular dwarf galaxy with a less pronounced bar and an eastern extension, known as the SMC Wing \citep{1972..de..Vaucouleure..optical..center,2012..Smitha..NE-SW..SMC,2017..Ripepi..NE-SW..SMC,2019Dalal..MORPHOLOGY..smc}. The SMC is nearly ten and hundred times less massive (dynamical mass $\approx 10^9 M_\odot$) than the LMC and the MW, respectively \citep{2004..stanimirovi..SMC..mass,van_der_Marel_2008,2014..van..der..Marel...LMC..mass}.
Therefore, morphology, dynamics and evolution of the SMC have been significantly shaped by the gravitational interactions with both the LMC and the MW
\citep{1998..mw-lmc-smc-inter,2020..Massana..SMC..int,Tatton..2021..inter}. Different stellar populations of the SMC show distinct morphology \citep[e.g.][]{ 1992..Gardiner..diff..pop,2004..active..sf..dust...BOT, 2019Dalal..MORPHOLOGY..smc,2021..Gaia}. The young population in the SMC is unevenly distributed, with a higher concentration in the central region and the SMC Wing that connects to the Magellanic Bridge \citep{2000..Zaritsky..SMC..morphology,2015MNRAS.449..639Rubele..SFH,2019Dalal..MORPHOLOGY..smc}. In the case of older stellar population, their spatial distribution appears more uniform and spheroidal\slash ellipsoidal \citep[e.g.][]{2000..Zaritsky..SMC..morphology,2004..stanimirovi..SMC..mass,2004AJ....127.1531Harris...SFH..SMC,2012AJ....144..106Haschke..3d..SMC,2015MNRAS.449..639Rubele..SFH,2019Dalal..MORPHOLOGY..smc}. Studies found that the SMC is elongated along the northeast-southwest (NE--SW) axis and NE of the SMC is closer to us \citep{2012..Smitha..NE-SW..SMC,2016..Scowcroft..NE-SW..SMC,2017..Ripepi..NE-SW..SMC,2024..Murray..SMC.dwarf..two..superimposed}. 

The SMC outskirts are home to several stellar sub-structures \citep{2017..Pieres..smc..ss,2018..Mackey..SMC..SS,2019..Belokurov..smc..ss,2020..Massana..SMC..int} such as northeastern SMC shell \citep[younger than 500 Myr;][]{2019..martinez..shell,2024..Hota..SMC..Shell,2024..Sakowska..shell}, the existence of the tidal counterpart of the Magellanic Bridge \citep[Counter Bridge;][]{2021..dias..CB}, the West Halo, a structure moving away from the SMC as confirmed in the western outskirts of the SMC \citep[][]{2016..Dias..WH,2018..Niederhofer..WH,2018..Zivic..MCs..PM,2022..Dias..WH}. \citet{2023..Cullinanesmc_south_ss} identified two different populations in the extreme south of the SMC outskirts. One population exhibits the properties of the SMC outskirts, while the other is the debris from the SMC tidally stripped due to interaction with the LMC. \citet{2021..abinaya} found a foreground stellar population in the eastern SMC outskirts, which they linked to the counterpart of the gaseous Magellanic Bridge.

Observing the SMC across various wavebands is essential to trace its morphology as a function of the stellar population, which is a proxy to probe the evolution of the SMC. Near-infrared (NIR) observations of the SMC were conducted by the Two Micron All Sky Survey \citep[2MASS;][]{2003..Cohen..2MASS} and by Visible and Infrared Survey Telescope for Astronomy (VISTA) survey of the Magellanic Clouds system \citep[VMC;][]{2011..VMC..survey..M-R-L..Cioni}. Medium and far-infrared (FIR) surveys were carried out by Spitzer \citep{2011..Gordon..SAGE}. Gaia has provided optical photometry and astrometry observations of the SMC \citep{2016..Gaia..Mission,2021..Gaia,2021GaiaEDR3,2023GaiaDR3}, and the Survey of the MAgellanic Stellar History has supplied deep optical data of the SMC \citep[SMASH;][]{2017..SMASH..Nidever,2021..Nidever..SMASH..DR2}. To understand the recent interaction and its influence on the spatial distribution of the young population, the far-ultraviolet (FUV) band is essential as it is a significant tracer of young stars that exhibit a peak in their flux in the FUV waveband \citep{1997..Cornett..UIT,2023..UVIT..SMC..Devaraj,2024..Hota..SMC..Shell}. The Hubble Space Telescope (HST) provides high resolution images in multiple passbands, including the FUV, but the coverage is limited due to a small field of view (FOV). Other instruments, such as the Ultraviolet Imaging Telescope \citep[UIT;][]{1994..UIT..cornett,1997..Cornett..UIT} and Swift/UVOT \citep[UltraViolet and Optical Telescope;][]{Gehrels_2004..UVOT} with spatial resolution greater than 2 $''$ have also observed the SMC. The SMC was partially observed by the Galaxy Evolution Explorer \citep[GALEX;][]{SIMGALEX} in the near-UV (NUV) and FUV bands with a spatial resolution of $\sim$ 5$''$. However, some areas of the SMC lack FUV observations by GALEX. The Ultra Violet Imaging Telescope \cite[UVIT;][]{2017June..TondonUVIT} with its superior spatial resolution ($\sim$ 1.4$''$) to GALEX, UIT, and UVOT, is capable of providing a good FUV coverage of the SMC, that has been lacking so far.

In this work, we present a point source catalog of the SMC in the FUV band based on the UVIT images of 39 fields in the SMC. The catalog is cross-matched with the VMC and Gaia Data Release 3 \citep[DR3;][]{2023GaiaDR3} catalogs to provide the corresponding IDs. We also present the morphology and kinematics of various populations of the SMC that are detected in the FUV. This paper is arranged as follows: Section~\ref{sec:data} covers the data used in this work and their analysis. We show the location of different types of populations on the color-magnitude diagrams (CMDs; UV--optical and IR CMDs) in Section~\ref{sec: cmd}. Section~\ref{sec: spatial_distribution} and Section~\ref{sec:kinematics} explore their morphology and kinematics, respectively. Discussion is in Section~\ref{sec:Discussion} and the summary and conclusions are in Section~\ref{sec:Summary}.

\section{Data and Analysis}
\label{sec:data}

\subsection{Observation and data reduction}
\label{subsec:obs_data_reduction}

\begin{figure*}       
\includegraphics[width=\textwidth]{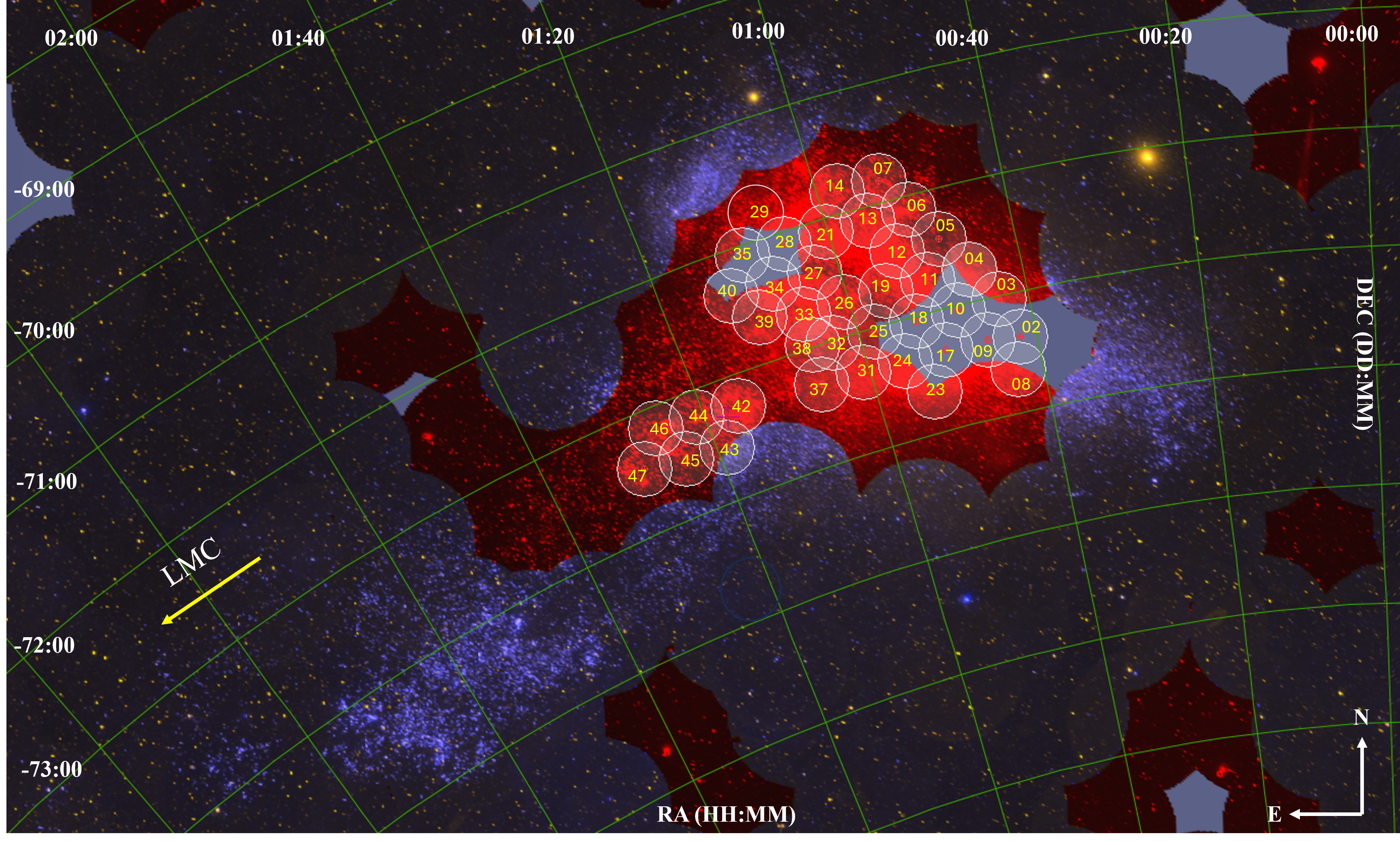}
\caption{The SMC: UVIT observed fields in white circles are displayed over the GALEX background. The background is a two-color combined image of GALEX, where the blue color represents FUV observations, whereas the red color denotes NUV observations (encompassing both GR6 and GR7 data releases). Yellow arrow points in the direction of the LMC \citep{2017..Belokurov..Magellanic..system,2020..De..Leo..tidal..scars..SMC}.
\label{fig:Galex}}
\end{figure*}

UVIT comprises two distinct telescopes of 38 cm in diameter; one is dedicated to the FUV (130 -- 180 nm) channel, while the other one is for NUV (200 -- 300 nm) and visible (VIS: 350 -- 550 nm) channels. In each of these channels, multiple filters are also present \citep{2020TondonUVITfilters}. The VIS observations are used primarily for drift correction due to the spacecraft motion \citep{2017June..TondonUVIT}. UVIT has a field of view of $\sim$ 28$'$ with a spatial resolution of $\sim$ 1.4$''$. The comprehensive information regarding the instrument and calibration of UVIT are available in \citet{2011PostemaDetectorPC,2012kumarUVIT,10.1117/12.2235271,2017Sep..TondonUVIT}. In this work, we utilized FUV images of thirty-nine (39) UVIT pointings in the medium passband filter; F172M \citep[silica: 1717$\pm$125 \AA;][]{2020TondonUVITfilters}. The Level 1 (L1) data for the UVIT observed fields are publicly available\footnote{\url{https://astrobrowse.issdc.gov.in/astro_archive/archive/Home.jsp}}. The 39 SMC fields used in this study are shown in \autoref{fig:Galex}, where pointings are overlaid on the GALEX image, with each circle having the UVIT field of view. The observational details of the SMC fields are provided in Table~\ref{Tab: data_table}. We note from Table~\ref{Tab: data_table} that out of 39 SMC fields, 25 SMC fields have exposure time $>$ 900 sec, 9 fields have exposure time between 400--830 sec, while 3 fields have exposure time between 300 to 400 sec. Two SMC fields, SMC-02 and SMC-42, exhibit shorter exposure time ($<$ 200 sec). SMC-02 is situated on the periphery of the SMC's main body, whereas SMC-42 is located within the inner region of the SMC Wing. We note that this study does not include the regions of the SMC covered by \citet{2024..Hota..SMC..Shell} and \citet{2023..UVIT..SMC..Devaraj}.

The science-ready images of the SMC fields are created by processing the L1 data obtained by UVIT using the CCDLAB pipeline as described in \citet{2017Postmaccdlab,2021Postmaccdlab}. This pipeline autonomously applies corrections such as flat fielding, fixed pattern noise, and spacecraft drift. Astrometry of these science-ready images are done using the same pipeline with WCS (World Coordinate System) reference from Gaia DR3 data \citep{2023GaiaDR3}. Consequently, we obtain science-ready images accompanied by astrometry information of the SMC fields.

\begin{deluxetable}{lccccBcccccBcccc}
\label{Tab: data_table}
\tabletypesize{\scriptsize}
\tablewidth{0pt} 
\tablecaption{Details of the UVIT observed SMC fields. RA and Dec are the central coordinates of the observed fields.}
\tablehead{
\colhead{Observed Field} & \colhead{RA [deg]}& \colhead{Dec [deg]}&
\colhead{t$_{\mathrm{exp}}$ [sec]}}
\startdata 
    SMC-02 & 10.47 & $-73.38$ & 153.0 \\ 
    SMC-03 & 10.84 & $-73.03$  & 954.2 \\ 
    SMC-04 & 11.49 & $-72.73$  & 953.9 \\  
    SMC-05 & 12.11 & $-72.43$  & 954.9 \\   
    SMC-06 & 12.72 & $-72.13$  &  445.3 \\   
    SMC-07 & 13.30 & $-71.82$  & 955.1 \\   
    SMC-08 & 10.76 & $-73.64$  & 954.6 \\   
    SMC-09 & 11.43 & $-73.34$  & 954.3 \\   
    SMC-10 & 12.08 & $-73.04$  & 954.8 \\   
    SMC-11 & 12.70 & $-72.74$  & 954.9\\    
    SMC-12 & 13.31 & $-72.43$  & 954.4\\    
    SMC-13 & 13.90 & $-72.13$  & 951.6 \\    
    SMC-14 & 14.46 & $-71.82$  & 951.7 \\    
    SMC-17 & 12.69 & $-73.35$  & 1022.3 \\    
    SMC-18 & 13.32 & $-73.04$  & 965.3 \\     
    SMC-19 & 13.92 & $-72.74$  & 957.2 \\     
    SMC-21 & 15.07 & $-72.12$  & 955.8 \\    
    SMC-23 & 13.32 & $-73.65$ & 952.7 \\      
    SMC-24 & 13.95 & $-73.35$  & 951.0 \\    
    SMC-25 & 14.55 & $-73.04$  & 952.6 \\       
    SMC-26 & 15.14 & $-72.73$  & 634.3 \\       
    SMC-27 & 15.70 & $-72.42$  & 952.0 \\       
    SMC-28 & 16.24 & $-72.10$  & 953.7 \\       
    SMC-29 & 16.77 & $-71.79$  & 953.9 \\
    SMC-31 & 15.21 & $-73.34$  & 951.3 \\    
    SMC-32 & 15.79 & $-73.03$  & 951.3 \\    
    SMC-33 & 16.35 & $-72.71$  & 825.8 \\    
    SMC-34 & 16.89 & $-72.40$  & 472.0\\        
    SMC-35 & 17.42 & $-72.08$  & 485.3 \\      
    SMC-37 & 16.46 & $-73.32$  & 377.0 \\       
    SMC-38 & 16.37 & $-72.98$  & 371.0 \\       
    SMC-39 & 17.54 & $-72.62$  & 395.4 \\       
    SMC-40 & 18.08 & $-72.37$  & 471.8 \\        
    SMC-42 & 18.97 & $-73.27$  & 187.8 \\        
    SMC-43 & 19.71 & $-73.55$  & 951.7 \\
    SMC-44 & 20.21 & $-73.23$  & 951.6 \\
    SMC-45 & 20.97 & $-73.51$  & 470.6 \\        
    SMC-46 & 21.45 & $-73.18$  & 472.0 \\      
    SMC-47 & 22.23 & $-73.46$  & 471.5\\      
\enddata

\end{deluxetable}

\subsection{PSF photometry}
\label{subsec:psf_phot}

\begin{figure*}[htpb]    
\includegraphics[width=\textwidth,trim={0.5cm 0 0.5cm 0.5cm},clip]{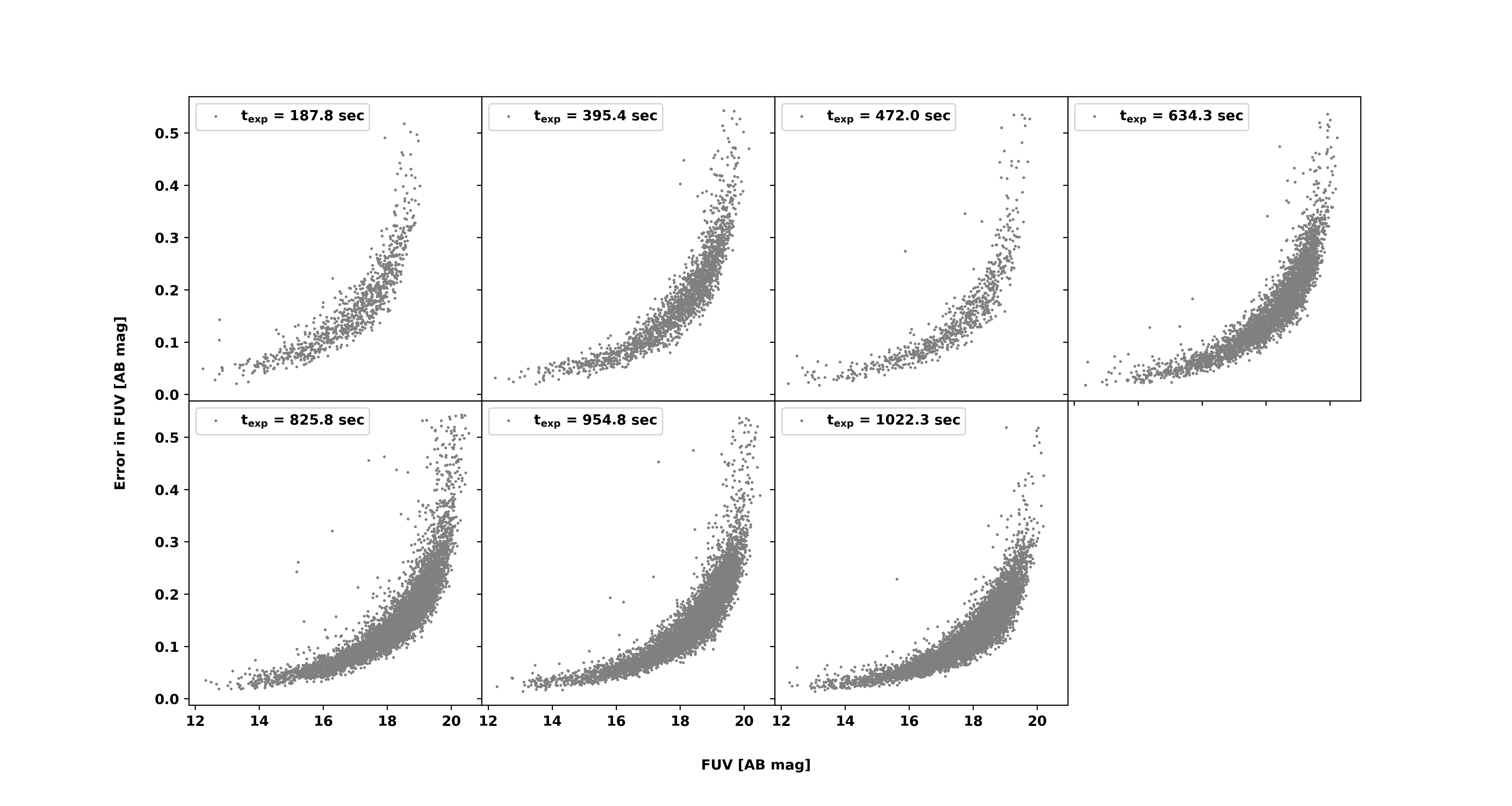}
\caption{FUV magnitude vs error in magnitude (PSF fit error) for various fields of the SMC with different exposure times. 
\label{fig:mag_vs_merr}}
\end{figure*}

To determine the magnitudes of individual sources in the science-ready images of the SMC fields, we performed point spread function (PSF) photometry. The PSF photometry involves the following major steps: detecting sources, performing aperture photometry, creating a model PSF, and applying the model PSF to all the detected sources. We conducted PSF photometry using the DAOPHOT package of IRAF \citep[Image Reduction And Analysis Facility;][]{1987..Stetson..IRAF} tool. First, we used the IMEXAM routine to estimate the background noise (sigma) and the full-width half-maxima (FWHM) of the sources in the image. Subsequently, the DAOFIND task is utilized for source detection, employing a threshold set at five times the background level. During this process, main parameters such as the exposure time of the observed field, the estimated FWHM of the sources, and the zero point of the filter are imposed. Following that, aperture photometry was carried out using the PHOT task, and a set of isolated bright stars was chosen to construct a model PSF using the PSF SELECT task. The PSF model was created by running the PSF task in IRAF and subsequently applied to all detected stars to determine the PSF magnitude simultaneously through the ALLSTAR task. PSF correction was performed to convert PSF magnitude to aperture magnitude. Afterwards, an aperture correction was applied to all detected sources using the curve of growth for a few isolated bright stars in the images. We transformed the instrumental magnitude to the AB magnitude scale using the relation \citep{2017June..TondonUVIT},

\begin{equation}
   m_{\text{AB}} = -2.5 \log(\text{CPS}) + ZP
\end{equation}

where CPS is the counts per second, and ZP is the zero-point given by

\begin{equation}
ZP = \left(-2.5 \log(UC) \cdot (\lambda_{\text{mean}})^2 \right) - 2.407
\end{equation}

The zero point value of the F172M filter is 16.274 as specified in \citet{2020TondonUVITfilters}. Saturation correction was applied for all the detected sources, as detailed in \citet{2017Sep..TondonUVIT}. The variation of PSF fit error with the obtained FUV magnitude and exposure time is shown in \autoref{fig:mag_vs_merr} using seven sample SMC fields: SMC-42 (187.8 sec), SMC-39 (395.4 sec), SMC-46 (472.0 sec), SMC-26 (634.3 sec), SMC-33 (825.8 sec), SMC-10 (954.8 sec), and SMC-17 (1022.3 sec). Other observed fields of the SMC exhibit a similar trend. To avoid the edge effect, sources within a $\sim$1$'$ annulus near the edges of the observed fields are excluded.

\subsection{Completeness factor}
\label{subsec:completeness}

\begin{figure}%[htpb]
\includegraphics[width=\columnwidth]{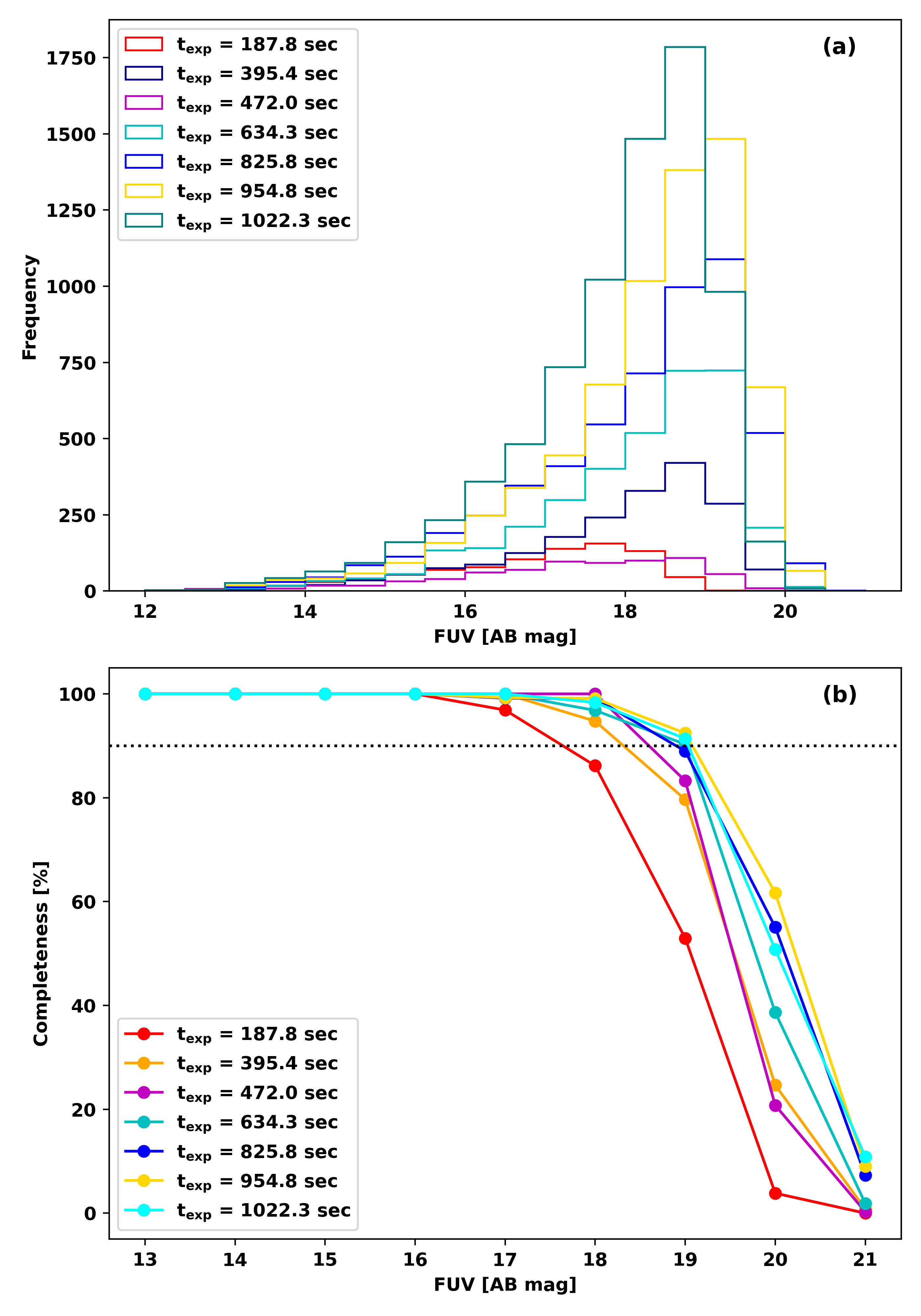}
 \caption{(a)The FUV magnitude distribution and (b)completeness in percentage for seven UVIT observed SMC fields with different exposure times. The black dotted line represents 90\% of completeness.}
 \label{fig:hist_mag}
\end{figure}

The peak of the magnitude distribution of an observed field is a pointer to the completeness in the data \citep{2020..Laehy..UVIT..M31,2023..UVIT..SMC..Devaraj}. In \autoref{fig:hist_mag}(a), we have shown the distribution of FUV magnitude of SMC fields with seven different exposure times (same as in \autoref{fig:mag_vs_merr}). The distribution shows peaks in magnitudes between 17.5 to 19.5 mag, and we note a sudden drop in the number of fainter stars along with an increase in the FUV magnitude error, as shown in \autoref{fig:mag_vs_merr}. This points to a range in incompleteness in the observed fields that need to be quantified.

In order to quantitatively estimate the incompleteness in the data across the observed fields, we conducted artificial star tests (ASTs) on the above mentioned seven fields to determine the completeness factor (CF).  We performed the AST using the DAOPHOT routine in IRAF. We randomly added artificial stars ranging between $\sim$ 10\% to 15\% of the detected stars in the individual images. More details of the AST and CF can be found in \citet{1991..Ram..Sagar..completeness,2011..Hu..Yi..AST,2022..sahu..AST..PSF}. Then, we followed the same procedure for the PSF photometry as in Subsection~\ref{subsec:psf_phot} and recovered the added stars. The CF is estimated as the ratio of the number of recovered stars to the number of added stars.

We executed AST twice for the magnitude range of 12 -- 21 mag and thrice for 15 -- 21 mag, 18 -- 21 mag, 19 -- 21 mag, and 20 -- 21 mag on an image. Then, the average CF is estimated from the multiple ASTs for a field, and the percentage of completeness is estimated. The same procedure of AST is followed for all the images. \autoref{fig:hist_mag}(b) shows the variation of completeness fraction with the FUV magnitude for seven SMC fields. Extending these CFs to all fields, we conclude that the observed fields exhibit 100\% completeness at $\sim$ 16 mag, and beyond which the completeness starts to reduce. For artificial stars up to 18 mag, all considered images (with different exposures and crowding) show a recovery rate of around 90\%. Therefore, we consider a completeness of 90\% or more at $\sim$ 18 mag (FUV) for all the observed SMC fields.

\subsection{FUV catalog and Cross-matching}
\label{subsec:Cross-match}

\begin{deluxetable*}{lcccccccBcccccccBccccccc}
\label{Tab: catalog}
\tabletypesize{\scriptsize}
\tablewidth{0pt} 
\tablecaption{FUV catalog of the sources detected in the direction of the SMC. Columns 1 to 8 of the catalog table present UVIT\_ID, spatial coordinates, magnitude in the F172M filter, corresponding fit error in magnitude, VMC\_PSF\_Source\_ID, Gaia\_DR3\_Source\_ID and probable SMC member (PSM) respectively. The complete table is accessible as supplementary online material.
\label{tab:UV_catalog}}
\tablehead{
\colhead{UVIT\_ID} & \colhead{RA [deg]}& \colhead{Dec [deg]} & \colhead{FUV [AB mag]} &
\colhead{Error [AB mag]} & \colhead{VMC\_PSF\_Source\_ID} & \colhead{Gaia\_DR3\_Source\_ID} & \colhead{PSM}}
\startdata 
    UVIT00000 & 10.315324 & $-73.277606$ & 16.116 & 0.098 & 558371534268 & 4688850813948992896 & Y\\
    UVIT00001 & 09.940444 & $-73.294990$ & 16.586 & 0.163
 & 558371499893 & 4688862942940210304 & Y\\
 UVIT00002 & 10.548706 & $-73.303662$ & 17.022 & 0.154 & 558371482809 & 4688848855413218688 & Y \\
    UVIT00003 & 10.034966 & $-73.309045$ & 16.083 & 0.129
 & 558371472160 & 4688851054467630464 & Y\\
 UVIT00004 & 10.368390 & $-73.330106$ & 17.078 & 0.134
 & 558371430415 & 4688847446664173568 & N \\
\enddata
 \tablecomments{Total detected FUV stars: $\sim$ 76800; cross-matched with Gaia: 73395; cross-matched with VMC: 72790; cross-matched with both Gaia and VMC: $\sim$ 70900; not cross-matched with VMC and Gaia: 1456. PSM identified by the truncated-optimal NN with RUWE $<$ 1.4 (62,901 sources) are presented in the last column of the table where Y and N stand for ``Yes'' and ``No'' respectively.}
\end{deluxetable*}

To produce the FUV catalog, we combined the photometric data of all the 39 UVIT observed fields. The presence of  data duplication in the overlapping regions are addressed by considering the source with better signal-to-noise ratio and/or photometric error.  For the catalog, we selected FUV sources with a magnitude error of $\le$ 0.2 mag (for signal-to-noise ratio; S/N $\ge$ 5) and an FUV magnitude $>$ 13.2 mag as the FUV sources brighter than $\sim$ 13.2 mag may be saturated \citep{2017Sep..TondonUVIT}. With the above criteria, a catalog with $\sim$ 76800 FUV sources are listed in Table~\ref{Tab: catalog}. Though this catalog covers the inner SMC, we point out the presence of a gap in the observed area, centered around the coordinates (${\mathrm{\alpha}}$,${\mathrm{\delta}}$):($14^{\circ}$, $-72.5^{\circ}$) as shown in \autoref{fig:Galex}.

To better characterize the detected FUV sources in the SMC, we cross-matched these FUV sources with IR and optical data. We obtained the Gaia DR3 data \citep[][]{2023GaiaDR3} of the SMC with ADQL query in the Gaia archive \footnote{\url{https://gea.esac.esa.int/archive/}} as mentioned in \cite{2021..Gaia}. The number of cross-matched UVIT-Gaia sources within a search radius of 1$''$ is 73395, and their corresponding Gaia\_DR3\_Source\_IDs are listed in the seventh column of Table~\ref{Tab: catalog}. The PSF photometry catalog of the SMC in IR band \citep{2018..Rubele..MCs..int} as part of the VMC, from the European Southern Observatory (ESO) archive \footnote{\url{https://archive.eso.org/cms.html}} is cross-matched with the detected FUV sources within a search radius of 1$''$ to obtain 72790 UVIT-VMC cross-matched sources (VMC\_PSF\_Source\_ID is listed in 6th column of Table~\ref{Tab: catalog}). We note that $\sim$ 70900 FUV sources have both optical (Gaia) and IR (VMC) data, while $\sim$ 1450 FUV sources do not have either Gaia or VMC cross-matches.

We performed a cross-match of the FUV catalog (see Table~\ref{Tab: catalog}) with the SIMBAD database using a search radius of 1$''$.
This analysis revealed that 9133 stars were matched with the SIMBAD database, and we note that $\sim$ 30\% of cross-matched sources are eclipsing binaries, and another $\sim$ 30\% are classified as "star" (no spectral classification). We also found some UVIT-SIMBAD stars are classified as Supergiants ($\sim$ 200), white dwarfs ($\sim$ 200), Be stars ($\sim$ 650), and high mass X-ray binaries ($\sim$ 100).
We also cross-matched our catalog with the catalog of 5324 massive stars in the SMC by \citet{2010..Bonanos..Massive_stars_SMC}, using a search radius of 1$''$ (UVIT observed fields cover only 65\% of the catalog), and 2388 stars were identified. The majority of these cross-matched stars are of OB type (O-type: $\sim$ 132, B-type: $\sim$ 2000).

\subsection{Most Probable FUV sources of the SMC}
\label{subsec: most_probable_smc_stars}

\begin{figure*}[htpb]
\includegraphics[width=\columnwidth,trim={0.3cm 2.0cm 0.3cm 1.5cm},clip]{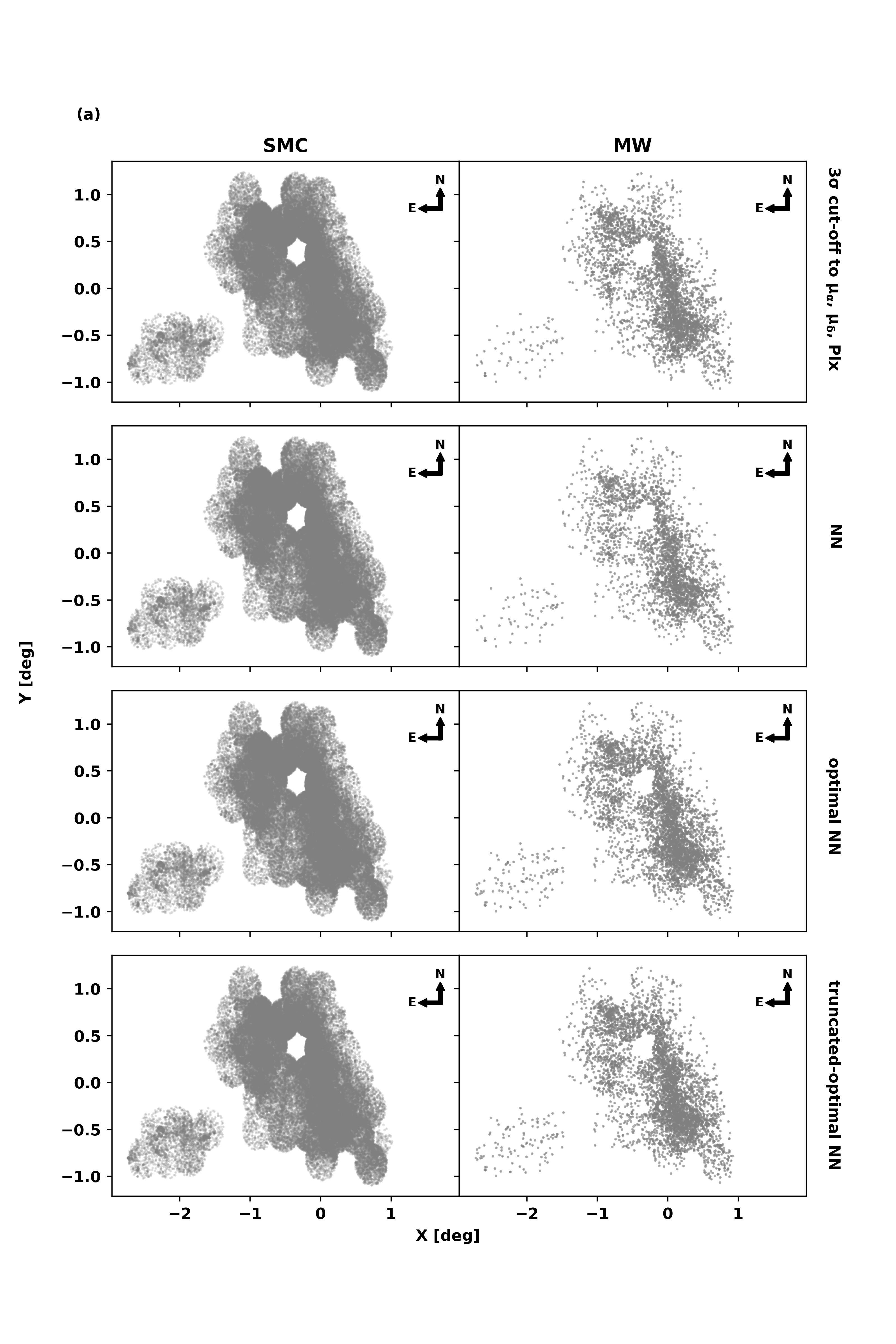}
\includegraphics[width=\columnwidth,trim={0.3cm 2.0cm 0.3cm 1.5cm},clip]{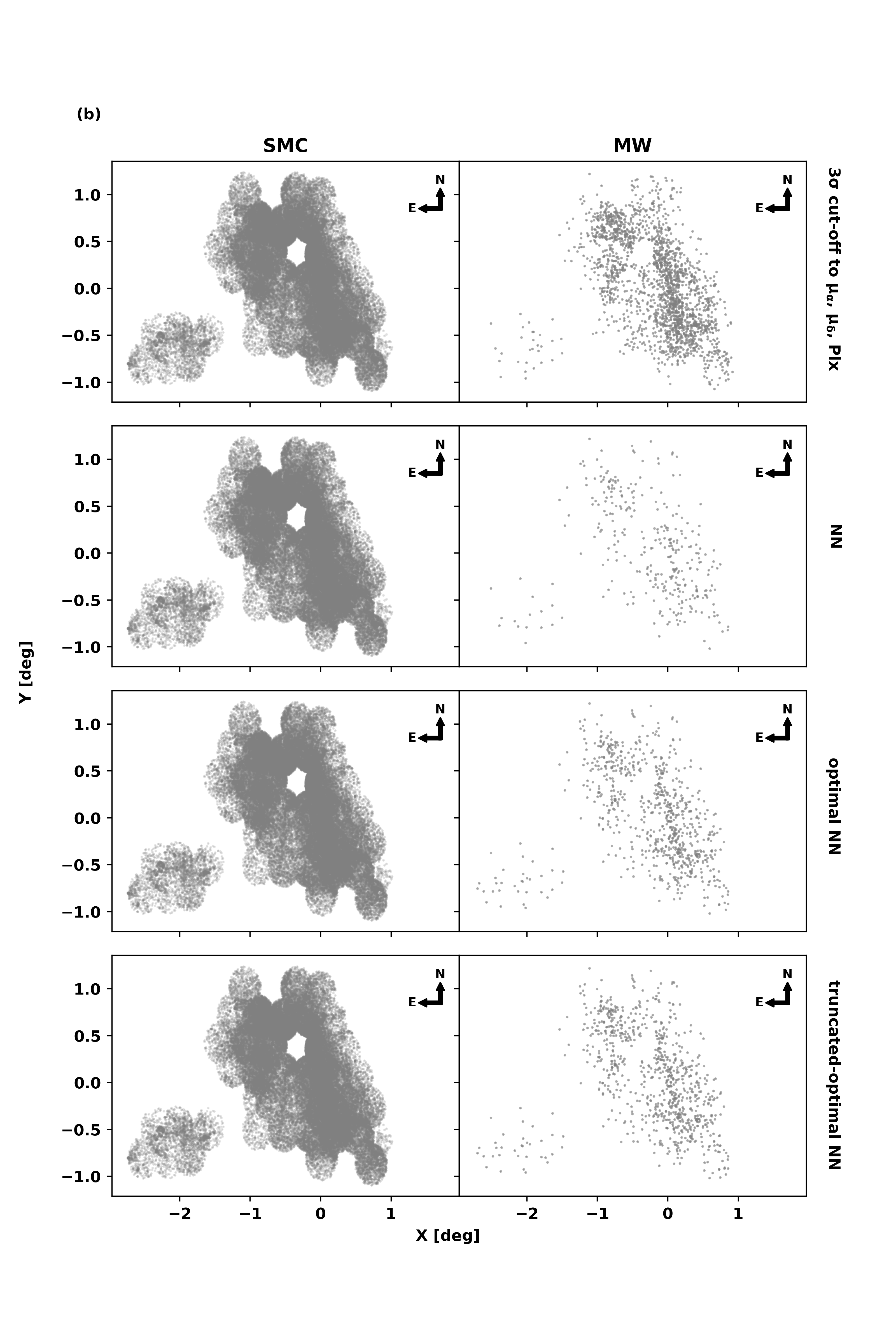}
\caption{(a) and (b) represent the spatial distribution of probable FUV stars of the SMC and the MW obtained from different methods under two distinct conditions: without RUWE constraints and with a RUWE $<$ 1.4, respectively.
\label{fig:comparison_ruwe_cutoff}}
\end{figure*}

\begin{deluxetable*}{lcccccBcccccBccccc}
  \label{Tab: ruwe}
   \tabletypesize{\scriptsize}
   \tablewidth{0pt} 
   \tablecaption{Number of probable FUV member sources of the SMC and the MW obtained using different classification methods.}
    \tablehead{
    \colhead{Methods} & \colhead{SMC }& \colhead{MW}& \colhead{SMC (RUWE $<$1.4)} & \colhead{MW (RUWE $<$1.4)}}
     \startdata 
 3$\sigma$ clipping to $<\mu_{\alpha}cos\delta>,<\mu_{\delta}>,<Plx>$ & 69389 & 4006 & 61624 & 2141\\
NN complete* & 70319 & 3076 & 63444 & 321 \\
 Optimal NN* & 68621 & 4774 & 62980 & 785 \\
Truncated-Optimal NN* & 68540 & 4855 & 62901 & 864 \\
    \enddata
    \tablecomments{* \citet{2023..Gaia_catalog_SMC_Prob_Jimenez}}
\end{deluxetable*}
To find the most probable FUV sources in the SMC, we considered only those FUV sources, that have Gaia counterparts. The spherical coordinates of the SMC are projected onto the XY-plane using the zenithal equidistant projection method, where we followed the X and Y conversion as defined in \citet{2001AJ..van..der..Marel..&..Cioni..SMC..projection}. For this, we considered the optical center of the SMC at $\alpha_{\mathrm{SMC}} = 00^{\mathrm{h}} 52^{\mathrm{m}} 12^{\mathrm{s}}.5$ and $\delta_{\mathrm{SMC}}$ = $-72 ^{\circ} $ 49$'$ 43$''$ \citep[J2000;][]{1972..de..Vaucouleure..optical..center}. Here, we followed two criteria for selecting the most probable SMC sources: (1) We implemented a 3$\sigma$ cut-off around the mean value of the parallax (Plx = 0.014 mas) and proper motion (PM) in both Right Ascension (RA) and Declination (Dec) ($\mu_\alpha \cos\delta$ = 0.74 mas/yr, $\mu_\delta = -1.25$ mas/yr) for the detected FUV sources. The resulting number of the most probable FUV sources of the SMC is 69389. Still, there is a possibility that these FUV sources may be contaminated with the background galaxies; (2) We cross-matched the FUV sources with the Gaia probability catalog of the SMC sources provided by \citet{2023..Gaia_catalog_SMC_Prob_Jimenez}. They obtained three samples of the SMC based on probability (P$_{\mathrm{cut}}$) and G mag cut-off. We followed the same cut-offs to obtain three samples of the most probable SMC stars: (i) NN complete (P$_{\mathrm{cut}}=$ 0.01): 70,319 FUV sources, (ii) NN optimal (P$_{\mathrm{cut}} =$ 0.31): 68,621 FUV sources and (iii) NN truncated-optimal sample (P$_{\mathrm{cut}} =$ 0.31 and G $<$ 19.5 mag): 68,540 FUV sources (see Table~\ref{Tab: ruwe}). The spatial distribution of the SMC and MW sources, obtained from different methods, are shown in \autoref{fig:comparison_ruwe_cutoff}(a). We note that the MW FUV stars show a similar spatial distribution as the SMC.

There is a possibility that Gaia sources are non-single or problematic for astrometric solutions \citep[][]{2018..Lindegren}. To identify and remove such UVIT-Gaia sources (see Table~\ref{Tab: catalog}), first, we gave a constraint on Renormalised Unit Weight Error $< 1.4$ \citep[RUWE;][]{2018..Lindegren}. Then, we followed again all the above two criteria for selecting the most probable SMC stars (see Table~\ref{Tab: ruwe}). In \autoref{fig:comparison_ruwe_cutoff}(b), we note that the MW stars obtained using only the first criteria ( 3$\sigma$ cut-off to the means of Plx: 0.013 mas, $\mu_{\mathrm{\alpha}cos\delta}$: 0.74 mas/yr and $\mu_{\mathrm{\delta}}$: $-1.25$ mas/yr) show a spatial distribution like the SMC even after RUWE constraint. But the MW stars obtained based on probability and\slash or G mag cutoff \citep{2023..Gaia_catalog_SMC_Prob_Jimenez} show more or less uniform distribution, which is the expected spatial distribution of the foreground stars. The numbers of most probable stars obtained after RUWE constraint in NN complete (63444), optimal NN (62980), and truncated-optimal NN (62901) samples are similar. Hereafter, we consider truncated-optimal NN samples (i.e., Gaia cross-matched FUV stars with probability $>$ 31\%, G $<$ 19.5 mag, and RUWE $< 1.4$, and listed as probable SMC members (PSMs) in the 8th column of Table~\ref{tab:UV_catalog}) for further analysis.

\section{Color-Magnitude Diagrams}
\label{sec: cmd}

\begin{figure*}
  
        \includegraphics[width=\textwidth,trim={1.2cm 1.2cm 1.5cm 1.2cm},clip]{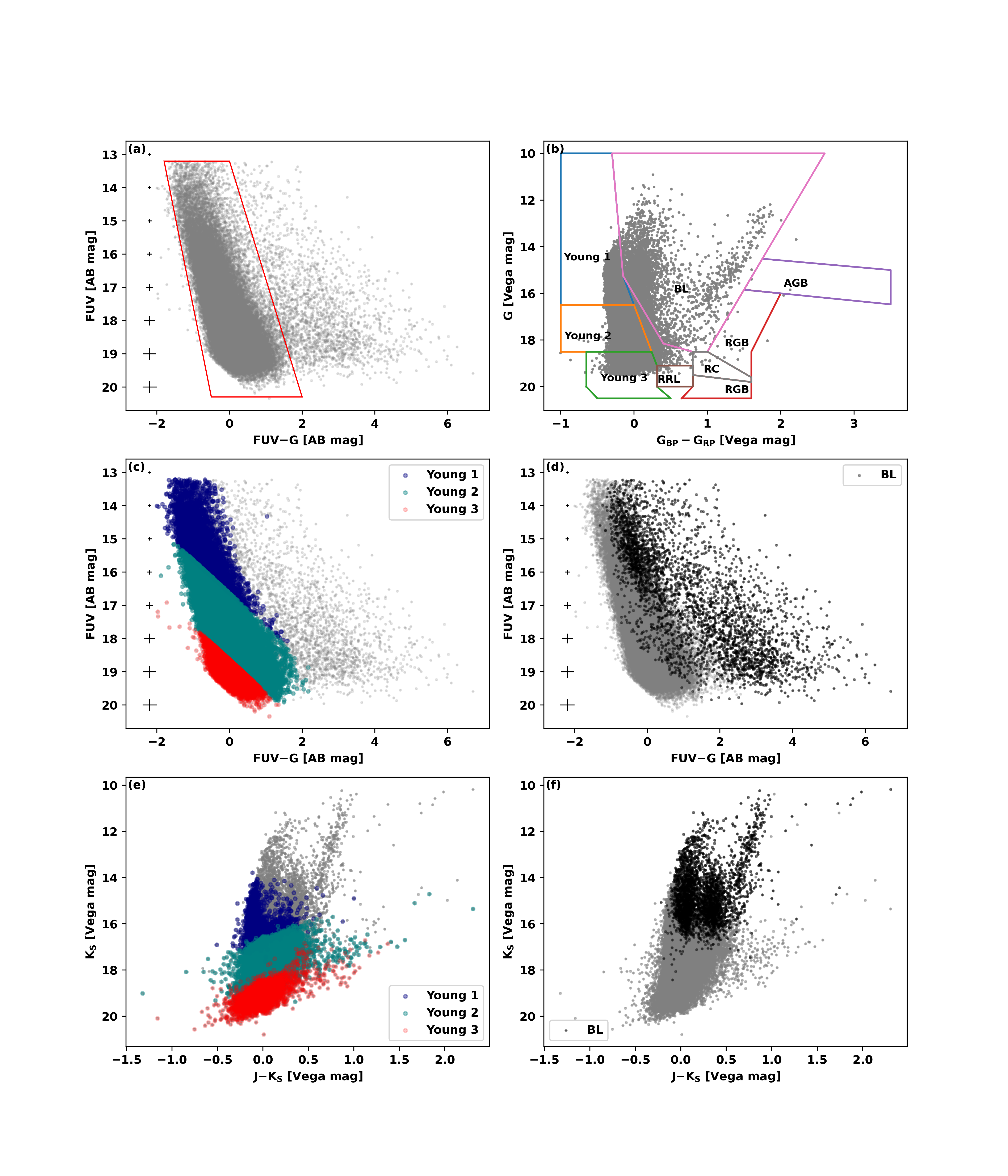}
       \caption{(a) FUV--optical CMD of the SMC sources. The main-sequence branch is shown within a red polygon. (b) Polygons on the optical CMD of the SMC sources representing different populations of the SMC \citep[see right panel of their Fig. 2, ][]{2021..Gaia}. The locations of Young 1, Young 2, Young 3, and BL stars on FUV--optical and IR CMDs are shown in (c) \& (d) and (e) \& (f), respectively. The error bars (median value) for FUV magnitude and FUV--optical color are shown in black for the FUV--optical CMDs. Gray points are the SMC FUV sources.}
       \label{fig:FUV_Gaia_CMDs}
\end{figure*}

A color-magnitude diagram (CMD) is a useful tool to trace the evolutionary paths of different stellar populations. Our goal is to identify various stellar populations within the SMC based on their location in the FUV--optical CMD. As the FUV is in AB magnitude, we converted G mag from the Vega magnitude system to the AB magnitude system\footnote{\url{https://gea.esac.esa.int/archive/documentation/GDR2/Data_processing/chap_cu5pho/sec_cu5pho_calibr/ssec_cu5pho_calibr_extern.html}} to created the FUV--optical CMD, as shown in \autoref{fig:FUV_Gaia_CMDs}(a). To identify various populations, we plotted the Gaia optical CMD (G vs. G$_{\mathrm{BP}}-{\mathrm{G}}_{\mathrm{RP}}$) for the 62901 probable SMC FUV members (see Subsection~\ref{subsec: most_probable_smc_stars}) as shown in \autoref{fig:FUV_Gaia_CMDs}(b) and identified different populations as described by \citet{2021..Gaia}. Here, we considered only the young populations (Young 1, Young 2, Young 3, Blue Loop: BL) as they are detected in large numbers (see \autoref{fig:FUV_Gaia_CMDs}(b)), as listed in Table~\ref{tab:diff_pop}. As the detected number of  Red Giant Branch (RGB), Asymptotic Giant Branch (AGB), Red Clump (RC), and RR-Lyrae (RRL) are significantly low, we did not consider them for further analysis. \autoref{fig:FUV_Gaia_CMDs}(c) and (d) show the position of the Young 1, Young 2, Young 3, and BL stars on the FUV--optical CMD of the SMC. We note that Young 1, Young 2, and Young 3 have distinct locations on the main-sequence branch with an increasing FUV mag (\autoref{fig:FUV_Gaia_CMDs}(c)), whereas BL stars majorly fall on the redder part of the FUV--optical CMD, with some located on the main-sequence branch (\autoref{fig:FUV_Gaia_CMDs}(d)).
 
The young populations defined from Gaia optical CMD (\autoref{fig:FUV_Gaia_CMDs}(b)) are also plotted in the IR CMD (K$\mathrm{_s}$ vs. J--K$\mathrm{_s}$) as shown in \autoref{fig:FUV_Gaia_CMDs}(e) \& (f), after restricting sources with a maximum photometric error of 0.2 mag in the two IR filters. We compared the positions of the populations in the IR CMD with a similar study by \citet{2019Dalal..MORPHOLOGY..smc} (see Figure 1 in their paper). Our analysis revealed that the BL stars align with the positions of supergiant and giant stars, while the Young 1, Young 2, and Young 3 populations are mainly located on the main-sequence and subgiant branches. Therefore, we conclude that the detected SMC FUV stars primarily belong to the main-sequence and supergiant/giant stars.

\begin{deluxetable*}{lcccBcccBccc}
\label{tab:diff_pop}
\tabletypesize{\scriptsize}
\tablewidth{0pt} 
\tablecaption{Description and star count of four young populations in our catalog.}
\tablehead{
\colhead{Population} & \colhead{Description$^*$}& \colhead{N}}
\startdata 
    Young 1 & very young main-sequence (ages $<$ 50 Myr) & 4924\\
     Young 2 & young main-sequence ( 50 $<$ age $<$ 400 Myr) & 40484\\
    Young 3 & intermediate-age main-sequence population (mixed
 ages reaching up 1--2 Gyr) & 11885\\
    BL & blue loop (including classical Cepheids) & 3584\\
 \enddata
 \tablecomments{*The descriptions of the populations are taken from \citet{2021..Gaia}.}
 \end{deluxetable*}

\section{Spatial distribution}
\label{sec: spatial_distribution}

\begin{figure}[htpb]
             
      \includegraphics[width=\columnwidth]{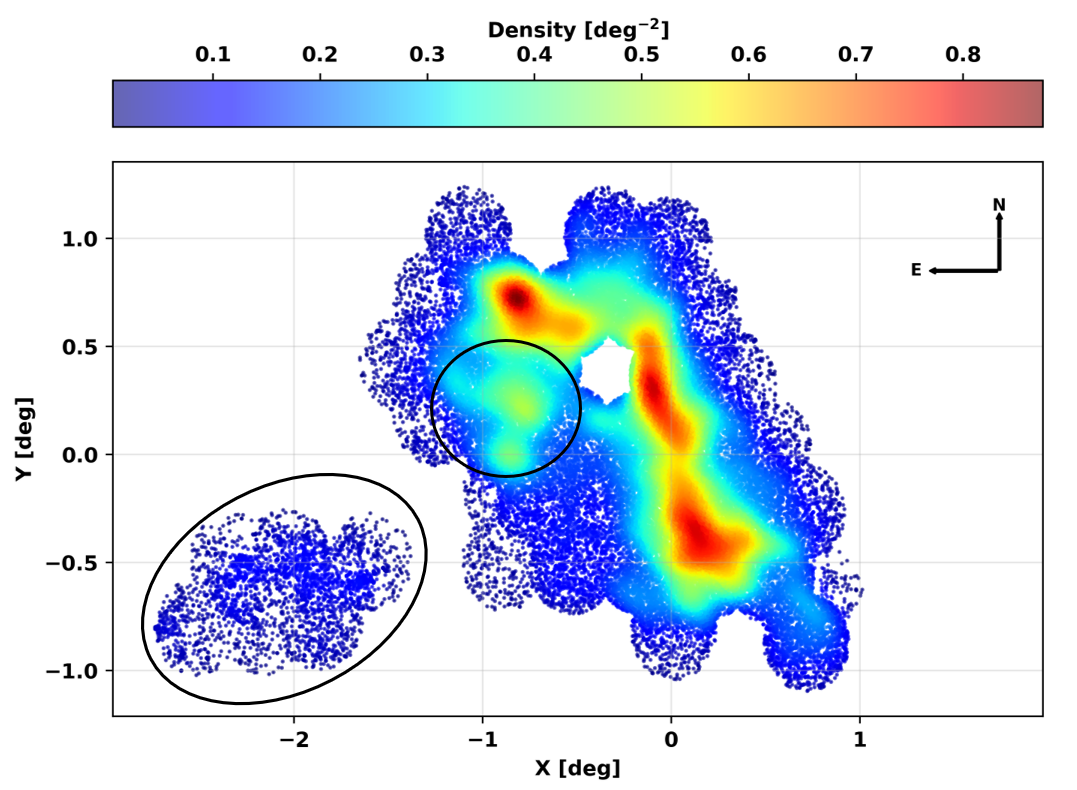}
         \caption{Surface density distribution (KDE; kernel width = 0.1 deg) of the SMC. The color bar represents the density in the unit of deg$^{-2}$. The black ellipse and circle represent the inner SMC Wing and shell-like structure, respectively.}
         \label{fig:kde_plot_SMC}
\end{figure}

The morphological features of the young population in the SMC are examined using the spatial distribution of the FUV stars. It is important to note that the SMC fields exhibit a range of completeness with the FUV magnitude, based on their exposure times and crowding from ~\autoref{fig:hist_mag}(b). Except two, all fields show their peak FUV magnitude around 19 mag before declining (~\autoref{fig:hist_mag}(a)) where completeness is more than $\sim$ 90\%. This suggests that the impact of incompleteness in the morphology based on the FUV stellar density is negligible. The surface stellar density of the SMC is shown in \autoref{fig:kde_plot_SMC}, based on a kernel density estimation (KDE) with a kernel width of 0.1 deg, that is sufficient to identify the dense regions. The distribution as seen in \autoref{fig:kde_plot_SMC} is irregular, with an asymmetric young bar and an eastern extension, the inner SMC Wing at ( $-2.75$ $<$ $\Delta\alpha$ $<$ $-1.25$, $-1$ $<$ $\Delta\delta$ $<$ $-0.25$) deg. The young bar has a relatively higher density segregated in three distinct regions in the inner SMC. When there is a continuity between the SW high-density region to the central high-density region, we note a discontinuity in density at $\sim$ ($-$0.5, 0.5) deg in the NE direction. This stellar density in the young bar is broken (i.e. density is not distributed continuously throughout the SMC bar) and bends towards the east direction. A shell-like structure is detected at ($-$ 0.75, 0.25) deg in the east direction towards the SMC Wing.

\begin{figure*}[htpb]
            
     \includegraphics[width=\linewidth]{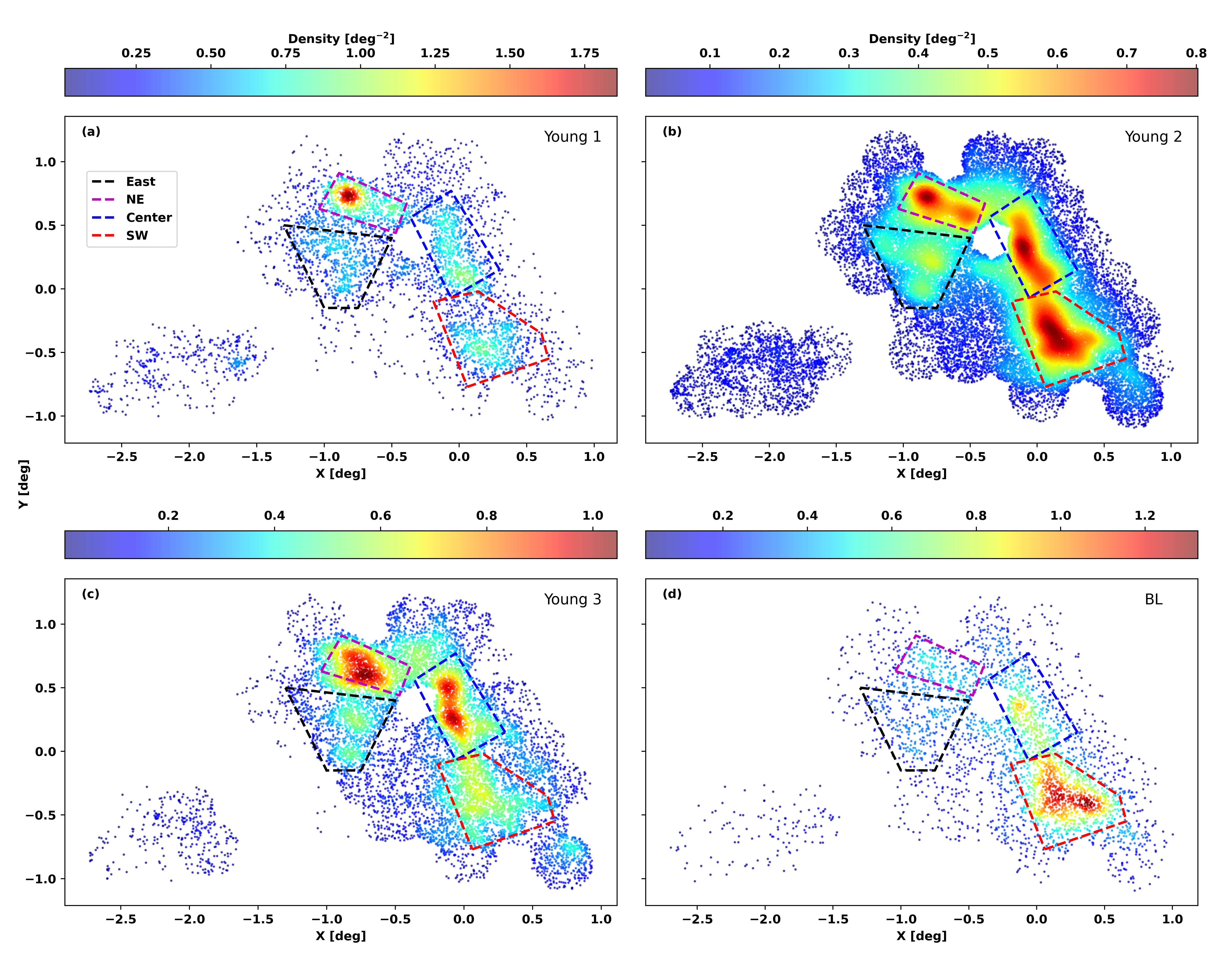}
        \caption{Surface density distribution (KDE; kernel width = 0.1 deg) of Young 1, Young 2, Young 3, and BL populations in (a), (b), (c), and (d), respectively. The color bar represents the density in units of deg$^{-2}$. Dashed polygons mark four distinct dense regions named east, NE, center, and SW. The labeling in panel (a) is consistent across all panels.}
        \label{fig:kde_plot_diff_pop}
 \end{figure*}

In order to trace the morphology, the KDE map of the stellar surface density of Young 1, Young 2, Young 3, and BL stars are shown in  \autoref{fig:kde_plot_diff_pop}. These show 4 distinct dense regions; visually identified and marked as east, NE, center, and SW on the basis of direction,  as shown in \autoref{fig:kde_plot_diff_pop}. The relative density in each region varies with population and the Young 2 population is found to closely resemble the overall density map as shown in \autoref{fig:kde_plot_SMC}. The maximum range in stellar surface density is seen for the Young 1, whereas the other 3 populations have a reduced but similar range of stellar density values. The densest region of the Young 1 population is in the NE of the bar (\autoref{fig:kde_plot_diff_pop}(a)), whereas the densest BL population is found in the SW (\autoref{fig:kde_plot_diff_pop}(d)) of the bar. The broken bar and the shell-like features are predominantly seen in Young 2 population. The densest regions of Young 3 are found in the NE as well as at the central part of the bar (see \autoref{fig:kde_plot_diff_pop}(c)). 

 \section{Kinematics}
\label{sec:kinematics}

As the SMC is disturbed due to interactions, the imprint of interactions could be traceable in the motion of the young stars through a kinematic analysis. We performed a 2-D kinematic analysis of 4 populations as well as the 5 regions using their PM values, and the results are presented here.

\begin{figure*}[htpb]
       
     \includegraphics[width=\linewidth]{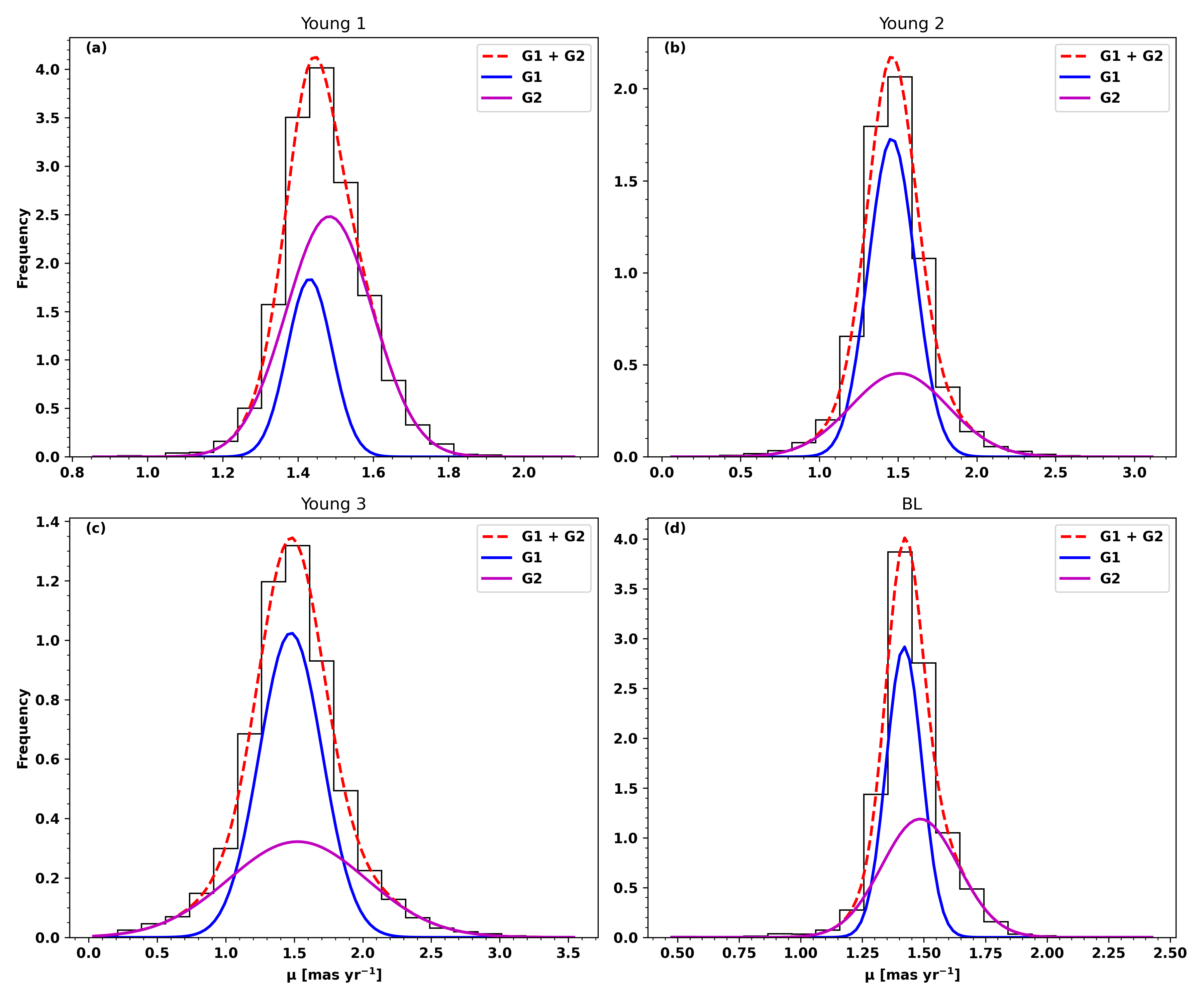}
        \caption{PM distribution of four young populations. Gaussian 1 (G1), Gaussian 2 (G2), and their sum (G1 $+$ G2) curves are shown in blue, magenta, and red, respectively.}
        \label{fig:PM_hist_gaussian_fit}
\end{figure*}

The PM distribution of all four populations and the fitted Gaussian curves are shown in \autoref{fig:PM_hist_gaussian_fit}. The fit parameters, such as the peak and standard deviation values, along with their corresponding errors from the covariance matrix, are listed in Table~\ref{Tab: Double_Gaussian_diff_Pop}. All the PM distributions had to be fitted with two Gaussian profiles. The Gaussian profiles fitted to Young 1 and BL have slightly differing peaks and widths, whereas those fitted to the Young 2 and 3 populations have significantly differing widths. The width of the Young 2 and 3 populations suggest that there is a cool component corresponding to a PM distribution with reduced dispersion and a hot component corresponding to a PM distribution with a relatively large dispersion. Interestingly, the two sub-populations of BL and Young 1 exhibit relatively less widths, when compared to both the widths of Young 2 and Young 3 profiles. Next, we set out to check if both components of the transverse motion (PM) are contributing to the above-observed differences.

\begin{figure*}[htpb]
            
     \includegraphics[width=\linewidth,trim={0.8cm 2.5cm 0.5cm 0.8cm},clip]{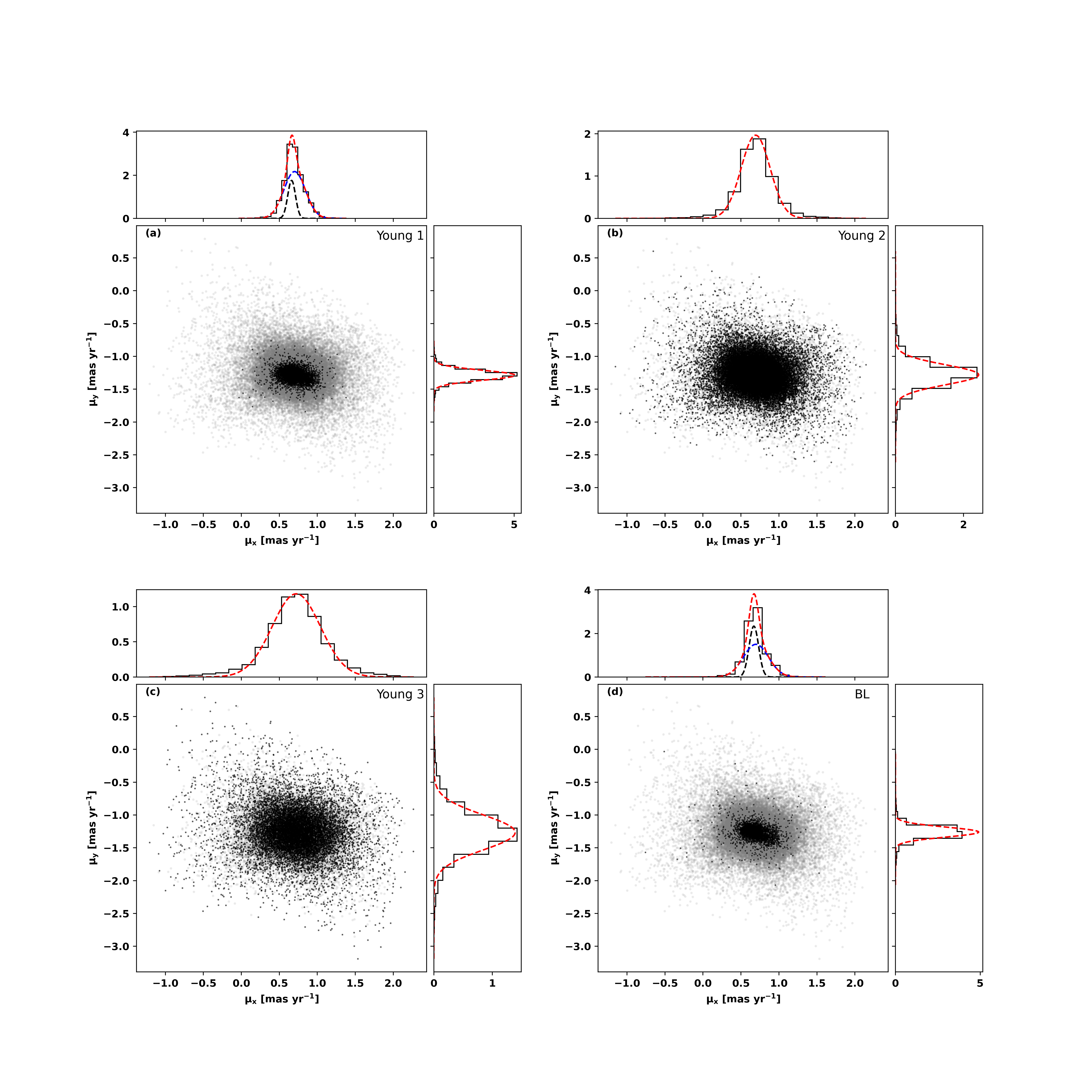}
        \caption{VPDs of four young populations. Distributions of PMRA and PMDec are shown at the top and right. Gaussian (sum) is shown in red for all panels, while Gaussian 1 and Gaussian 2 are shown in black and blue, respectively, in (a) and (d). Gray points represent the SMC FUV stars.}
        \label{fig:vpd_diff_pop}
\end{figure*}

\begin{figure*}[htpb]
       
\includegraphics[width=\textwidth]{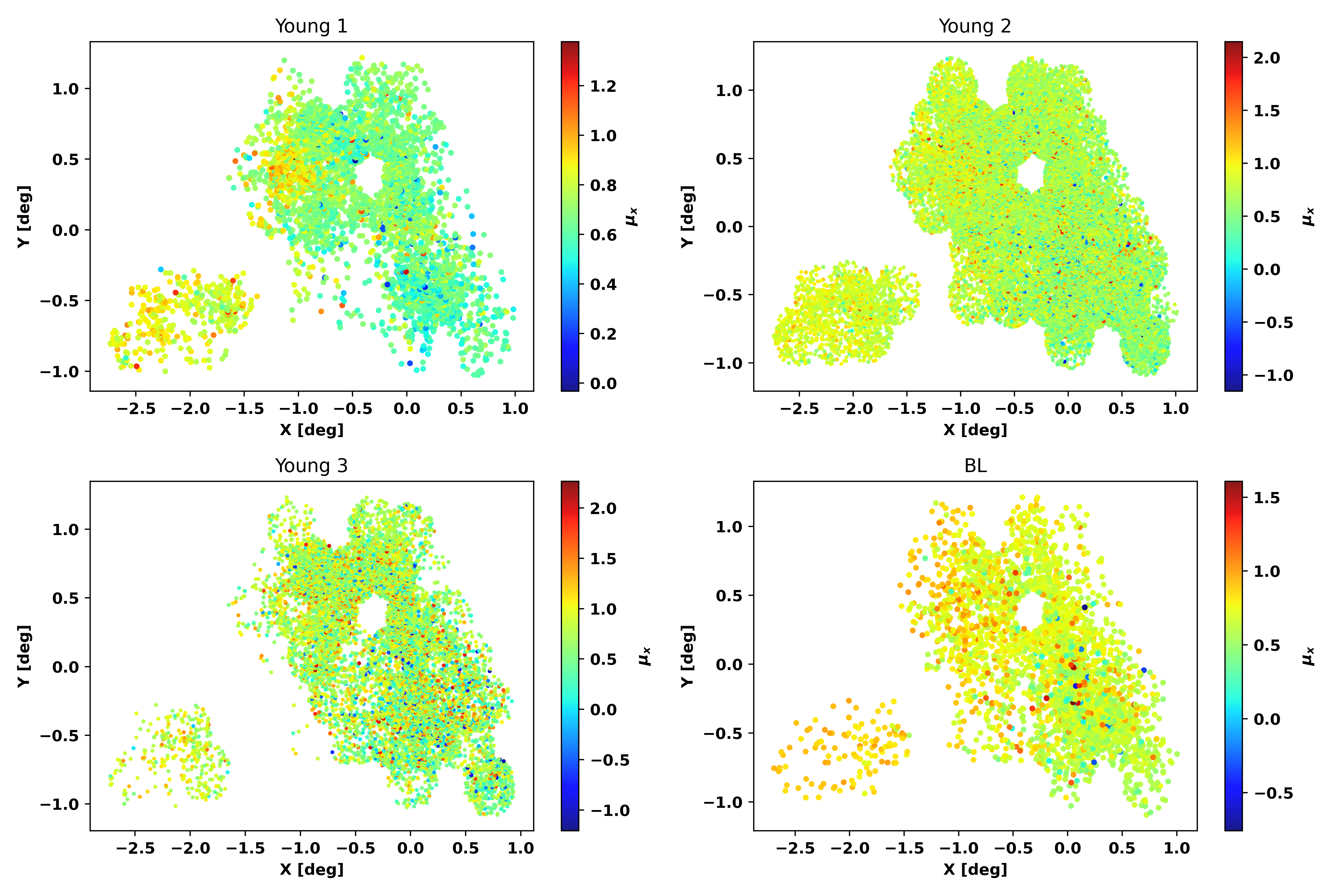}
\caption{Spatial distribution of PMRA of four young populations.}
\label{fig:PMRA_dist}
\end{figure*}

We transformed ($\mu_{\alpha}cos\delta$, $\mu_{\delta}$) into the Cartesian plane ($\mu_{x}$, $\mu_{y}$) using the conversion equation as defined by \citet[][Equation 2]{2021..Gaia}. After this transformation, we conducted further analysis on the kinematics using the obtained PM values on the X-Y plane, with $\mu_{x}$ and $\mu_{y}$ representing PMRA and PMDec, respectively. 
Then, we created vector point diagrams (VPDs) for four populations as shown in \autoref{fig:vpd_diff_pop}. VPDs of Young 1 and BL stars  (\autoref{fig:vpd_diff_pop}) show a spread in PMRA, whereas the PMDec distributions do not.  The distributions of Young 2 and Young 3 populations have broad profiles in both PMRA and PMDec distributions. Double Gaussians are well-fitted for the PMRA distributions of Young 1 and BL, while the single Gaussian curves are well-fitted for the rest. The peak and standard deviation values of these Gaussian fits, along with their corresponding errors obtained from the covariance matrix, are listed in Table~\ref{Tab: Double_Gaussian_diff_Pop} (4th to 6th column). The peak PMRA and PMDec values of all the population are similar to each other. The PMRA distribution widths of the Young 1 and BL populations are dissimilar with the broad distribution showing about thrice the width of the narrow one. We note an increasing width of the PMRA and PMDec distribution from Young 2 to Young 3, which are even larger than the width of the PMRA distribution of Young 1  population and BL stars. We checked the spatial distribution of PMRA for these four populations as shown in \autoref{fig:PMRA_dist}. We observed a noticeable PMRA gradient, increasing towards the east and NE direction for the Young 1 population in contrast to Young 2 and Young 3. The spatial distribution of PMRA for the BL stars also indicates a similar, though less pronounced, gradient compared to Young 1 population. We note a distinct pocket of relatively high PMRA in the eastern part of the SMC in the case of the Young 1 population, near the shell-like structure and the inner SMC wing.

\begin{figure}[htpb]
            
     \includegraphics[width=\linewidth]{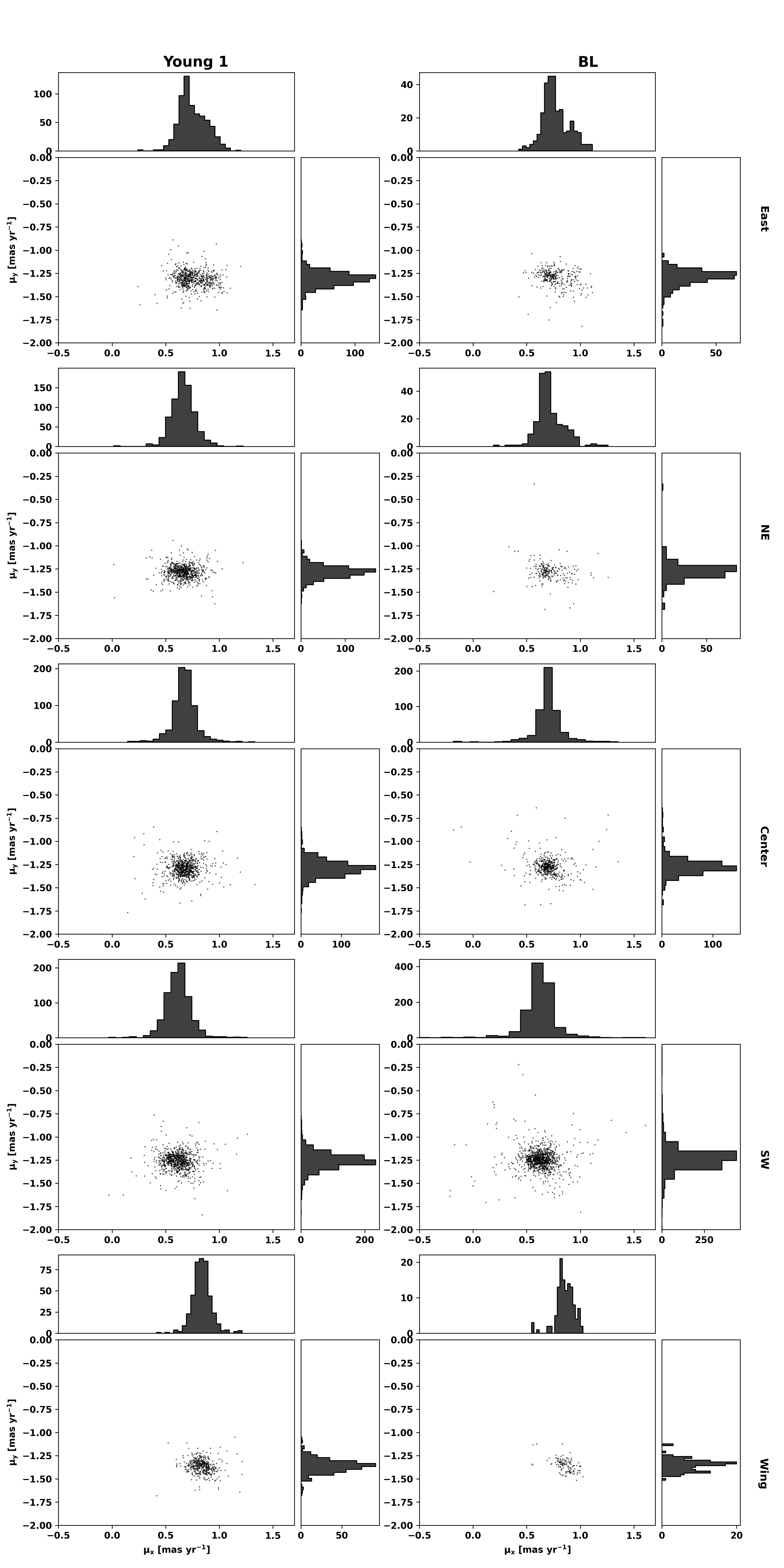}
        \caption{VPDs of four dense and inner Wing regions of Young 1 and BL stars. Distributions of PMRA and PMDec are shown at the top and right, respectively.}
        \label{fig: VPDs_blobs_diff_pop}
\end{figure}

We have found two sub-populations in BL and Young 1 based on PMRA distribution (see \autoref{fig:vpd_diff_pop}). It is quite interesting to examine whether the sub-populations are located all over the observed SMC or confined to a specific region. To do so, we plotted their VPDs separately for four dense regions and the inner Wing region, as depicted in \autoref{fig: VPDs_blobs_diff_pop}. We clearly see that there are two peaks of PMRA distribution of BL in the east and NE parts of the SMC, although BL stars are more concentrated in the SW part. In the case of the Young 1 population, two possible peaks of PMRA distribution are in the east and NE direction. We note a higher PMRA value in the inner Wing region for both Young 1 and BL populations. We conclude that the two sub-populations with differing PMRA distribution found for Young 1 and BL are mainly located in the east, i.e., the shell-like structure and inner SMC wing. In summary, the analysis points to kinematic differences in PMRA for the young population in east and NE, without any noticeable difference in PMDec. 
\begin{table*}
\centering
\scriptsize
\setlength{\tabcolsep}{1.4pt}
\caption{Peak and standard deviation values (in the unit of mas yr$^{-1}$) of Gaussian curve fit on PM, PMRA, and PMDec distributions, along with fit error for different young populations.}
\label{Tab: Double_Gaussian_diff_Pop}
\begin{tabular}{lcccccccccc}
\hline
\multirow{2}{*}{Populations} & \multicolumn{2}{c}{Gaussian 1 (PM)} & \multicolumn{2}{c}{Gaussian 2 (PM)} & \multicolumn{2}{c}{Gaussian 1 (PMRA)} & \multicolumn{2}{c}{Gaussian 2 (PMRA)} & \multicolumn{2}{c}{Gaussian (PMDec)} \\ \cline{2-11} 
 & Peak & Std & Peak & Std & Peak & Std & Peak & Std & Peak & Std \\ \hline
Young 1 & 1.431$\pm$0.002 & 0.060$\pm$0.002 & 1.483$\pm$0.003 & 0.115$\pm$0.002 & 0.662$\pm$0.001 & 0.051$\pm$0.002 & 0.701$\pm$0.002 & 0.142$\pm$0.002 & $-$1.288$\pm$0.001 & 0.076$\pm$0.001 \\ 
Young 2 & 1.460$\pm$0.002 & 0.148$\pm$0.004 & 1.508$\pm$0.012 & 0.308$\pm$0.019 & 0.693$\pm$0.004 & 0.190$\pm$0.004 & -- & -- & $-$1.281$\pm$0.003 & 0.152$\pm$0.003 \\ 
Young 3 & 1.475$\pm$0.002 & 0.230$\pm$0.005 & 1.525$\pm$0.011 & 0.504$\pm$0.021 & 0.723$\pm$0.006 & 0.318$\pm$0.006 & -- & -- & $-$1.266$\pm$0.006 & 0.268$\pm$0.006 \\ 
BL & 1.420$\pm$0.001 & 0.072$\pm$0.001 & 1.485$\pm$0.003 & 0.155$\pm$0.002 & 0.670$\pm$0.003 & 0.063$\pm$0.006 & 0.697$\pm$0.008 & 0.166$\pm$0.013 & $-$1.261$\pm$0.002 & 0.076$\pm$0.002 \\ 
\hline
\end{tabular}
\end{table*}

\section{Discussion}
\label{sec:Discussion}

\begin{figure}[htpb]
\includegraphics[width=\columnwidth,trim={0.8cm 0.6cm 0.5cm 0.8cm},clip]{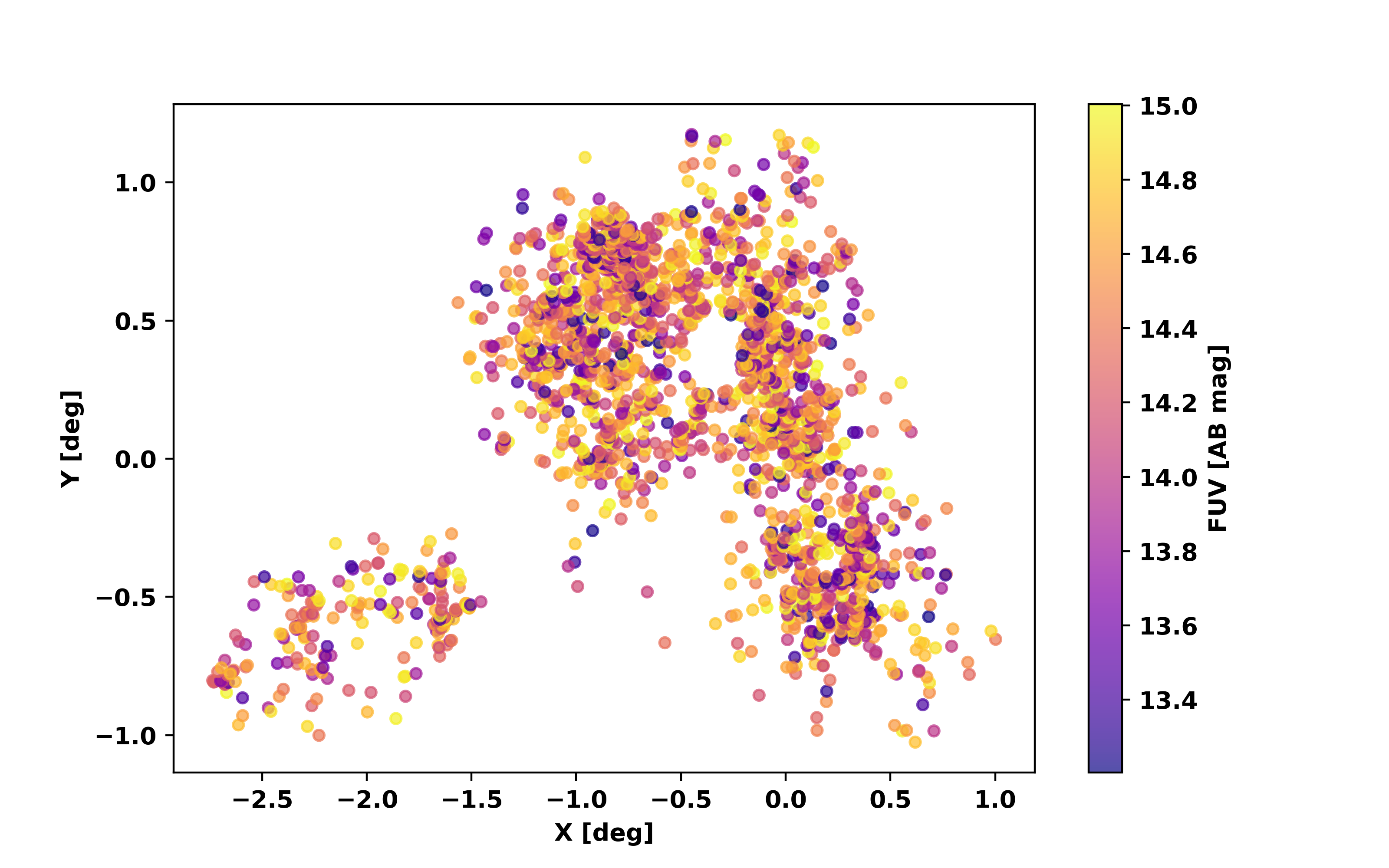}
\caption{Spatial distribution of the SMC FUV stars brightern than 15 magnitude corresponding to a photometric mass $>$ 8\(\textup{M}_\odot\).}
\label{fig:Spatial_8_solar_mass}
\end{figure}

\begin{figure*}[htpb]
\includegraphics[width=\linewidth,trim={0.8cm 2.5cm 0.5cm 0.8cm},clip]{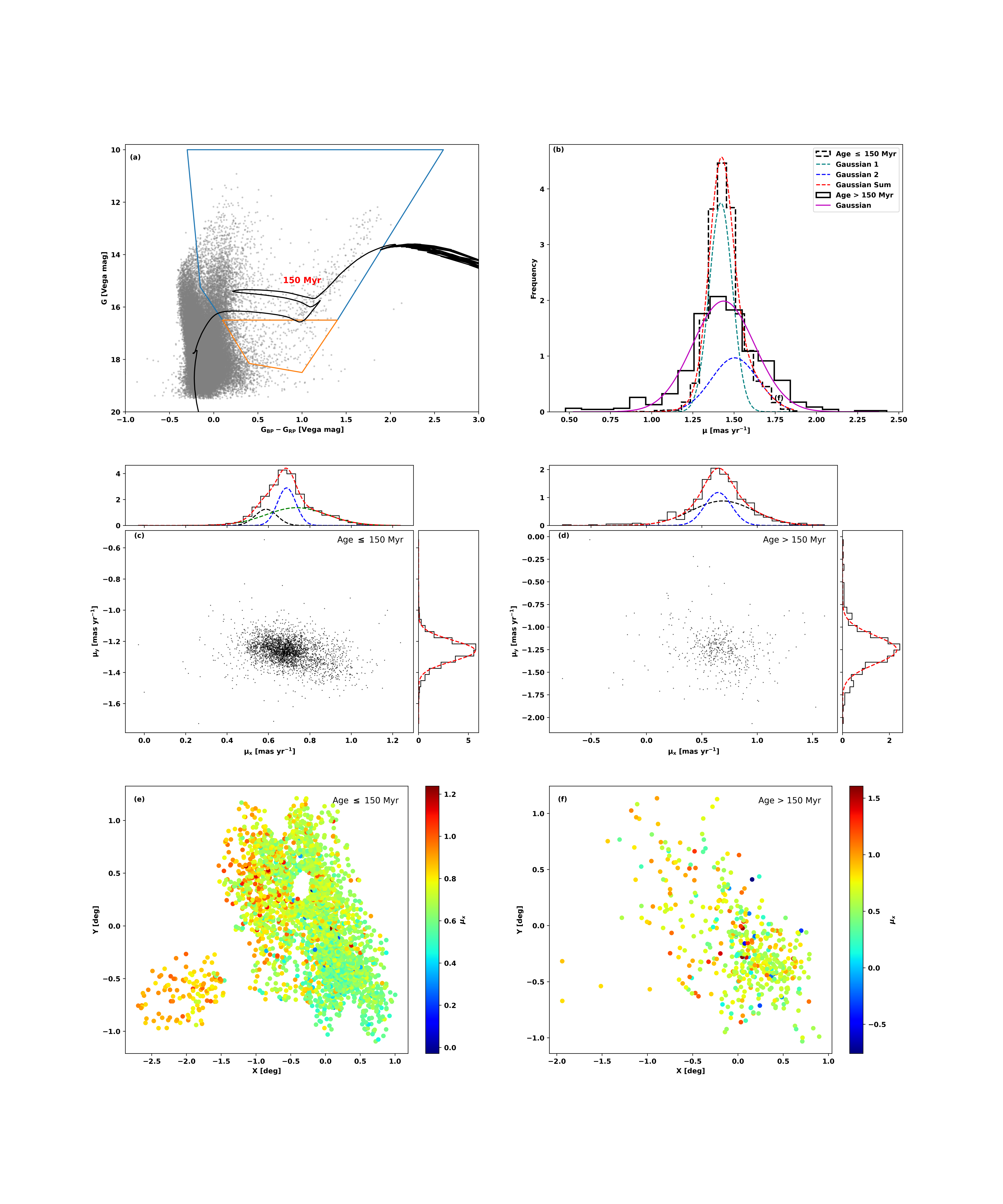}
\caption{(a) BL stars on the optical CMD. Blue and orange polygons denote the BL stars with age $\le$ 150 Myr and $>$ 150 Myr, respectively. (b) Gaussian curve fits the PM distribution. (c) and (d) VPDs with PMRA and PMDec distribution at the top and right, respectively. (e) and (f) present the spatial distribution of PMRA of two BL sub-populations (age $\leq$ 150 Myr and $>$ 150 Myr)}
\label{fig: BL}
\end{figure*}

\begin{table*}
\scriptsize
\caption{Peak and standard deviation values of Gaussian curve fit on PMRA distributions, along with fit error for different BL populations.}
\label{Tab:BL}
\begin{tabular}{lcccccc}
\hline
\multirow{2}{*}{BL population} & \multicolumn{2}{c}{Gaussian 1} & \multicolumn{2}{c}{Gaussian 2} & \multicolumn{2}{c}{Gaussian 3} \\ \cline{2-7}
& Peak [mas yr$^{-1}$] & Std [mas yr$^{-1}$] & Peak [mas yr$^{-1}$] & Std [mas yr$^{-1}$] & Peak [mas yr$^{-1}$] & Std [mas yr$^{-1}$] \\
\hline
age $\leq$ 150 Myr & 0.588$\pm$0.013 & 0.053$\pm$0.008 & 0.688$\pm$0.004 & 0.045$\pm$0.002  & 0.736$\pm$0.008 & 0.148$\pm$0.004 \\ 
age $>$ 150 Myr & 0.688$\pm$0.023 & 0.283$\pm$0.036 & 0.647$\pm$0.010 & 0.120$\pm$0.018 & -- & --  \\ 
\hline
\end{tabular}
\end{table*}

The interactions of the SMC with the LMC play a vital role in shaping the morphology of the SMC. In this study, we created an FUV catalog of the SMC obtained from UVIT observations and examined the spatial distribution and kinematics. We found an irregular and clumpy surface density distribution of the young population of the SMC, which is consistent with the previous studies \citep[e.g.][]{2000..Zaritsky..SMC..morphology,2000...Cioni..SMC..morphology, 2019Dalal..MORPHOLOGY..smc}.

\citet{2022MNRAS..Patrick..RSG..SMC} studied the binary systems that contain red supergiant (RSG) stars of the SMC based on FUV data of 88 red supergiants (RSG)  from UVIT. We found a cross-match of 61 FUV sources from our catalog (Table~\ref{Tab: catalog}) with their RSGs (see their appendix, Table A2), and the FUV magnitudes of these cross-matched sources are within the error. The RSGs, which are not cross-matched, are either found to be faint or associated with larger errors in FUV magnitude (the cutoff error in FUV magnitude in this study is $\leq$ 0.2 mag). 

\citet{2001..Maragoudaki..SMC..Morphology} investigated the spatial distribution of different populations of the SMC. They found that the young population exhibited an asymmetric distribution, unlike the old populations. They attributed this irregularity to the recent interaction between the LMC and the SMC at $\sim$ 0.2 -- 0.4 Gyr ago. \citet{2009..Gonidakis..Str..SMC} mapped the spatial distribution of four age groups of the SMC stellar populations and provided the isopleth contour maps using 2MASS data. The spatial distribution of their B and A-type stars are similar to the distribution of the Young 2 population. In a study by \citet{2011..Belcheva..Spatial..dist..MCs}, iso-density contour maps were used to analyze the morphology of different populations of the Magellanic Clouds. Their spatial distribution for the young stars (see their Figure 4) of the SMC is similar to our result. We also note a shell-like structure at $\sim$ ($-0.8$, 0.25) deg towards the SMC Wing  (\autoref{fig:kde_plot_SMC}) that is already mentioned by \citet{2001..Maragoudaki..SMC..Morphology}.

The surface density of the SMC, as well as that of all the four young populations, depict a broken young bar which shows a discontinuity in density at $\sim$ ($-$0.5, 0.5) deg (see \autoref{fig:kde_plot_SMC} \& \ref{fig:kde_plot_diff_pop}). The presence of the broken bar has previously been reported by \citet[][see their Figure 6]{2019Dalal..MORPHOLOGY..smc} and they mention the bending of the bar by about $\sim$ 30 deg to the east. In the density distribution of the BL stars, we note a denser region towards the SW of the bar, which is also found in the density distribution of the supergiant and giant stars by \citet{2019Dalal..MORPHOLOGY..smc}. \citet{2000...Cioni..SMC..morphology} also found an asymmetric distribution of the younger populations (main-sequence and supergiant stars) with protuberance.

In this study, we find that the eastern region and the northeastern regions are mainly populated by Young 1, 2, and 3. The central region predominantly has the Young 2 and 3 populations, whereas the SW has BL stars, Young 2 and 3. These can be used to trace the recent star formation history of these regions (age $<$ 400 Myr) and study stars of various mass, age, and evolutionary phases.

This catalog has stars up to a mass of $\sim$ 30 to 45 M$_{\odot}$ (photometric mass obtained from a 1 Myr Padova-PARSEC isochrone \citep[][]{2012...Bressan..Padova..PARSEC..iso} using a distance modulus of m$-$M = 18.96,  \citep{2015..de..Grijs} and a metallicity of Z = 0.002 dex \citep[][]{1992ApJ...384..508R,2008A&A...488..731R,2017A&A...608A..85L}) corresponding to an FUV magnitude of $\sim$ 13.2. Here, we have used the extinction law from \citet{1989...Cardelli}, R$_V$ value of 3.13 \citep{2024...Goradon..extinction} and a range of color excess values; $ E (B-V)$ = 0.05 \citep[mean value;][]{2011AJ....141..158H} and 0.1 \citep[a higher value around OB stars;][]{2024...Goradon..extinction}. As massive stars (M $>$ 8 \(\textup{M}_\odot\)) are likely progenitors of supernovae, we show the spatial distribution of stars brighter than 15 mag in FUV corresponding to a mass of $\sim$ 8 \(\textup{M}_\odot\) as shown in ~\autoref{fig:Spatial_8_solar_mass}. The spatial distribution of these massive stars ($\sim$ 2200) is similar to the distribution of H$\alpha$ of the SMC \citep{2001..Gaustad..H_alpha.,2018..H_alpha}. These are the sites of high-mass star formation in the inner SMC. These FUV bright stars, as well as their location, will be of interest to various ongoing studies such as binary at low metallicity by \citet{2024..Shenar..BLOeM}, B-type supergiants by \citet{2024..Parsons..B} and many other studies of massive stars.

The PMRA and PMDec values estimated for the young populations are well-matched with the previous studies \citep{2009Costa..MCs..PM,2016van..der..Marel.MCs..PM,2018..Zivic..MCs..PM,2021..Niederhofer..MCs..PM}. 
From the PMRA distribution, we found that Young 1 and BL stars have two types of sub-populations, which are mainly located in the east and NE direction of the SMC (see \autoref{fig:vpd_diff_pop}, ~\ref{fig: VPDs_blobs_diff_pop}). However, the width of the PM distribution of both the populations are relatively low, indicating that these sub-populations are relatively undisturbed (as seen in \autoref{fig:PM_hist_gaussian_fit} and Table~\ref{Tab: Double_Gaussian_diff_Pop}). We also note an increase in PMRA (not found in PMDec) for BL and Young 1 populations in the eastern SMC. A similar pattern in the PM of young stars in the SMC was reported by \citet{2021..Niederhofer..MCs..PM}. However, our findings confirm that this stretching in PM is predominantly due to the motion of young stars in the RA direction. Additionally, \citet{2020..De..Leo..tidal..scars..SMC}, in their analysis (see Figure 10), also observed increased PMRA values for RGB stars in the eastern region of the SMC. \citet{2019..Murray..Gas..Kinematic} identified a similar gradient in the residual PM of 143 massive stars, whose radial velocities are aligned with HI gas peaks. A recent study by \citet{2024..Almeida..kinematics} found a similar pattern in the radial velocity and proper motion of RGB stars. The study suggests that stars located on the eastern side of the SMC are part of material pulled out from the central SMC due to its tidal interaction with the LMC. However, these higher PMRA values are found within the 2 deg of the SMC, and hence unlikely to be connected to the foreground population found by \citet{2021..abinaya} and \citet{2021..James..SS..RV} (using RC and RGB tracers, respectively) in the east of the SMC beyond 2 deg from the SMC center. Overall, our result, the PMRA gradient observed in the Young 1 and BL populations, indicates that stars younger than 250 Myr are being stretched towards the LMC, likely as a consequence of the recent LMC-SMC interaction.

In order to check whether we can trace the impact of the recent LMC-SMC interaction at 150--300 Myr ago \citep{1985..Irwin..MB,2012Besla...first..MCs, 2022..Choi..MCs..interactions} in the motion of the young population, we overlaid a 150 Myr Padova-PARSEC isochrone \citep{2012...Bressan..Padova..PARSEC..iso} considering a distance modulus of m$-$M = 18.96 \citep{2015..de..Grijs} on the Gaia optical CMD as shown in ~\autoref{fig: BL}(a), to separate populations born before and after the interaction. We can see that BL stars have both types of populations: (1) before interaction (age $\leq$ 150 Myr) and (2) during\slash after the interaction (age $>$ 150 Myr). From the Gaussian curve fit to the PM distribution of these two populations of BL stars as shown in ~\autoref{fig: BL}(b), we determined the peak and sigma values as follows: for the BL population with age $\leq$ 150 Myr, ($\mu_1$, $\sigma_1$) = (1.420$\pm$0.002, 0.070$\pm$0.003) and ($\mu_2$, $\sigma_2$) = (1.504$\pm$0.024, 0.142$\pm$0.011) and for the BL population with age $>$ 150 Myr, ($\mu$, $\sigma$) = (1.436$\pm$0.010, 0.189$\pm$0.010). BL stars with age $\leq$ 150 Myr show a peaked narrow distribution that has contributions from two populations with different transverse motions. The broader distribution of the above two appears to be similar to that of BL population with age $>$ 150 Myr. Since the number of stars of the older BL population is not statistically significant, we will not discuss this population further but provide its VPD as shown in ~\autoref{fig: BL}(d). Peak and standard deviation of PMDec distribution for the two BL populations with age $\leq$ 150 Myr and $>$ 150 Myr are ($-1.260$$\pm$0.002, 0.067$\pm$0.002) and ($-1.252$$\pm$0.006, 0.157$\pm$0.006), respectively. The Gaussian curve fits for PMRA distribution to these are listed in Table~\ref{Tab:BL}. We note from PMRA distribution as shown in ~\autoref{fig: BL}(c) that BL stars younger than 150 Myr have three sub-populations with slightly different peaks and widths in PMRA (Table~\ref{Tab:BL}). We checked the spatial distribution of PMRA of the BL stars (age $\leq$ 150 Myr and $>$ 150 Myr) as shown in ~\autoref{fig: BL}(e) \& (f). We found that there is a gradient in PMRA for the BL stars with age $\leq$ 150 Myr, unlike the older BL stars. The gradient in PMRA is in the direction from SW to NE and east.  We already found that the Young 1 population that is also formed after the recent interaction shows a very similar kinematic signature.  We conclude that the younger sub-population of BL stars and the Young 1 population bear the kinematic signature of the recent LMC-SMC interaction. This signature is in the PMRA distribution and we do not detect any perturbation in the PMDec distribution. This is an important pointer that can constrain the models of LMC-SMC interaction.

\begin{figure}[htpb]
\includegraphics[width=\columnwidth]{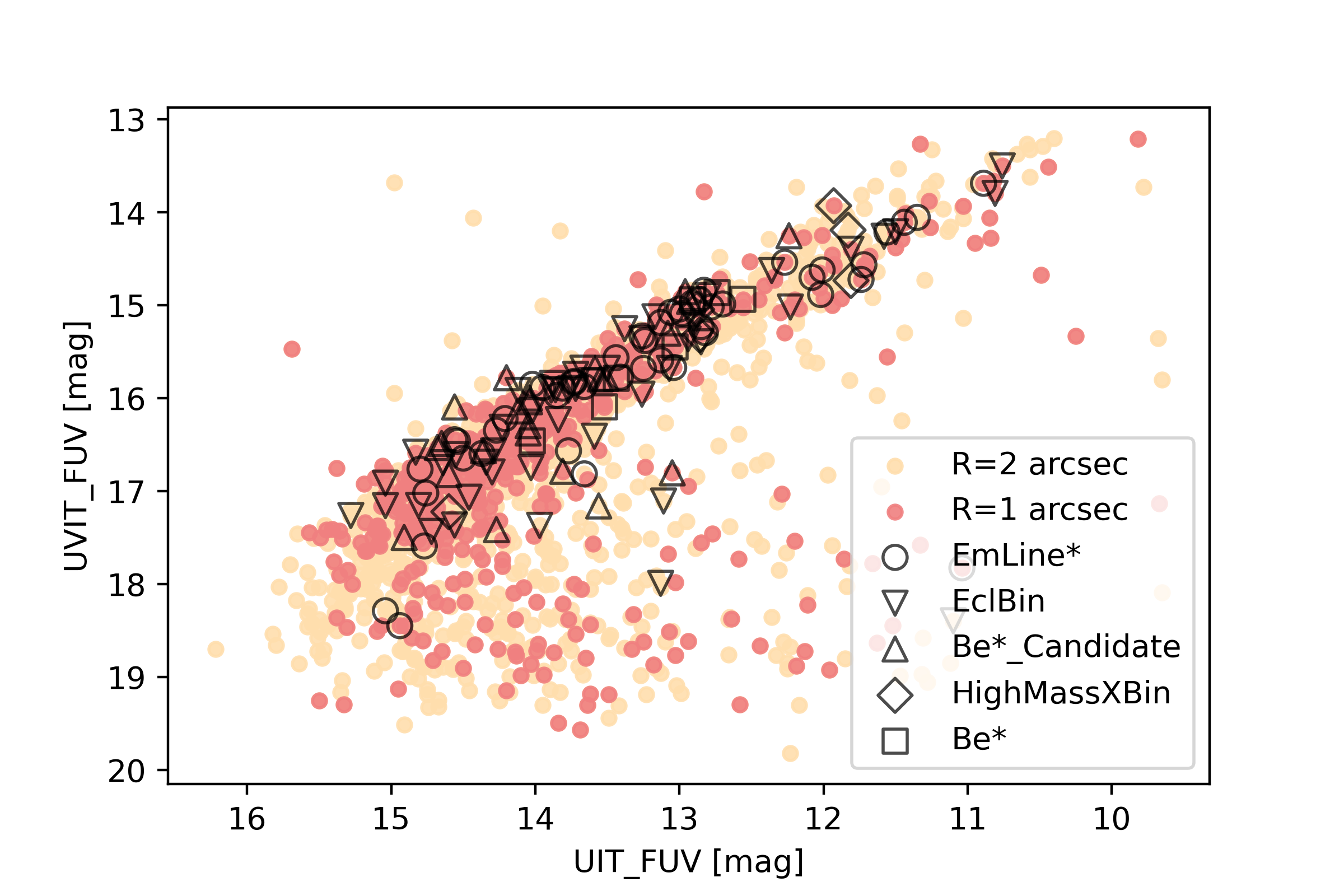}
\caption{UVIT and UIT FUV magnitudes of common sources (for search radii of 2$''$ and 1$''$). Different markers represent UVIT-UIT-SIMBAD stars classified as emission-line stars, eclipsing binaries,Be star candidates, high-mass X-ray binaries, and Be stars.}
\label{fig:UIT}
\end{figure}

Four regions of the SMC (bar) were observed  by the UIT\footnote{Of the four UIT fields, three overlap with UVIT fields, while one lies in a gap within the UVIT observed area.} \citep{1997..Cornett..UIT}. We have cross-matched the UVIT catalog with the UIT catalog (FUV: 1620 \AA) of the SMC with search radii of 2$''$ and 1$''$(spatial resolution of UIT is 3$''$) and found $\sim$ 1700 and $\sim$ 550 common sources, respectively. \autoref{fig:UIT} shows the relation between the UVIT and UIT magnitudes and their fluxes are found to be well matched in a study by \citet{Subramaniam_2016}.  Cross-matching the common sources with the SIMBAD database revealed that many of these sources are ecliping binaries, emission line stars etc., and are marked in the figure. Differences in the magnitudes between the two catalogs arise from variations in filter wavelengths and widths, magnitude system, and the spatial resolution of UVIT and UIT and variable nature of some sources.

The SMC FUV sources are younger than 400 Myr \citep{2024..Hota..SMC..Shell}. Various studies have indicated that massive stars are often found in binary or higher multiple systems \citep{1998..Mason..Binary,2001..Garc..Binary,2014..Sana..Binary,2017..Moe..Binary}. It is estimated that approximately 70\% of close binary systems \citep{2012..Sana..Binary..70,2014..Kobulnicky..binary..70} are likely to interact, impacting stellar evolution, apart from leading to the end stages of a massive star, such as a supernova explosion \citep{1992..Podsiadlowski..evolution,2013..Mink..evolution,2017..De..Marco..evolution}. They also have the potential to form double compact object binaries \citep{2017..Marchant..DCOB}. In our future work, we plan to identify binaries through multi-band photometry and characterize them, apart from estimating the properties of single stars. The catalog presented in this work will be a very useful resource to study the FUV properties of various exotic populations such as the high-mass X-ray binaries, Wolf-Rayet stars, O/B emission line stars, Be-X ray binaries, etc. This photometric catalog will also be quite a useful resource to plan photometric/spectroscopic observations of future space-based observatories, such as UVEX \citep[Ultraviolet Explorer;][]{2021..UVEX} and INSIST \citep[INdian Spectroscopic and Imaging Space Telescope;][]{2022JApA..INSIST,2023..INSIST}.

\section{Summary and Conclusions}
\label{sec:Summary}

In this study, we present a Far-UV catalog of the SMC obtained using the UVIT\slash AstroSat, based on observations of 39 pointings of UVIT in the filter F172M. The catalog contains $\sim$ 76800 sources along with their IDs in Gaia and VMC. This catalog will be an excellent resource for various studies of young and massive stellar populations in the SMC. We also provide the data completeness as a function of magnitude for the observed regions.

 Applying different classification methods to get the most probable SMC stars, we found that RUWE constraint is required for better MW decontamination. We adopt the Truncated-Optimal NN of \citet{2023..Gaia_catalog_SMC_Prob_Jimenez} and RUWE$<$1.4 cutoff to identify the probable SMC population consisting of 62901 stars. The spatial as well as kinematical properties of this population as found in this study are summarized below:

 \begin{itemize}
     \item  FUV stars in the catalog are mainly main-sequence, giant, and subgiant stars as found from the CMDs (Gaia optical, FUV--optical, and VMC IR CMDs). 
     \item There are about $\sim$ 2200 stars that are brighter than 15 mag in FUV, corresponding to a photometric mass of M$>$ 8M$_\odot$. The catalog is therefore an important resource to study massive stars in low metallicity environment.
     \item Stellar surface density distribution is found to be clumpy. The distribution traces three morphological features such as the inner SMC Wing, a broken bar, and a shell-like structure. We also find a high concentration of FUV stars in the SW of the bar. 
     \item The morphology as a function of stellar population is studied through KDE plots. From the KDE plots of Young 1, Young 2, Young 3, and BL stars, we find that the morphology varies with population. The density distribution of Young 2 has a close resemblance to the overall density while Young 1 and BL stars are concentrated at the NE and SW of the bar, respectively. The densest regions of Young 3 are located in the northeast and at the central part of the bar.
     \item The kinematic properties of the above population suggest that all populations show two kinematically distinct sub-populations. The Young 2 and 3 show a kinematically cool, less dispersed (with a narrower width in the PM distribution) and a hot, highly dispersed (with a broad PM distribution) sub-populations, whereas the Young 1 and BL stars show kinematically distinct two cool populations.
     \item The VPDs suggest that the kinematically distinct sub-populations of Young 1 and BL stars differ in PMRA, but have similar PMDec distribution.
      The spatial distribution of PMRA of Young 1 and BL stars show a gradient with increasing towards the eastern SMC, mainly located in the shell-like structure (east) and NE part of the bar.
      \item The kinematic analysis of this study points to a specific disturbance for stars younger than $\sim$ 150 Myr in the PMRA, with no significant disturbance in the PMDec. This can place strong constraints on the details of the recent LMC-SMC interaction.
      \item The probable SMC members ($\sim$ 62900) with FUV magnitudes will be very useful to the study of massive young stars in the low metallicity and kinematically perturbed environment of the SMC.
 \end{itemize}

\section*{acknowledgement}
We want to thank the anonymous referee for the insightful review that helped in the improvement of the paper. AS acknowledges support from SERB for the POWER fellowship. PKN acknowledges support from the Centro de Astrofisica y Tecnologias Afines (CATA) fellowship via grant Agencia Nacional de Investigacion y Desarrollo (ANID), BASAL FB210003. This publication uses the data from the AstroSat mission of the ISRO, archived at the Indian Space Science Data Centre (ISSDC). The optical and PM data utilized in this work have been obtained from the European Space Agency (ESA) space mission Gaia (\href{https://www.cosmos.esa.int/gaia}{https://www.cosmos.esa.int/gaia}). We are grateful to the Gaia Data Processing and Analysis Consortium (DPAC, \href{https://www.cosmos.esa.int/web/gaia/dpac/consortium}{https://www.cosmos.esa.int/web/gaia/dpac/consortium}) for their ongoing efforts in processing Gaia data. The DPAC's work is made possible through funding provided by national institutions, with a special acknowledgment to the institutions participating in the Gaia MultiLateral Agreement (MLA). Infra-red data are created from observations collected at the European Organisation for Astronomical Research in the Southern Hemisphere under ESO programme 179.B-2003. This research has made use of the services of the ESO Science Archive Facility.

Software: PYTHON packages, like NUMPY \citep{2020Natur.585..357H}, ASTROPY \citep{2013A&A...558A..33A}, MATPLOTLIB \citep{2007CSE.....9...90H} and SCIPY \citep{2020NatMe..17..261V}.

%\bibliography{Reference}{}

\begin{thebibliography}{}
\expandafter\ifx\csname natexlab\endcsname\relax\def\natexlab#1{#1}\fi
\providecommand{\url}[1]{\href{#1}{#1}}
\providecommand{\dodoi}[1]{doi:~\href{http://doi.org/#1}{\nolinkurl{#1}}}
\providecommand{\doeprint}[1]{\href{http://ascl.net/#1}{\nolinkurl{http://ascl.net/#1}}}
\providecommand{\doarXiv}[1]{\href{https://arxiv.org/abs/#1}{\nolinkurl{https://arxiv.org/abs/#1}}}

\bibitem[{{Almeida} {et~al.}(2024){Almeida}, {Majewski}, {Nidever}, {Olsen},
  {Monachesi}, {Kallivayalil}, {Hasselquist}, {Choi}, {Povick}, {Wilson},
  {Geisler}, {Lane}, {Nitschelm}, {Sobeck}, \&
  {Stringfellow}}]{2024..Almeida..kinematics}
{Almeida}, A., {Majewski}, S.~R., {Nidever}, D.~L., {et~al.} 2024, \mnras, 529,
  3858, \dodoi{10.1093/mnras/stae373}

\bibitem[{{Astropy Collaboration} {et~al.}(2013){Astropy Collaboration},
  {Robitaille}, {Tollerud}, {Greenfield}, {Droettboom}, {Bray}, {Aldcroft},
  {Davis}, {Ginsburg}, {Price-Whelan}, {Kerzendorf}, {Conley}, {Crighton},
  {Barbary}, {Muna}, {Ferguson}, {Grollier}, {Parikh}, {Nair}, {Unther},
  {Deil}, {Woillez}, {Conseil}, {Kramer}, {Turner}, {Singer}, {Fox}, {Weaver},
  {Zabalza}, {Edwards}, {Azalee Bostroem}, {Burke}, {Casey}, {Crawford},
  {Dencheva}, {Ely}, {Jenness}, {Labrie}, {Lim}, {Pierfederici}, {Pontzen},
  {Ptak}, {Refsdal}, {Servillat}, \& {Streicher}}]{2013A&A...558A..33A}
{Astropy Collaboration}, {Robitaille}, T.~P., {Tollerud}, E.~J., {et~al.} 2013,
  \aap, 558, A33, \dodoi{10.1051/0004-6361/201322068}

\bibitem[{{Belcheva} {et~al.}(2011){Belcheva}, {Livanou}, {Kontizas},
  {Nikolov}, \& {Kontizas}}]{2011..Belcheva..Spatial..dist..MCs}
{Belcheva}, M.~K., {Livanou}, E., {Kontizas}, M., {Nikolov}, G.~B., \&
  {Kontizas}, E. 2011, \aap, 527, A31, \dodoi{10.1051/0004-6361/201015835}

\bibitem[{{Belokurov} {et~al.}(2017){Belokurov}, {Erkal}, {Deason}, {Koposov},
  {De Angeli}, {Evans}, {Fraternali}, \&
  {Mackey}}]{2017..Belokurov..Magellanic..system}
{Belokurov}, V., {Erkal}, D., {Deason}, A.~J., {et~al.} 2017, \mnras, 466,
  4711, \dodoi{10.1093/mnras/stw3357}

\bibitem[{{Belokurov} \& {Erkal}(2019)}]{2019..Belokurov..smc..ss}
{Belokurov}, V.~A., \& {Erkal}, D. 2019, \mnras, 482, L9,
  \dodoi{10.1093/mnrasl/sly178}

\bibitem[{Besla {et~al.}(2007)Besla, Kallivayalil, Hernquist, Robertson, Cox,
  van~der Marel, \& Alcock}]{2007..Besla..first..passage..MCs}
Besla, G., Kallivayalil, N., Hernquist, L., {et~al.} 2007, The Astrophysical
  Journal, 668, 949

\bibitem[{{Besla} {et~al.}(2012){Besla}, {Kallivayalil}, {Hernquist}, {van der
  Marel}, {Cox}, \& {Kere{\v{s}}}}]{2012Besla...first..MCs}
{Besla}, G., {Kallivayalil}, N., {Hernquist}, L., {et~al.} 2012, \mnras, 421,
  2109, \dodoi{10.1111/j.1365-2966.2012.20466.x}

\bibitem[{{Bonanos} {et~al.}(2010){Bonanos}, {Lennon}, {K{\"o}hlinger}, {van
  Loon}, {Massa}, {Sewilo}, {Evans}, {Panagia}, {Babler}, {Block}, {Bracker},
  {Engelbracht}, {Gordon}, {Hora}, {Indebetouw}, {Meade}, {Meixner}, {Misselt},
  {Robitaille}, {Shiao}, \& {Whitney}}]{2010..Bonanos..Massive_stars_SMC}
{Bonanos}, A.~Z., {Lennon}, D.~J., {K{\"o}hlinger}, F., {et~al.} 2010, \aj,
  140, 416, \dodoi{10.1088/0004-6256/140/2/416}

\bibitem[{{Bot} {et~al.}(2004){Bot}, {Boulanger}, {Lagache}, {Cambr{\'e}sy}, \&
  {Egret}}]{2004..active..sf..dust...BOT}
{Bot}, C., {Boulanger}, F., {Lagache}, G., {Cambr{\'e}sy}, L., \& {Egret}, D.
  2004, \aap, 423, 567, \dodoi{10.1051/0004-6361:20035918}

\bibitem[{{Bressan} {et~al.}(2012){Bressan}, {Marigo}, {Girardi}, {Salasnich},
  {Dal Cero}, {Rubele}, \& {Nanni}}]{2012...Bressan..Padova..PARSEC..iso}
{Bressan}, A., {Marigo}, P., {Girardi}, L., {et~al.} 2012, \mnras, 427, 127,
  \dodoi{10.1111/j.1365-2966.2012.21948.x}

\bibitem[{{Cardelli} {et~al.}(1989){Cardelli}, {Clayton}, \&
  {Mathis}}]{1989...Cardelli}
{Cardelli}, J.~A., {Clayton}, G.~C., \& {Mathis}, J.~S. 1989, \apj, 345, 245,
  \dodoi{10.1086/167900}

\bibitem[{{Chandra} {et~al.}(2023){Chandra}, {Naidu}, {Conroy}, {Bonaca},
  {Zaritsky}, {Cargile}, {Caldwell}, {Johnson}, {Han}, \&
  {Ting}}]{2023chandra..first..passage..MCs}
{Chandra}, V., {Naidu}, R.~P., {Conroy}, C., {et~al.} 2023, \apj, 956, 110,
  \dodoi{10.3847/1538-4357/acf7bf}

\bibitem[{{Choi} {et~al.}(2022){Choi}, {Olsen}, {Besla}, {van der Marel},
  {Zivick}, {Kallivayalil}, \& {Nidever}}]{2022..Choi..MCs..interactions}
{Choi}, Y., {Olsen}, K. A.~G., {Besla}, G., {et~al.} 2022, \apj, 927, 153,
  \dodoi{10.3847/1538-4357/ac4e90}

\bibitem[{{Cioni} {et~al.}(2000){Cioni}, {Habing}, \&
  {Israel}}]{2000...Cioni..SMC..morphology}
{Cioni}, M. R.~L., {Habing}, H.~J., \& {Israel}, F.~P. 2000, \aap, 358, L9,
  \dodoi{10.48550/arXiv.astro-ph/0005057}

\bibitem[{{Cioni} {et~al.}(2011){Cioni}, {Clementini}, {Girardi}, {Guandalini},
  {Gullieuszik}, {Miszalski}, {Moretti}, {Ripepi}, {Rubele}, {Bagheri},
  {Bekki}, {Cross}, {de Blok}, {de Grijs}, {Emerson}, {Evans}, {Gibson},
  {Gonzales-Solares}, {Groenewegen}, {Irwin}, {Ivanov}, {Lewis}, {Marconi},
  {Marquette}, {Mastropietro}, {Moore}, {Napiwotzki}, {Naylor}, {Oliveira},
  {Read}, {Sutorius}, {van Loon}, {Wilkinson}, \&
  {Wood}}]{2011..VMC..survey..M-R-L..Cioni}
{Cioni}, M. R.~L., {Clementini}, G., {Girardi}, L., {et~al.} 2011, \aap, 527,
  A116, \dodoi{10.1051/0004-6361/201016137}

\bibitem[{{Cohen} {et~al.}(2003){Cohen}, {Wheaton}, \&
  {Megeath}}]{2003..Cohen..2MASS}
{Cohen}, M., {Wheaton}, W.~A., \& {Megeath}, S.~T. 2003, \aj, 126, 1090,
  \dodoi{10.1086/376474}

\bibitem[{{Cornett} {et~al.}(1994){Cornett}, {Hill}, {Bohlin}, {O'Connell},
  {Roberts}, {Smith}, \& {Stecher}}]{1994..UIT..cornett}
{Cornett}, R.~H., {Hill}, J.~K., {Bohlin}, R.~C., {et~al.} 1994, \apjl, 430,
  L117, \dodoi{10.1086/187452}

\bibitem[{{Cornett} {et~al.}(1997){Cornett}, {Greason}, {Hill}, {Parker},
  {Waller}, {Bohlin}, {Cheng}, {Neff}, {O'Connell}, {Roberts}, {Smith}, \&
  {Stecher}}]{1997..Cornett..UIT}
{Cornett}, R.~H., {Greason}, M.~R., {Hill}, J.~K., {et~al.} 1997, \aj, 113,
  1011, \dodoi{10.1086/118317}

\bibitem[{{Costa} {et~al.}(2009){Costa}, {M{\'e}ndez}, {Pedreros}, {Moyano},
  {Gallart}, {No{\"e}l}, {Baume}, \& {Carraro}}]{2009Costa..MCs..PM}
{Costa}, E., {M{\'e}ndez}, R.~A., {Pedreros}, M.~H., {et~al.} 2009, \aj, 137,
  4339, \dodoi{10.1088/0004-6256/137/5/4339}

\bibitem[{{Cullinane} {et~al.}(2023){Cullinane}, {Mackey}, {Da Costa},
  {Koposov}, \& {Erkal}}]{2023..Cullinanesmc_south_ss}
{Cullinane}, L.~R., {Mackey}, A.~D., {Da Costa}, G.~S., {Koposov}, S.~E., \&
  {Erkal}, D. 2023, \mnras, 518, L25, \dodoi{10.1093/mnrasl/slac129}

\bibitem[{{de Grijs} \& {Bono}(2015)}]{2015..de..Grijs}
{de Grijs}, R., \& {Bono}, G. 2015, \aj, 149, 179,
  \dodoi{10.1088/0004-6256/149/6/179}

\bibitem[{{de Grijs} {et~al.}(2014){de Grijs}, {Wicker}, \&
  {Bono}}]{2014..de..Grijs}
{de Grijs}, R., {Wicker}, J.~E., \& {Bono}, G. 2014, \aj, 147, 122,
  \dodoi{10.1088/0004-6256/147/5/122}

\bibitem[{{De Leo} {et~al.}(2020){De Leo}, {Carrera}, {No{\"e}l}, {Read},
  {Erkal}, \& {Gallart}}]{2020..De..Leo..tidal..scars..SMC}
{De Leo}, M., {Carrera}, R., {No{\"e}l}, N. E.~D., {et~al.} 2020, \mnras, 495,
  98, \dodoi{10.1093/mnras/staa1122}

\bibitem[{{De Marco} \& {Izzard}(2017)}]{2017..De..Marco..evolution}
{De Marco}, O., \& {Izzard}, R.~G. 2017, \pasa, 34, e001,
  \dodoi{10.1017/pasa.2016.52}

\bibitem[{{de Mink} {et~al.}(2013){de Mink}, {Langer}, {Izzard}, {Sana}, \& {de
  Koter}}]{2013..Mink..evolution}
{de Mink}, S.~E., {Langer}, N., {Izzard}, R.~G., {Sana}, H., \& {de Koter}, A.
  2013, \apj, 764, 166, \dodoi{10.1088/0004-637X/764/2/166}

\bibitem[{{de Vaucouleurs} \&
  {Freeman}(1972)}]{1972..de..Vaucouleure..optical..center}
{de Vaucouleurs}, G., \& {Freeman}, K.~C. 1972, Vistas in Astronomy, 14, 163,
  \dodoi{10.1016/0083-6656(72)90026-8}

\bibitem[{{Devaraj} {et~al.}(2023){Devaraj}, {Joseph}, {Stalin}, {Tandon}, \&
  {Ghosh}}]{2023..UVIT..SMC..Devaraj}
{Devaraj}, A., {Joseph}, P., {Stalin}, C.~S., {Tandon}, S.~N., \& {Ghosh},
  S.~K. 2023, \apj, 946, 65, \dodoi{10.3847/1538-4357/acba9c}

\bibitem[{{Dias} {et~al.}(2016){Dias}, {Kerber}, {Barbuy}, {Bica}, \&
  {Ortolani}}]{2016..Dias..WH}
{Dias}, B., {Kerber}, L., {Barbuy}, B., {Bica}, E., \& {Ortolani}, S. 2016,
  \aap, 591, A11, \dodoi{10.1051/0004-6361/201527558}

\bibitem[{{Dias} {et~al.}(2021){Dias}, {Angelo}, {Oliveira}, {Maia}, {Parisi},
  {De Bortoli}, {Souza}, {Katime Santrich}, {Bassino}, {Barbuy}, {Bica},
  {Geisler}, {Kerber}, {P{\'e}rez-Villegas}, {Quint}, {Sanmartim}, {Santos}, \&
  {Westera}}]{2021..dias..CB}
{Dias}, B., {Angelo}, M.~S., {Oliveira}, R.~A.~P., {et~al.} 2021, \aap, 647,
  L9, \dodoi{10.1051/0004-6361/202040015}

\bibitem[{{Dias} {et~al.}(2022){Dias}, {Parisi}, {Angelo}, {Maia}, {Oliveira},
  {Souza}, {Kerber}, {Santos}, {P{\'e}rez-Villegas}, {Sanmartim}, {Quint},
  {Fraga}, {Barbuy}, {Bica}, {Santrich}, {Hernandez-Jimenez}, {Geisler},
  {Minniti}, {De B{\'o}rtoli}, {Bassino}, \& {Rocha}}]{2022..Dias..WH}
{Dias}, B., {Parisi}, M.~C., {Angelo}, M., {et~al.} 2022, \mnras, 512, 4334,
  \dodoi{10.1093/mnras/stac259}

\bibitem[{{D'Onghia} \& {Fox}(2016)}]{2016A..Onghia..MS}
{D'Onghia}, E., \& {Fox}, A.~J. 2016, \araa, 54, 363,
  \dodoi{10.1146/annurev-astro-081915-023251}

\bibitem[{{El Youssoufi} {et~al.}(2019){El Youssoufi}, {Cioni}, {Bell},
  {Rubele}, {Bekki}, {de Grijs}, {Girardi}, {Ivanov}, {Matijevic},
  {Niederhofer}, {Oliveira}, {Ripepi}, {Subramanian}, \& {van
  Loon}}]{2019Dalal..MORPHOLOGY..smc}
{El Youssoufi}, D., {Cioni}, M.-R.~L., {Bell}, C. P.~M., {et~al.} 2019, \mnras,
  490, 1076, \dodoi{10.1093/mnras/stz2400}

\bibitem[{{Gaia Collaboration} {et~al.}(2016){Gaia Collaboration}, {Prusti},
  {de Bruijne}, {Brown}, {Vallenari}, {Babusiaux}, {Bailer-Jones}, {Bastian},
  {Biermann}, {Evans}, {Eyer}, {Jansen}, {Jordi}, {Klioner}, {Lammers},
  {Lindegren}, {Luri}, {Mignard}, {Milligan}, {Panem}, {Poinsignon},
  {Pourbaix}, {Randich}, {Sarri}, {Sartoretti}, {Siddiqui}, {Soubiran},
  {Valette}, {van Leeuwen}, {Walton}, {Aerts}, {Arenou}, {Cropper}, {Drimmel},
  {H{\o}g}, {Katz}, {Lattanzi}, {O'Mullane}, {Grebel}, {Holland}, {Huc},
  {Passot}, {Bramante}, {Cacciari}, {Casta{\~n}eda}, {Chaoul}, {Cheek}, {De
  Angeli}, {Fabricius}, {Guerra}, {Hern{\'a}ndez}, {Jean-Antoine-Piccolo},
  {Masana}, {Messineo}, {Mowlavi}, {Nienartowicz}, {Ord{\'o}{\~n}ez-Blanco},
  {Panuzzo}, {Portell}, {Richards}, {Riello}, {Seabroke}, {Tanga},
  {Th{\'e}venin}, {Torra}, {Els}, {Gracia-Abril}, {Comoretto},
  {Garcia-Reinaldos}, {Lock}, {Mercier}, {Altmann}, {Andrae}, {Astraatmadja},
  {Bellas-Velidis}, {Benson}, {Berthier}, {Blomme}, {Busso}, {Carry},
  {Cellino}, {Clementini}, {Cowell}, {Creevey}, {Cuypers}, {Davidson}, {De
  Ridder}, {de Torres}, {Delchambre}, {Dell'Oro}, {Ducourant}, {Fr{\'e}mat},
  {Garc{\'\i}a-Torres}, {Gosset}, {Halbwachs}, {Hambly}, {Harrison}, {Hauser},
  {Hestroffer}, {Hodgkin}, {Huckle}, {Hutton}, {Jasniewicz}, {Jordan},
  {Kontizas}, {Korn}, {Lanzafame}, {Manteiga}, {Moitinho}, {Muinonen},
  {Osinde}, {Pancino}, {Pauwels}, {Petit}, {Recio-Blanco}, {Robin}, {Sarro},
  {Siopis}, {Smith}, {Smith}, {Sozzetti}, {Thuillot}, {van Reeven}, {Viala},
  {Abbas}, {Abreu Aramburu}, {Accart}, {Aguado}, {Allan}, {Allasia},
  {Altavilla}, {{\'A}lvarez}, {Alves}, {Anderson}, {Andrei}, {Anglada Varela},
  {Antiche}, {Antoja}, {Ant{\'o}n}, {Arcay}, {Atzei}, {Ayache}, {Bach},
  {Baker}, {Balaguer-N{\'u}{\~n}ez}, {Barache}, {Barata}, {Barbier}, {Barblan},
  {Baroni}, {Barrado y Navascu{\'e}s}, {Barros}, {Barstow}, {Becciani},
  {Bellazzini}, {Bellei}, {Bello Garc{\'\i}a}, {Belokurov}, {Bendjoya},
  {Berihuete}, {Bianchi}, {Bienaym{\'e}}, {Billebaud}, {Blagorodnova},
  {Blanco-Cuaresma}, {Boch}, {Bombrun}, {Borrachero}, {Bouquillon}, {Bourda},
  {Bouy}, {Bragaglia}, {Breddels}, {Brouillet}, {Br{\"u}semeister},
  {Bucciarelli}, {Budnik}, {Burgess}, {Burgon}, {Burlacu}, {Busonero}, {Buzzi},
  {Caffau}, {Cambras}, {Campbell}, {Cancelliere}, {Cantat-Gaudin}, {Carlucci},
  {Carrasco}, {Castellani}, {Charlot}, {Charnas}, {Charvet}, {Chassat},
  {Chiavassa}, {Clotet}, {Cocozza}, {Collins}, {Collins}, {Costigan}, {Crifo},
  {Cross}, {Crosta}, {Crowley}, {Dafonte}, {Damerdji}, {Dapergolas}, {David},
  {David}, {De Cat}, {de Felice}, {de Laverny}, {De Luise}, {De March}, {de
  Martino}, {de Souza}, {Debosscher}, {del Pozo}, {Delbo}, {Delgado},
  {Delgado}, {di Marco}, {Di Matteo}, {Diakite}, {Distefano}, {Dolding}, {Dos
  Anjos}, {Drazinos}, {Dur{\'a}n}, {Dzigan}, {Ecale}, {Edvardsson}, {Enke},
  {Erdmann}, {Escolar}, {Espina}, {Evans}, {Eynard Bontemps}, {Fabre},
  {Fabrizio}, {Faigler}, {Falc{\~a}o}, {Farr{\`a}s Casas}, {Faye}, {Federici},
  {Fedorets}, {Fern{\'a}ndez-Hern{\'a}ndez}, {Fernique}, {Fienga}, {Figueras},
  {Filippi}, {Findeisen}, {Fonti}, {Fouesneau}, {Fraile}, {Fraser}, {Fuchs},
  {Furnell}, {Gai}, {Galleti}, {Galluccio}, {Garabato}, {Garc{\'\i}a-Sedano},
  {Gar{\'e}}, {Garofalo}, {Garralda}, {Gavras}, {Gerssen}, {Geyer}, {Gilmore},
  {Girona}, {Giuffrida}, {Gomes}, {Gonz{\'a}lez-Marcos},
  {Gonz{\'a}lez-N{\'u}{\~n}ez}, {Gonz{\'a}lez-Vidal}, {Granvik}, {Guerrier},
  {Guillout}, {Guiraud}, {G{\'u}rpide}, {Guti{\'e}rrez-S{\'a}nchez}, {Guy},
  {Haigron}, {Hatzidimitriou}, {Haywood}, {Heiter}, {Helmi}, {Hobbs},
  {Hofmann}, {Holl}, {Holland}, {Hunt}, {Hypki}, {Icardi}, {Irwin}, {Jevardat
  de Fombelle}, {Jofr{\'e}}, {Jonker}, {Jorissen}, {Julbe}, {Karampelas},
  {Kochoska}, {Kohley}, {Kolenberg}, {Kontizas}, {Koposov}, {Kordopatis},
  {Koubsky}, {Kowalczyk}, {Krone-Martins}, {Kudryashova}, {Kull}, {Bachchan},
  {Lacoste-Seris}, {Lanza}, {Lavigne}, {Le Poncin-Lafitte}, {Lebreton},
  {Lebzelter}, {Leccia}, {Leclerc}, {Lecoeur-Taibi}, {Lemaitre}, {Lenhardt},
  {Leroux}, {Liao}, {Licata}, {Lindstr{\o}m}, {Lister}, {Livanou}, {Lobel},
  {L{\"o}ffler}, {L{\'o}pez}, {Lopez-Lozano}, {Lorenz}, {Loureiro},
  {MacDonald}, {Magalh{\~a}es Fernandes}, {Managau}, {Mann}, {Mantelet},
  {Marchal}, {Marchant}, {Marconi}, {Marie}, {Marinoni}, {Marrese},
  {Marschalk{\'o}}, {Marshall}, {Mart{\'\i}n-Fleitas}, {Martino}, {Mary},
  {Matijevi{\v{c}}}, {Mazeh}, {McMillan}, {Messina}, {Mestre}, {Michalik},
  {Millar}, {Miranda}, {Molina}, {Molinaro}, {Molinaro}, {Moln{\'a}r},
  {Moniez}, {Montegriffo}, {Monteiro}, {Mor}, {Mora}, {Morbidelli}, {Morel},
  {Morgenthaler}, {Morley}, {Morris}, {Mulone}, {Muraveva}, {Musella},
  {Narbonne}, {Nelemans}, {Nicastro}, {Noval}, {Ord{\'e}novic},
  {Ordieres-Mer{\'e}}, {Osborne}, {Pagani}, {Pagano}, {Pailler}, {Palacin},
  {Palaversa}, {Parsons}, {Paulsen}, {Pecoraro}, {Pedrosa}, {Pentik{\"a}inen},
  {Pereira}, {Pichon}, {Piersimoni}, {Pineau}, {Plachy}, {Plum}, {Poujoulet},
  {Pr{\v{s}}a}, {Pulone}, {Ragaini}, {Rago}, {Rambaux}, {Ramos-Lerate},
  {Ranalli}, {Rauw}, {Read}, {Regibo}, {Renk}, {Reyl{\'e}}, {Ribeiro},
  {Rimoldini}, {Ripepi}, {Riva}, {Rixon}, {Roelens}, {Romero-G{\'o}mez},
  {Rowell}, {Royer}, {Rudolph}, {Ruiz-Dern}, {Sadowski}, {Sagrist{\`a}
  Sell{\'e}s}, {Sahlmann}, {Salgado}, {Salguero}, {Sarasso}, {Savietto},
  {Schnorhk}, {Schultheis}, {Sciacca}, {Segol}, {Segovia}, {Segransan},
  {Serpell}, {Shih}, {Smareglia}, {Smart}, {Smith}, {Solano}, {Solitro},
  {Sordo}, {Soria Nieto}, {Souchay}, {Spagna}, {Spoto}, {Stampa}, {Steele},
  {Steidelm{\"u}ller}, {Stephenson}, {Stoev}, {Suess}, {S{\"u}veges}, {Surdej},
  {Szabados}, {Szegedi-Elek}, {Tapiador}, {Taris}, {Tauran}, {Taylor},
  {Teixeira}, {Terrett}, {Tingley}, {Trager}, {Turon}, {Ulla}, {Utrilla},
  {Valentini}, {van Elteren}, {Van Hemelryck}, {van Leeuwen}, {Varadi},
  {Vecchiato}, {Veljanoski}, {Via}, {Vicente}, {Vogt}, {Voss}, {Votruba},
  {Voutsinas}, {Walmsley}, {Weiler}, {Weingrill}, {Werner}, {Wevers},
  {Whitehead}, {Wyrzykowski}, {Yoldas}, {{\v{Z}}erjal}, {Zucker}, {Zurbach},
  {Zwitter}, {Alecu}, {Allen}, {Allende Prieto}, {Amorim},
  {Anglada-Escud{\'e}}, {Arsenijevic}, {Azaz}, {Balm}, {Beck}, {Bernstein},
  {Bigot}, {Bijaoui}, {Blasco}, {Bonfigli}, {Bono}, {Boudreault}, {Bressan},
  {Brown}, {Brunet}, {Bunclark}, {Buonanno}, {Butkevich}, {Carret}, {Carrion},
  {Chemin}, {Ch{\'e}reau}, {Corcione}, {Darmigny}, {de Boer}, {de Teodoro}, {de
  Zeeuw}, {Delle Luche}, {Domingues}, {Dubath}, {Fodor}, {Fr{\'e}zouls},
  {Fries}, {Fustes}, {Fyfe}, {Gallardo}, {Gallegos}, {Gardiol}, {Gebran},
  {Gomboc}, {G{\'o}mez}, {Grux}, {Gueguen}, {Heyrovsky}, {Hoar}, {Iannicola},
  {Isasi Parache}, {Janotto}, {Joliet}, {Jonckheere}, {Keil}, {Kim},
  {Klagyivik}, {Klar}, {Knude}, {Kochukhov}, {Kolka}, {Kos}, {Kutka}, {Lainey},
  {LeBouquin}, {Liu}, {Loreggia}, {Makarov}, {Marseille}, {Martayan},
  {Martinez-Rubi}, {Massart}, {Meynadier}, {Mignot}, {Munari}, {Nguyen},
  {Nordlander}, {Ocvirk}, {O'Flaherty}, {Olias Sanz}, {Ortiz}, {Osorio},
  {Oszkiewicz}, {Ouzounis}, {Palmer}, {Park}, {Pasquato}, {Peltzer}, {Peralta},
  {P{\'e}turaud}, {Pieniluoma}, {Pigozzi}, {Poels}, {Prat}, {Prod'homme},
  {Raison}, {Rebordao}, {Risquez}, {Rocca-Volmerange}, {Rosen}, {Ruiz-Fuertes},
  {Russo}, {Sembay}, {Serraller Vizcaino}, {Short}, {Siebert}, {Silva},
  {Sinachopoulos}, {Slezak}, {Soffel}, {Sosnowska}, {Strai{\v{z}}ys}, {ter
  Linden}, {Terrell}, {Theil}, {Tiede}, {Troisi}, {Tsalmantza}, {Tur},
  {Vaccari}, {Vachier}, {Valles}, {Van Hamme}, {Veltz}, {Virtanen}, {Wallut},
  {Wichmann}, {Wilkinson}, {Ziaeepour}, \& {Zschocke}}]{2016..Gaia..Mission}
{Gaia Collaboration}, {Prusti}, T., {de Bruijne}, J.~H.~J., {et~al.} 2016,
  \aap, 595, A1, \dodoi{10.1051/0004-6361/201629272}

\bibitem[{{Gaia Collaboration} {et~al.}(2021{\natexlab{a}}){Gaia
  Collaboration}, {Luri}, {Chemin}, {Clementini}, {Delgado}, {McMillan},
  {Romero-G{\'o}mez}, {Balbinot}, {Castro-Ginard}, {Mor}, {Ripepi}, {Sarro},
  {Cioni}, {Fabricius}, {Garofalo}, {Helmi}, {Muraveva}, {Brown}, {Vallenari},
  {Prusti}, {de Bruijne}, {Babusiaux}, {Biermann}, {Creevey}, {Evans}, {Eyer},
  {Hutton}, {Jansen}, {Jordi}, {Klioner}, {Lammers}, {Lindegren}, {Mignard},
  {Panem}, {Pourbaix}, {Randich}, {Sartoretti}, {Soubiran}, {Walton}, {Arenou},
  {Bailer-Jones}, {Bastian}, {Cropper}, {Drimmel}, {Katz}, {Lattanzi}, {van
  Leeuwen}, {Bakker}, {Casta{\~n}eda}, {De Angeli}, {Ducourant}, {Fouesneau},
  {Fr{\'e}mat}, {Guerra}, {Guerrier}, {Guiraud}, {Jean-Antoine Piccolo},
  {Masana}, {Messineo}, {Mowlavi}, {Nicolas}, {Nienartowicz}, {Pailler},
  {Panuzzo}, {Riclet}, {Roux}, {Seabroke}, {Sordo}, {Tanga}, {Th{\'e}venin},
  {Gracia-Abril}, {Portell}, {Teyssier}, {Altmann}, {Andrae}, {Bellas-Velidis},
  {Benson}, {Berthier}, {Blomme}, {Brugaletta}, {Burgess}, {Busso}, {Carry},
  {Cellino}, {Cheek}, {Damerdji}, {Davidson}, {Delchambre}, {Dell'Oro},
  {Fern{\'a}ndez-Hern{\'a}ndez}, {Galluccio}, {Garc{\'\i}a-Lario},
  {Garcia-Reinaldos}, {Gonz{\'a}lez-N{\'u}{\~n}ez}, {Gosset}, {Haigron},
  {Halbwachs}, {Hambly}, {Harrison}, {Hatzidimitriou}, {Heiter},
  {Hern{\'a}ndez}, {Hestroffer}, {Hodgkin}, {Holl}, {Jan{\ss}en}, {Jevardat de
  Fombelle}, {Jordan}, {Krone-Martins}, {Lanzafame}, {L{\"o}ffler}, {Lorca},
  {Manteiga}, {Marchal}, {Marrese}, {Moitinho}, {Mora}, {Muinonen}, {Osborne},
  {Pancino}, {Pauwels}, {Recio-Blanco}, {Richards}, {Riello}, {Rimoldini},
  {Robin}, {Roegiers}, {Rybizki}, {Siopis}, {Smith}, {Sozzetti}, {Ulla},
  {Utrilla}, {van Leeuwen}, {van Reeven}, {Abbas}, {Abreu Aramburu}, {Accart},
  {Aerts}, {Aguado}, {Ajaj}, {Altavilla}, {{\'A}lvarez}, {{\'A}lvarez
  Cid-Fuentes}, {Alves}, {Anderson}, {Anglada Varela}, {Antoja}, {Audard},
  {Baines}, {Baker}, {Balaguer-N{\'u}{\~n}ez}, {Balog}, {Barache}, {Barbato},
  {Barros}, {Barstow}, {Bartolom{\'e}}, {Bassilana}, {Bauchet},
  {Baudesson-Stella}, {Becciani}, {Bellazzini}, {Bernet}, {Bertone}, {Bianchi},
  {Blanco-Cuaresma}, {Boch}, {Bombrun}, {Bossini}, {Bouquillon}, {Bragaglia},
  {Bramante}, {Breedt}, {Bressan}, {Brouillet}, {Bucciarelli}, {Burlacu},
  {Busonero}, {Butkevich}, {Buzzi}, {Caffau}, {Cancelliere}, {C{\'a}novas},
  {Cantat-Gaudin}, {Carballo}, {Carlucci}, {Carnerero}, {Carrasco},
  {Casamiquela}, {Castellani}, {Castro Sampol}, {Chaoul}, {Charlot},
  {Chiavassa}, {Comoretto}, {Cooper}, {Cornez}, {Cowell}, {Crifo}, {Crosta},
  {Crowley}, {Dafonte}, {Dapergolas}, {David}, {David}, {de Laverny}, {De
  Luise}, {De March}, {De Ridder}, {de Souza}, {de Teodoro}, {de Torres}, {del
  Peloso}, {del Pozo}, {Delgado}, {Delisle}, {Di Matteo}, {Diakite}, {Diener},
  {Distefano}, {Dolding}, {Eappachen}, {Enke}, {Esquej}, {Fabre}, {Fabrizio},
  {Faigler}, {Fedorets}, {Fernique}, {Fienga}, {Figueras}, {Fouron},
  {Fragkoudi}, {Fraile}, {Franke}, {Gai}, {Garabato}, {Garcia-Gutierrez},
  {Garc{\'\i}a-Torres}, {Gavras}, {Gerlach}, {Geyer}, {Giacobbe}, {Gilmore},
  {Girona}, {Giuffrida}, {Gomez}, {Gonzalez-Santamaria}, {Gonz{\'a}lez-Vidal},
  {Granvik}, {Guti{\'e}rrez-S{\'a}nchez}, {Guy}, {Hauser}, {Haywood},
  {Hidalgo}, {Hilger}, {H{\l}adczuk}, {Hobbs}, {Holland}, {Huckle},
  {Jasniewicz}, {Jonker}, {Juaristi Campillo}, {Julbe}, {Karbevska},
  {Kervella}, {Khanna}, {Kochoska}, {Kontizas}, {Kordopatis}, {Korn},
  {Kostrzewa-Rutkowska}, {Kruszy{\'n}ska}, {Lambert}, {Lanza}, {Lasne}, {Le
  Campion}, {Le Fustec}, {Lebreton}, {Lebzelter}, {Leccia}, {Leclerc},
  {Lecoeur-Taibi}, {Liao}, {Licata}, {Lindstr{\o}m}, {Lister}, {Livanou},
  {Lobel}, {Madrero Pardo}, {Managau}, {Mann}, {Marchant}, {Marconi}, {Marcos
  Santos}, {Marinoni}, {Marocco}, {Marshall}, {Martin Polo},
  {Mart{\'\i}n-Fleitas}, {Masip}, {Massari}, {Mastrobuono-Battisti}, {Mazeh},
  {Messina}, {Michalik}, {Millar}, {Mints}, {Molina}, {Molinaro}, {Moln{\'a}r},
  {Montegriffo}, {Morbidelli}, {Morel}, {Morris}, {Mulone}, {Munoz}, {Murphy},
  {Musella}, {Noval}, {Ord{\'e}novic}, {Orr{\`u}}, {Osinde}, {Pagani},
  {Pagano}, {Palaversa}, {Palicio}, {Panahi}, {Pawlak}, {Pe{\~n}alosa
  Esteller}, {Penttil{\"a}}, {Piersimoni}, {Pineau}, {Plachy}, {Plum},
  {Poggio}, {Poretti}, {Poujoulet}, {Pr{\v{s}}a}, {Pulone}, {Racero},
  {Ragaini}, {Rainer}, {Raiteri}, {Rambaux}, {Ramos}, {Ramos-Lerate}, {Re
  Fiorentin}, {Regibo}, {Reyl{\'e}}, {Riva}, {Rixon}, {Robichon}, {Robin},
  {Roelens}, {Rohrbasser}, {Rowell}, {Royer}, {Rybicki}, {Sadowski},
  {Sagrist{\`a} Sell{\'e}s}, {Sahlmann}, {Salgado}, {Salguero}, {Samaras},
  {Gimenez}, {Sanna}, {Santove{\~n}a}, {Sarasso}, {Schultheis}, {Sciacca},
  {Segol}, {Segovia}, {S{\'e}gransan}, {Semeux}, {Siddiqui}, {Siebert},
  {Siltala}, {Slezak}, {Smart}, {Solano}, {Solitro}, {Souami}, {Souchay},
  {Spagna}, {Spoto}, {Steele}, {Steidelm{\"u}ller}, {Stephenson},
  {S{\"u}veges}, {Szabados}, {Szegedi-Elek}, {Taris}, {Tauran}, {Taylor},
  {Teixeira}, {Thuillot}, {Tonello}, {Torra}, {Torra}, {Turon}, {Unger},
  {Vaillant}, {van Dillen}, {Vanel}, {Vecchiato}, {Viala}, {Vicente},
  {Voutsinas}, {Weiler}, {Wevers}, {Wyrzykowski}, {Yoldas}, {Yvard}, {Zhao},
  {Zorec}, {Zucker}, {Zurbach}, \& {Zwitter}}]{2021..Gaia}
{Gaia Collaboration}, {Luri}, X., {Chemin}, L., {et~al.} 2021{\natexlab{a}},
  \aap, 649, A7, \dodoi{10.1051/0004-6361/202039588}

\bibitem[{{Gaia Collaboration} {et~al.}(2021{\natexlab{b}}){Gaia
  Collaboration}, {Brown}, {Vallenari}, {Prusti}, {de Bruijne}, {Babusiaux},
  {Biermann}, {Creevey}, {Evans}, {Eyer}, {Hutton}, {Jansen}, {Jordi},
  {Klioner}, {Lammers}, {Lindegren}, {Luri}, {Mignard}, {Panem}, {Pourbaix},
  {Randich}, {Sartoretti}, {Soubiran}, {Walton}, {Arenou}, {Bailer-Jones},
  {Bastian}, {Cropper}, {Drimmel}, {Katz}, {Lattanzi}, {van Leeuwen}, {Bakker},
  {Cacciari}, {Casta{\~n}eda}, {De Angeli}, {Ducourant}, {Fabricius},
  {Fouesneau}, {Fr{\'e}mat}, {Guerra}, {Guerrier}, {Guiraud}, {Jean-Antoine
  Piccolo}, {Masana}, {Messineo}, {Mowlavi}, {Nicolas}, {Nienartowicz},
  {Pailler}, {Panuzzo}, {Riclet}, {Roux}, {Seabroke}, {Sordo}, {Tanga},
  {Th{\'e}venin}, {Gracia-Abril}, {Portell}, {Teyssier}, {Altmann}, {Andrae},
  {Bellas-Velidis}, {Benson}, {Berthier}, {Blomme}, {Brugaletta}, {Burgess},
  {Busso}, {Carry}, {Cellino}, {Cheek}, {Clementini}, {Damerdji}, {Davidson},
  {Delchambre}, {Dell'Oro}, {Fern{\'a}ndez-Hern{\'a}ndez}, {Galluccio},
  {Garc{\'\i}a-Lario}, {Garcia-Reinaldos}, {Gonz{\'a}lez-N{\'u}{\~n}ez},
  {Gosset}, {Haigron}, {Halbwachs}, {Hambly}, {Harrison}, {Hatzidimitriou},
  {Heiter}, {Hern{\'a}ndez}, {Hestroffer}, {Hodgkin}, {Holl}, {Jan{\ss}en},
  {Jevardat de Fombelle}, {Jordan}, {Krone-Martins}, {Lanzafame},
  {L{\"o}ffler}, {Lorca}, {Manteiga}, {Marchal}, {Marrese}, {Moitinho}, {Mora},
  {Muinonen}, {Osborne}, {Pancino}, {Pauwels}, {Petit}, {Recio-Blanco},
  {Richards}, {Riello}, {Rimoldini}, {Robin}, {Roegiers}, {Rybizki}, {Sarro},
  {Siopis}, {Smith}, {Sozzetti}, {Ulla}, {Utrilla}, {van Leeuwen}, {van
  Reeven}, {Abbas}, {Abreu Aramburu}, {Accart}, {Aerts}, {Aguado}, {Ajaj},
  {Altavilla}, {{\'A}lvarez}, {{\'A}lvarez Cid-Fuentes}, {Alves}, {Anderson},
  {Anglada Varela}, {Antoja}, {Audard}, {Baines}, {Baker},
  {Balaguer-N{\'u}{\~n}ez}, {Balbinot}, {Balog}, {Barache}, {Barbato},
  {Barros}, {Barstow}, {Bartolom{\'e}}, {Bassilana}, {Bauchet},
  {Baudesson-Stella}, {Becciani}, {Bellazzini}, {Bernet}, {Bertone}, {Bianchi},
  {Blanco-Cuaresma}, {Boch}, {Bombrun}, {Bossini}, {Bouquillon}, {Bragaglia},
  {Bramante}, {Breedt}, {Bressan}, {Brouillet}, {Bucciarelli}, {Burlacu},
  {Busonero}, {Butkevich}, {Buzzi}, {Caffau}, {Cancelliere}, {C{\'a}novas},
  {Cantat-Gaudin}, {Carballo}, {Carlucci}, {Carnerero}, {Carrasco},
  {Casamiquela}, {Castellani}, {Castro-Ginard}, {Castro Sampol}, {Chaoul},
  {Charlot}, {Chemin}, {Chiavassa}, {Cioni}, {Comoretto}, {Cooper}, {Cornez},
  {Cowell}, {Crifo}, {Crosta}, {Crowley}, {Dafonte}, {Dapergolas}, {David},
  {David}, {de Laverny}, {De Luise}, {De March}, {De Ridder}, {de Souza}, {de
  Teodoro}, {de Torres}, {del Peloso}, {del Pozo}, {Delbo}, {Delgado},
  {Delgado}, {Delisle}, {Di Matteo}, {Diakite}, {Diener}, {Distefano},
  {Dolding}, {Eappachen}, {Edvardsson}, {Enke}, {Esquej}, {Fabre}, {Fabrizio},
  {Faigler}, {Fedorets}, {Fernique}, {Fienga}, {Figueras}, {Fouron},
  {Fragkoudi}, {Fraile}, {Franke}, {Gai}, {Garabato}, {Garcia-Gutierrez},
  {Garc{\'\i}a-Torres}, {Garofalo}, {Gavras}, {Gerlach}, {Geyer}, {Giacobbe},
  {Gilmore}, {Girona}, {Giuffrida}, {Gomel}, {Gomez}, {Gonzalez-Santamaria},
  {Gonz{\'a}lez-Vidal}, {Granvik}, {Guti{\'e}rrez-S{\'a}nchez}, {Guy},
  {Hauser}, {Haywood}, {Helmi}, {Hidalgo}, {Hilger}, {H{\l}adczuk}, {Hobbs},
  {Holland}, {Huckle}, {Jasniewicz}, {Jonker}, {Juaristi Campillo}, {Julbe},
  {Karbevska}, {Kervella}, {Khanna}, {Kochoska}, {Kontizas}, {Kordopatis},
  {Korn}, {Kostrzewa-Rutkowska}, {Kruszy{\'n}ska}, {Lambert}, {Lanza}, {Lasne},
  {Le Campion}, {Le Fustec}, {Lebreton}, {Lebzelter}, {Leccia}, {Leclerc},
  {Lecoeur-Taibi}, {Liao}, {Licata}, {Lindstr{\o}m}, {Lister}, {Livanou},
  {Lobel}, {Madrero Pardo}, {Managau}, {Mann}, {Marchant}, {Marconi}, {Marcos
  Santos}, {Marinoni}, {Marocco}, {Marshall}, {Martin Polo},
  {Mart{\'\i}n-Fleitas}, {Masip}, {Massari}, {Mastrobuono-Battisti}, {Mazeh},
  {McMillan}, {Messina}, {Michalik}, {Millar}, {Mints}, {Molina}, {Molinaro},
  {Moln{\'a}r}, {Montegriffo}, {Mor}, {Morbidelli}, {Morel}, {Morris},
  {Mulone}, {Munoz}, {Muraveva}, {Murphy}, {Musella}, {Noval}, {Ord{\'e}novic},
  {Orr{\`u}}, {Osinde}, {Pagani}, {Pagano}, {Palaversa}, {Palicio}, {Panahi},
  {Pawlak}, {Pe{\~n}alosa Esteller}, {Penttil{\"a}}, {Piersimoni}, {Pineau},
  {Plachy}, {Plum}, {Poggio}, {Poretti}, {Poujoulet}, {Pr{\v{s}}a}, {Pulone},
  {Racero}, {Ragaini}, {Rainer}, {Raiteri}, {Rambaux}, {Ramos}, {Ramos-Lerate},
  {Re Fiorentin}, {Regibo}, {Reyl{\'e}}, {Ripepi}, {Riva}, {Rixon}, {Robichon},
  {Robin}, {Roelens}, {Rohrbasser}, {Romero-G{\'o}mez}, {Rowell}, {Royer},
  {Rybicki}, {Sadowski}, {Sagrist{\`a} Sell{\'e}s}, {Sahlmann}, {Salgado},
  {Salguero}, {Samaras}, {Sanchez Gimenez}, {Sanna}, {Santove{\~n}a},
  {Sarasso}, {Schultheis}, {Sciacca}, {Segol}, {Segovia}, {S{\'e}gransan},
  {Semeux}, {Shahaf}, {Siddiqui}, {Siebert}, {Siltala}, {Slezak}, {Smart},
  {Solano}, {Solitro}, {Souami}, {Souchay}, {Spagna}, {Spoto}, {Steele},
  {Steidelm{\"u}ller}, {Stephenson}, {S{\"u}veges}, {Szabados}, {Szegedi-Elek},
  {Taris}, {Tauran}, {Taylor}, {Teixeira}, {Thuillot}, {Tonello}, {Torra},
  {Torra}, {Turon}, {Unger}, {Vaillant}, {van Dillen}, {Vanel}, {Vecchiato},
  {Viala}, {Vicente}, {Voutsinas}, {Weiler}, {Wevers}, {Wyrzykowski}, {Yoldas},
  {Yvard}, {Zhao}, {Zorec}, {Zucker}, {Zurbach}, \& {Zwitter}}]{2021GaiaEDR3}
{Gaia Collaboration}, {Brown}, A.~G.~A., {Vallenari}, A., {et~al.}
  2021{\natexlab{b}}, \aap, 649, A1, \dodoi{10.1051/0004-6361/202039657}

\bibitem[{{Gaia Collaboration} {et~al.}(2023){Gaia Collaboration}, {Vallenari},
  {Brown}, {Prusti}, {de Bruijne}, {Arenou}, {Babusiaux}, {Biermann},
  {Creevey}, {Ducourant}, {Evans}, {Eyer}, {Guerra}, {Hutton}, {Jordi},
  {Klioner}, {Lammers}, {Lindegren}, {Luri}, {Mignard}, {Panem}, {Pourbaix},
  {Randich}, {Sartoretti}, {Soubiran}, {Tanga}, {Walton}, {Bailer-Jones},
  {Bastian}, {Drimmel}, {Jansen}, {Katz}, {Lattanzi}, {van Leeuwen}, {Bakker},
  {Cacciari}, {Casta{\~n}eda}, {De Angeli}, {Fabricius}, {Fouesneau},
  {Fr{\'e}mat}, {Galluccio}, {Guerrier}, {Heiter}, {Masana}, {Messineo},
  {Mowlavi}, {Nicolas}, {Nienartowicz}, {Pailler}, {Panuzzo}, {Riclet}, {Roux},
  {Seabroke}, {Sordo}, {Th{\'e}venin}, {Gracia-Abril}, {Portell}, {Teyssier},
  {Altmann}, {Andrae}, {Audard}, {Bellas-Velidis}, {Benson}, {Berthier},
  {Blomme}, {Burgess}, {Busonero}, {Busso}, {C{\'a}novas}, {Carry}, {Cellino},
  {Cheek}, {Clementini}, {Damerdji}, {Davidson}, {de Teodoro}, {Nu{\~n}ez
  Campos}, {Delchambre}, {Dell'Oro}, {Esquej}, {Fern{\'a}ndez-Hern{\'a}ndez},
  {Fraile}, {Garabato}, {Garc{\'\i}a-Lario}, {Gosset}, {Haigron}, {Halbwachs},
  {Hambly}, {Harrison}, {Hern{\'a}ndez}, {Hestroffer}, {Hodgkin}, {Holl},
  {Jan{\ss}en}, {Jevardat de Fombelle}, {Jordan}, {Krone-Martins}, {Lanzafame},
  {L{\"o}ffler}, {Marchal}, {Marrese}, {Moitinho}, {Muinonen}, {Osborne},
  {Pancino}, {Pauwels}, {Recio-Blanco}, {Reyl{\'e}}, {Riello}, {Rimoldini},
  {Roegiers}, {Rybizki}, {Sarro}, {Siopis}, {Smith}, {Sozzetti}, {Utrilla},
  {van Leeuwen}, {Abbas}, {{\'A}brah{\'a}m}, {Abreu Aramburu}, {Aerts},
  {Aguado}, {Ajaj}, {Aldea-Montero}, {Altavilla}, {{\'A}lvarez}, {Alves},
  {Anders}, {Anderson}, {Anglada Varela}, {Antoja}, {Baines}, {Baker},
  {Balaguer-N{\'u}{\~n}ez}, {Balbinot}, {Balog}, {Barache}, {Barbato},
  {Barros}, {Barstow}, {Bartolom{\'e}}, {Bassilana}, {Bauchet}, {Becciani},
  {Bellazzini}, {Berihuete}, {Bernet}, {Bertone}, {Bianchi}, {Binnenfeld},
  {Blanco-Cuaresma}, {Blazere}, {Boch}, {Bombrun}, {Bossini}, {Bouquillon},
  {Bragaglia}, {Bramante}, {Breedt}, {Bressan}, {Brouillet}, {Brugaletta},
  {Bucciarelli}, {Burlacu}, {Butkevich}, {Buzzi}, {Caffau}, {Cancelliere},
  {Cantat-Gaudin}, {Carballo}, {Carlucci}, {Carnerero}, {Carrasco},
  {Casamiquela}, {Castellani}, {Castro-Ginard}, {Chaoul}, {Charlot}, {Chemin},
  {Chiaramida}, {Chiavassa}, {Chornay}, {Comoretto}, {Contursi}, {Cooper},
  {Cornez}, {Cowell}, {Crifo}, {Cropper}, {Crosta}, {Crowley}, {Dafonte},
  {Dapergolas}, {David}, {David}, {de Laverny}, {De Luise}, {De March}, {De
  Ridder}, {de Souza}, {de Torres}, {del Peloso}, {del Pozo}, {Delbo},
  {Delgado}, {Delisle}, {Demouchy}, {Dharmawardena}, {Di Matteo}, {Diakite},
  {Diener}, {Distefano}, {Dolding}, {Edvardsson}, {Enke}, {Fabre}, {Fabrizio},
  {Faigler}, {Fedorets}, {Fernique}, {Fienga}, {Figueras}, {Fournier},
  {Fouron}, {Fragkoudi}, {Gai}, {Garcia-Gutierrez}, {Garcia-Reinaldos},
  {Garc{\'\i}a-Torres}, {Garofalo}, {Gavel}, {Gavras}, {Gerlach}, {Geyer},
  {Giacobbe}, {Gilmore}, {Girona}, {Giuffrida}, {Gomel}, {Gomez},
  {Gonz{\'a}lez-N{\'u}{\~n}ez}, {Gonz{\'a}lez-Santamar{\'\i}a},
  {Gonz{\'a}lez-Vidal}, {Granvik}, {Guillout}, {Guiraud},
  {Guti{\'e}rrez-S{\'a}nchez}, {Guy}, {Hatzidimitriou}, {Hauser}, {Haywood},
  {Helmer}, {Helmi}, {Sarmiento}, {Hidalgo}, {Hilger}, {H{\l}adczuk}, {Hobbs},
  {Holland}, {Huckle}, {Jardine}, {Jasniewicz}, {Jean-Antoine Piccolo},
  {Jim{\'e}nez-Arranz}, {Jorissen}, {Juaristi Campillo}, {Julbe}, {Karbevska},
  {Kervella}, {Khanna}, {Kontizas}, {Kordopatis}, {Korn}, {K{\'o}sp{\'a}l},
  {Kostrzewa-Rutkowska}, {Kruszy{\'n}ska}, {Kun}, {Laizeau}, {Lambert},
  {Lanza}, {Lasne}, {Le Campion}, {Lebreton}, {Lebzelter}, {Leccia}, {Leclerc},
  {Lecoeur-Taibi}, {Liao}, {Licata}, {Lindstr{\o}m}, {Lister}, {Livanou},
  {Lobel}, {Lorca}, {Loup}, {Madrero Pardo}, {Magdaleno Romeo}, {Managau},
  {Mann}, {Manteiga}, {Marchant}, {Marconi}, {Marcos}, {Marcos Santos},
  {Mar{\'\i}n Pina}, {Marinoni}, {Marocco}, {Marshall}, {Martin Polo},
  {Mart{\'\i}n-Fleitas}, {Marton}, {Mary}, {Masip}, {Massari},
  {Mastrobuono-Battisti}, {Mazeh}, {McMillan}, {Messina}, {Michalik}, {Millar},
  {Mints}, {Molina}, {Molinaro}, {Moln{\'a}r}, {Monari}, {Mongui{\'o}},
  {Montegriffo}, {Montero}, {Mor}, {Mora}, {Morbidelli}, {Morel}, {Morris},
  {Muraveva}, {Murphy}, {Musella}, {Nagy}, {Noval}, {Oca{\~n}a}, {Ogden},
  {Ordenovic}, {Osinde}, {Pagani}, {Pagano}, {Palaversa}, {Palicio},
  {Pallas-Quintela}, {Panahi}, {Payne-Wardenaar}, {Pe{\~n}alosa Esteller},
  {Penttil{\"a}}, {Pichon}, {Piersimoni}, {Pineau}, {Plachy}, {Plum}, {Poggio},
  {Pr{\v{s}}a}, {Pulone}, {Racero}, {Ragaini}, {Rainer}, {Raiteri}, {Rambaux},
  {Ramos}, {Ramos-Lerate}, {Re Fiorentin}, {Regibo}, {Richards}, {Rios Diaz},
  {Ripepi}, {Riva}, {Rix}, {Rixon}, {Robichon}, {Robin}, {Robin}, {Roelens},
  {Rogues}, {Rohrbasser}, {Romero-G{\'o}mez}, {Rowell}, {Royer}, {Ruz Mieres},
  {Rybicki}, {Sadowski}, {S{\'a}ez N{\'u}{\~n}ez}, {Sagrist{\`a} Sell{\'e}s},
  {Sahlmann}, {Salguero}, {Samaras}, {Sanchez Gimenez}, {Sanna},
  {Santove{\~n}a}, {Sarasso}, {Schultheis}, {Sciacca}, {Segol}, {Segovia},
  {S{\'e}gransan}, {Semeux}, {Shahaf}, {Siddiqui}, {Siebert}, {Siltala},
  {Silvelo}, {Slezak}, {Slezak}, {Smart}, {Snaith}, {Solano}, {Solitro},
  {Souami}, {Souchay}, {Spagna}, {Spina}, {Spoto}, {Steele},
  {Steidelm{\"u}ller}, {Stephenson}, {S{\"u}veges}, {Surdej}, {Szabados},
  {Szegedi-Elek}, {Taris}, {Taylor}, {Teixeira}, {Tolomei}, {Tonello}, {Torra},
  {Torra}, {Torralba Elipe}, {Trabucchi}, {Tsounis}, {Turon}, {Ulla}, {Unger},
  {Vaillant}, {van Dillen}, {van Reeven}, {Vanel}, {Vecchiato}, {Viala},
  {Vicente}, {Voutsinas}, {Weiler}, {Wevers}, {Wyrzykowski}, {Yoldas}, {Yvard},
  {Zhao}, {Zorec}, {Zucker}, \& {Zwitter}}]{2023GaiaDR3}
{Gaia Collaboration}, {Vallenari}, A., {Brown}, A.~G.~A., {et~al.} 2023, \aap,
  674, A1, \dodoi{10.1051/0004-6361/202243940}

\bibitem[{{Garc{\'\i}a} \& {Mermilliod}(2001)}]{2001..Garc..Binary}
{Garc{\'\i}a}, B., \& {Mermilliod}, J.~C. 2001, \aap, 368, 122,
  \dodoi{10.1051/0004-6361:20000528}

\bibitem[{{Gardiner} \& {Hatzidimitriou}(1992)}]{1992..Gardiner..diff..pop}
{Gardiner}, L.~T., \& {Hatzidimitriou}, D. 1992, \mnras, 257, 195,
  \dodoi{10.1093/mnras/257.2.195}

\bibitem[{Gardiner {et~al.}(1994)Gardiner, Sawa, \& Fujimoto}]{1994...Gardiner}
Gardiner, L.~T., Sawa, T., \& Fujimoto, M. 1994, Monthly Notices of the Royal
  Astronomical Society, 266, 567, \dodoi{10.1093/mnras/266.3.567}

\bibitem[{{Gaustad} {et~al.}(2001){Gaustad}, {McCullough}, {Rosing}, \& {Van
  Buren}}]{2001..Gaustad..H_alpha.}
{Gaustad}, J.~E., {McCullough}, P.~R., {Rosing}, W., \& {Van Buren}, D. 2001,
  \pasp, 113, 1326, \dodoi{10.1086/323969}

\bibitem[{Gehrels {et~al.}(2004)Gehrels, Chincarini, Giommi, Mason, Nousek,
  Wells, White, Barthelmy, Burrows, Cominsky, Hurley, Marshall, Mészáros,
  Roming, Angelini, Barbier, Belloni, Campana, Caraveo, Chester, Citterio,
  Cline, Cropper, Cummings, Dean, Feigelson, Fenimore, Frail, Fruchter,
  Garmire, Gendreau, Ghisellini, Greiner, Hill, Hunsberger, Krimm, Kulkarni,
  Kumar, Lebrun, Lloyd-Ronning, Markwardt, Mattson, Mushotzky, Norris, Osborne,
  Paczynski, Palmer, Park, Parsons, Paul, Rees, Reynolds, Rhoads, Sasseen,
  Schaefer, Short, Smale, Smith, Stella, Tagliaferri, Takahashi, Tashiro,
  Townsley, Tueller, Turner, Vietri, Voges, Ward, Willingale, Zerbi, \&
  Zhang}]{Gehrels_2004..UVOT}
Gehrels, N., Chincarini, G., Giommi, P., {et~al.} 2004, The Astrophysical
  Journal, 611, 1005, \dodoi{10.1086/422091}

\bibitem[{{Glatt} {et~al.}(2010){Glatt}, {Grebel}, \&
  {Koch}}]{2010..Glatt..MCs..Interaction}
{Glatt}, K., {Grebel}, E.~K., \& {Koch}, A. 2010, \aap, 517, A50,
  \dodoi{10.1051/0004-6361/201014187}

\bibitem[{{Gonidakis} {et~al.}(2009){Gonidakis}, {Livanou}, {Kontizas},
  {Klein}, {Kontizas}, {Belcheva}, {Tsalmantza}, \&
  {Karampelas}}]{2009..Gonidakis..Str..SMC}
{Gonidakis}, I., {Livanou}, E., {Kontizas}, E., {et~al.} 2009, \aap, 496, 375,
  \dodoi{10.1051/0004-6361/200809828}

\bibitem[{{Gordon} {et~al.}(2011){Gordon}, {Meixner}, {Meade}, {Whitney},
  {Engelbracht}, {Bot}, {Boyer}, {Lawton}, {Sewi{\l}o}, {Babler}, {Bernard},
  {Bracker}, {Block}, {Blum}, {Bolatto}, {Bonanos}, {Harris}, {Hora},
  {Indebetouw}, {Misselt}, {Reach}, {Shiao}, {Tielens}, {Carlson},
  {Churchwell}, {Clayton}, {Chen}, {Cohen}, {Fukui}, {Gorjian}, {Hony},
  {Israel}, {Kawamura}, {Kemper}, {Leroy}, {Li}, {Madden}, {Marble},
  {McDonald}, {Mizuno}, {Mizuno}, {Muller}, {Oliveira}, {Olsen}, {Onishi},
  {Paladini}, {Paradis}, {Points}, {Robitaille}, {Rubin}, {Sandstrom}, {Sato},
  {Shibai}, {Simon}, {Smith}, {Srinivasan}, {Vijh}, {Van Dyk}, {van Loon}, \&
  {Zaritsky}}]{2011..Gordon..SAGE}
{Gordon}, K.~D., {Meixner}, M., {Meade}, M.~R., {et~al.} 2011, \aj, 142, 102,
  \dodoi{10.1088/0004-6256/142/4/102}

\bibitem[{{Gordon} {et~al.}(2024){Gordon}, {Fitzpatrick}, {Massa}, {Bohlin},
  {Chastenet}, {Murray}, {Clayton}, {Lennon}, {Misselt}, \&
  {Sandstrom}}]{2024...Goradon..extinction}
{Gordon}, K.~D., {Fitzpatrick}, E.~L., {Massa}, D., {et~al.} 2024, \apj, 970,
  51, \dodoi{10.3847/1538-4357/ad4be1}

\bibitem[{{Graczyk} {et~al.}(2020){Graczyk}, {Pietrzy{\'n}ski}, {Thompson},
  {Gieren}, {Zgirski}, {Villanova}, {G{\'o}rski}, {Wielg{\'o}rski},
  {Karczmarek}, {Narloch}, {Pilecki}, {Taormina}, {Smolec}, {Suchomska},
  {Gallenne}, {Nardetto}, {Storm}, {Kudritzki}, {Ka{\l}uszy{\'n}ski}, \&
  {Pych}}]{2020ApJ...904...13Graczyk..SMC..dist}
{Graczyk}, D., {Pietrzy{\'n}ski}, G., {Thompson}, I.~B., {et~al.} 2020, \apj,
  904, 13, \dodoi{10.3847/1538-4357/abbb2b}

\bibitem[{{Harris} {et~al.}(2020){Harris}, {Millman}, {van der Walt},
  {Gommers}, {Virtanen}, {Cournapeau}, {Wieser}, {Taylor}, {Berg}, {Smith},
  {Kern}, {Picus}, {Hoyer}, {van Kerkwijk}, {Brett}, {Haldane}, {del R{\'\i}o},
  {Wiebe}, {Peterson}, {G{\'e}rard-Marchant}, {Sheppard}, {Reddy}, {Weckesser},
  {Abbasi}, {Gohlke}, \& {Oliphant}}]{2020Natur.585..357H}
{Harris}, C.~R., {Millman}, K.~J., {van der Walt}, S.~J., {et~al.} 2020, \nat,
  585, 357, \dodoi{10.1038/s41586-020-2649-2}

\bibitem[{{Harris} \& {Zaritsky}(2004)}]{2004AJ....127.1531Harris...SFH..SMC}
{Harris}, J., \& {Zaritsky}, D. 2004, \aj, 127, 1531, \dodoi{10.1086/381953}

\bibitem[{{Haschke} {et~al.}(2011){Haschke}, {Grebel}, \&
  {Duffau}}]{2011AJ....141..158H}
{Haschke}, R., {Grebel}, E.~K., \& {Duffau}, S. 2011, \aj, 141, 158,
  \dodoi{10.1088/0004-6256/141/5/158}

\bibitem[{{Haschke} {et~al.}(2012){Haschke}, {Grebel}, \&
  {Duffau}}]{2012AJ....144..106Haschke..3d..SMC}
---. 2012, \aj, 144, 106, \dodoi{10.1088/0004-6256/144/4/106}

\bibitem[{{Hindman} {et~al.}(1963){Hindman}, {Kerr}, \&
  {McGee}}]{1963..Hindman..MB}
{Hindman}, J.~V., {Kerr}, F.~J., \& {McGee}, R.~X. 1963, Australian Journal of
  Physics, 16, 570, \dodoi{10.1071/PH630570}

\bibitem[{{Hota} {et~al.}(2024){Hota}, {Subramaniam}, {Dhanush}, {Cioni}, \&
  {Subramanian}}]{2024..Hota..SMC..Shell}
{Hota}, S., {Subramaniam}, A., {Dhanush}, S.~R., {Cioni}, M.-R.~L., \&
  {Subramanian}, S. 2024, \mnras, 532, 322, \dodoi{10.1093/mnras/stae1438}

\bibitem[{{Hu} {et~al.}(2011){Hu}, {Deng}, {de Grijs}, \&
  {Liu}}]{2011..Hu..Yi..AST}
{Hu}, Y., {Deng}, L., {de Grijs}, R., \& {Liu}, Q. 2011, \pasp, 107,
  \dodoi{10.1086/658162}

\bibitem[{{Hunter}(2007)}]{2007CSE.....9...90H}
{Hunter}, J.~D. 2007, Computing in Science and Engineering, 9, 90,
  \dodoi{10.1109/MCSE.2007.55}

\bibitem[{{Indu} \& {Subramaniam}(2011)}]{2011..Indu..MCs..int}
{Indu}, G., \& {Subramaniam}, A. 2011, \aap, 535, A115,
  \dodoi{10.1051/0004-6361/201117298}

\bibitem[{{Irwin} {et~al.}(1985){Irwin}, {Kunkel}, \&
  {Demers}}]{1985..Irwin..MB}
{Irwin}, M.~J., {Kunkel}, W.~E., \& {Demers}, S. 1985, \nat, 318, 160,
  \dodoi{10.1038/318160a0}

\bibitem[{{James} {et~al.}(2021){James}, {Subramanian}, {Omkumar}, {Mary},
  {Bekki}, {Cioni}, {de Grijs}, {El Youssoufi}, {Kartha}, {Niederhofer}, \&
  {van Loon}}]{2021..James..SS..RV}
{James}, D., {Subramanian}, S., {Omkumar}, A.~O., {et~al.} 2021, \mnras, 508,
  5854, \dodoi{10.1093/mnras/stab2873}

\bibitem[{{Jim{\'e}nez-Arranz} {et~al.}(2023){Jim{\'e}nez-Arranz},
  {Romero-G{\'o}mez}, {Luri}, \&
  {Masana}}]{2023..Gaia_catalog_SMC_Prob_Jimenez}
{Jim{\'e}nez-Arranz}, {\'O}., {Romero-G{\'o}mez}, M., {Luri}, X., \& {Masana},
  E. 2023, \aap, 672, A65, \dodoi{10.1051/0004-6361/202245720}

\bibitem[{{Joshi} \& {Panchal}(2019)}]{2019..Joshi..MCs..int}
{Joshi}, Y.~C., \& {Panchal}, A. 2019, \aap, 628, A51,
  \dodoi{10.1051/0004-6361/201834574}

\bibitem[{{Kallivayalil} {et~al.}(2006){Kallivayalil}, {van der Marel}, \&
  {Alcock}}]{2006..Kallivayali..firts_passage_MCs}
{Kallivayalil}, N., {van der Marel}, R.~P., \& {Alcock}, C. 2006, \apj, 652,
  1213, \dodoi{10.1086/508014}

\bibitem[{{Kobulnicky} {et~al.}(2014){Kobulnicky}, {Kiminki}, {Lundquist},
  {Burke}, {Chapman}, {Keller}, {Lester}, {Rolen}, {Topel}, {Bhattacharjee},
  {Smullen}, {Vargas {\'A}lvarez}, {Runnoe}, {Dale}, \&
  {Brotherton}}]{2014..Kobulnicky..binary..70}
{Kobulnicky}, H.~A., {Kiminki}, D.~C., {Lundquist}, M.~J., {et~al.} 2014,
  \apjs, 213, 34, \dodoi{10.1088/0067-0049/213/2/34}

\bibitem[{{Kulkarni} {et~al.}(2021){Kulkarni}, {Harrison}, {Grefenstette},
  {Earnshaw}, {Andreoni}, {Berg}, {Bloom}, {Cenko}, {Chornock}, {Christiansen},
  {Coughlin}, {Wuollet Criswell}, {Darvish}, {Das}, {De}, {Dessart}, {Dixon},
  {Dorsman}, {El-Badry}, {Evans}, {Ford}, {Fremling}, {Gansicke}, {Gezari},
  {Goetberg}, {Green}, {Graham}, {Heida}, {Ho}, {Jaodand}, {Johns-Krull},
  {Kasliwal}, {Lazzarini}, {Lu}, {Margutti}, {Martin}, {Masters}, {McKernan},
  {Naze}, {Nissanke}, {Parazin}, {Perley}, {Phinney}, {Piro}, {Raaijmakers},
  {Rauw}, {Rodriguez}, {Sana}, {Senchyna}, {Singer}, {Spake}, {Stassun},
  {Stern}, {Teplitz}, {Weisz}, \& {Yao}}]{2021..UVEX}
{Kulkarni}, S.~R., {Harrison}, F.~A., {Grefenstette}, B.~W., {et~al.} 2021,
  arXiv e-prints, arXiv:2111.15608, \dodoi{10.48550/arXiv.2111.15608}

\bibitem[{{Kumar} {et~al.}(2012){Kumar}, {Ghosh}, {Hutchings}, {Kamath},
  {Kathiravan}, {Mahesh}, {Murthy}, {Nagbhushana}, {Pati}, {Rao}, {Rao},
  {Sriram}, \& {Tandon}}]{2012kumarUVIT}
{Kumar}, A., {Ghosh}, S.~K., {Hutchings}, J., {et~al.} 2012, in Society of
  Photo-Optical Instrumentation Engineers (SPIE) Conference Series, Vol. 8443,
  Space Telescopes and Instrumentation 2012: Ultraviolet to Gamma Ray, ed.
  T.~{Takahashi}, S.~S. {Murray}, \& J.-W.~A. {den Herder}, 84431N,
  \dodoi{10.1117/12.924507}

\bibitem[{{Leahy} {et~al.}(2020){Leahy}, {Postma}, {Chen}, \&
  {Buick}}]{2020..Laehy..UVIT..M31}
{Leahy}, D.~A., {Postma}, J., {Chen}, Y., \& {Buick}, M. 2020, \apjs, 247, 47,
  \dodoi{10.3847/1538-4365/ab77a9}

\bibitem[{{Lemasle} {et~al.}(2017){Lemasle}, {Groenewegen}, {Grebel}, {Bono},
  {Fiorentino}, {Fran{\c{c}}ois}, {Inno}, {Kovtyukh}, {Matsunaga}, {Pedicelli},
  {Primas}, {Pritchard}, {Romaniello}, \& {da Silva}}]{2017A&A...608A..85L}
{Lemasle}, B., {Groenewegen}, M.~A.~T., {Grebel}, E.~K., {et~al.} 2017, \aap,
  608, A85, \dodoi{10.1051/0004-6361/201731370}

\bibitem[{{Lindegren} {et~al.}(2018){Lindegren}, {Hern{\'a}ndez}, {Bombrun},
  {Klioner}, {Bastian}, {Ramos-Lerate}, {de Torres}, {Steidelm{\"u}ller},
  {Stephenson}, {Hobbs}, {Lammers}, {Biermann}, {Geyer}, {Hilger}, {Michalik},
  {Stampa}, {McMillan}, {Casta{\~n}eda}, {Clotet}, {Comoretto}, {Davidson},
  {Fabricius}, {Gracia}, {Hambly}, {Hutton}, {Mora}, {Portell}, {van Leeuwen},
  {Abbas}, {Abreu}, {Altmann}, {Andrei}, {Anglada}, {Balaguer-N{\'u}{\~n}ez},
  {Barache}, {Becciani}, {Bertone}, {Bianchi}, {Bouquillon}, {Bourda},
  {Br{\"u}semeister}, {Bucciarelli}, {Busonero}, {Buzzi}, {Cancelliere},
  {Carlucci}, {Charlot}, {Cheek}, {Crosta}, {Crowley}, {de Bruijne}, {de
  Felice}, {Drimmel}, {Esquej}, {Fienga}, {Fraile}, {Gai}, {Garralda},
  {Gonz{\'a}lez-Vidal}, {Guerra}, {Hauser}, {Hofmann}, {Holl}, {Jordan},
  {Lattanzi}, {Lenhardt}, {Liao}, {Licata}, {Lister}, {L{\"o}ffler},
  {Marchant}, {Martin-Fleitas}, {Messineo}, {Mignard}, {Morbidelli}, {Poggio},
  {Riva}, {Rowell}, {Salguero}, {Sarasso}, {Sciacca}, {Siddiqui}, {Smart},
  {Spagna}, {Steele}, {Taris}, {Torra}, {van Elteren}, {van Reeven}, \&
  {Vecchiato}}]{2018..Lindegren}
{Lindegren}, L., {Hern{\'a}ndez}, J., {Bombrun}, A., {et~al.} 2018, \aap, 616,
  A2, \dodoi{10.1051/0004-6361/201832727}

\bibitem[{{Liu}(1992)}]{1992..Liu..MS}
{Liu}, Y.-Z. 1992, \aap, 257, 505

\bibitem[{{Lucchini} {et~al.}(2020){Lucchini}, {D'Onghia}, {Fox}, {Bustard},
  {Bland-Hawthorn}, \& {Zweibel}}]{2020..Luchin..MS}
{Lucchini}, S., {D'Onghia}, E., {Fox}, A.~J., {et~al.} 2020, \nat, 585, 203,
  \dodoi{10.1038/s41586-020-2663-4}

\bibitem[{{Mackey} {et~al.}(2018){Mackey}, {Koposov}, {Da Costa}, {Belokurov},
  {Erkal}, \& {Kuzma}}]{2018..Mackey..SMC..SS}
{Mackey}, D., {Koposov}, S., {Da Costa}, G., {et~al.} 2018, \apjl, 858, L21,
  \dodoi{10.3847/2041-8213/aac175}

\bibitem[{{Maragoudaki} {et~al.}(2001){Maragoudaki}, {Kontizas}, {Morgan},
  {Kontizas}, {Dapergolas}, \& {Livanou}}]{2001..Maragoudaki..SMC..Morphology}
{Maragoudaki}, F., {Kontizas}, M., {Morgan}, D.~H., {et~al.} 2001, \aap, 379,
  864, \dodoi{10.1051/0004-6361:20011454}

\bibitem[{{Marchant} {et~al.}(2017){Marchant}, {Langer}, {Podsiadlowski},
  {Tauris}, {de Mink}, {Mandel}, \& {Moriya}}]{2017..Marchant..DCOB}
{Marchant}, P., {Langer}, N., {Podsiadlowski}, P., {et~al.} 2017, \aap, 604,
  A55, \dodoi{10.1051/0004-6361/201630188}

\bibitem[{{Mart{\'\i}nez-Delgado} {et~al.}(2019){Mart{\'\i}nez-Delgado},
  {Vivas}, {Grebel}, {Gallart}, {Pieres}, {Bell}, {Zivick}, {Lemasle}, {Clifton
  Johnson}, {Carballo-Bello}, {No{\"e}l}, {Cioni}, {Choi}, {Besla}, {Schmidt},
  {Zaritsky}, {Gruendl}, {Seibert}, {Nidever}, {Monteagudo}, {Monelli}, {Hubl},
  {van der Marel}, {Ballesteros}, {Stringfellow}, {Walker}, {Blum}, {Bell},
  {Conn}, {Olsen}, {Martin}, {Chu}, {Inno}, {Boer}, {Kallivayalil}, {De Leo},
  {Beletsky}, {Neyer}, \& {Mu{\~n}oz}}]{2019..martinez..shell}
{Mart{\'\i}nez-Delgado}, D., {Vivas}, A.~K., {Grebel}, E.~K., {et~al.} 2019,
  \aap, 631, A98, \dodoi{10.1051/0004-6361/201936021}

\bibitem[{{Mason} {et~al.}(1998){Mason}, {Gies}, {Hartkopf}, {Bagnuolo}, {ten
  Brummelaar}, \& {McAlister}}]{1998..Mason..Binary}
{Mason}, B.~D., {Gies}, D.~R., {Hartkopf}, W.~I., {et~al.} 1998, \aj, 115, 821,
  \dodoi{10.1086/300234}

\bibitem[{{Massana} {et~al.}(2020){Massana}, {No{\"e}l}, {Nidever}, {Erkal},
  {de Boer}, {Choi}, {Majewski}, {Olsen}, {Monachesi}, {Gallart}, {Marel},
  {Ruiz-Lara}, {Zaritsky}, {Martin}, {Mu{\~n}oz}, {Cioni}, {Bell}, {Bell},
  {Stringfellow}, {Belokurov}, {Monelli}, {Walker}, {Mart{\'\i}nez-Delgado},
  {Vivas}, \& {Conn}}]{2020..Massana..SMC..int}
{Massana}, P., {No{\"e}l}, N. E.~D., {Nidever}, D.~L., {et~al.} 2020, \mnras,
  498, 1034, \dodoi{10.1093/mnras/staa2451}

\bibitem[{{McClure-Griffiths} {et~al.}(2018){McClure-Griffiths}, {D{\'e}nes},
  {Dickey}, {Stanimirovi{\'c}}, {}, {Staveley-Smith}, {Jameson}, {Di Teodoro},
  {Allison}, {Collier}, {Chippendale}, {Franzen}, {G{\"u}rkan}, {Heald},
  {Hotan}, {Kleiner}, {Lee-Waddell}, {McConnell}, {Popping}, {Rhee}, {Riseley},
  {Voronkov}, \& {Whiting}}]{2018..H_alpha}
{McClure-Griffiths}, N.~M., {D{\'e}nes}, H., {Dickey}, J.~M., {et~al.} 2018,
  Nature Astronomy, 2, 901, \dodoi{10.1038/s41550-018-0608-8}

\bibitem[{{Moe} \& {Di Stefano}(2017)}]{2017..Moe..Binary}
{Moe}, M., \& {Di Stefano}, R. 2017, \apjs, 230, 15,
  \dodoi{10.3847/1538-4365/aa6fb6}

\bibitem[{{Muller} \& {Bekki}(2007)}]{2007..Muller_MBR}
{Muller}, E., \& {Bekki}, K. 2007, \mnras, 381, L11,
  \dodoi{10.1111/j.1745-3933.2007.00356.x}

\bibitem[{{Murray} {et~al.}(2019){Murray}, {Peek}, {Di Teodoro},
  {McClure-Griffiths}, {Dickey}, \& {D{\'e}nes}}]{2019..Murray..Gas..Kinematic}
{Murray}, C.~E., {Peek}, J.~E.~G., {Di Teodoro}, E.~M., {et~al.} 2019, \apj,
  887, 267, \dodoi{10.3847/1538-4357/ab510f}

\bibitem[{{Murray} {et~al.}(2024){Murray}, {Hasselquist}, {Peek}, {Lindberg},
  {Almeida}, {Choi}, {Craig}, {D{\'e}nes}, {Dickey}, {Di Teodoro}, {Federrath},
  {Gerrard}, {Gibson}, {Leahy}, {Lee}, {Lynn}, {Ma}, {Marchal},
  {McClure-Griffiths}, {Nidever}, {Nguyen}, {Pingel}, {Tarantino}, {Uscanga},
  \& {van Loon}}]{2024..Murray..SMC.dwarf..two..superimposed}
{Murray}, C.~E., {Hasselquist}, S., {Peek}, J. E.~G., {et~al.} 2024, \apj, 962,
  120, \dodoi{10.3847/1538-4357/ad1591}

\bibitem[{{Nayak} {et~al.}(2018){Nayak}, {Subramaniam}, {Choudhury}, \&
  {Sagar}}]{2018..PK..Nayak}
{Nayak}, P.~K., {Subramaniam}, A., {Choudhury}, S., \& {Sagar}, R. 2018, \aap,
  616, A187, \dodoi{10.1051/0004-6361/201732227}

\bibitem[{Nidever {et~al.}(2008)Nidever, Majewski, \&
  Burton}]{Nidever_2008..MS}
Nidever, D.~L., Majewski, S.~R., \& Burton, W.~B. 2008, The Astrophysical
  Journal, 679, 432–459, \dodoi{10.1086/587042}

\bibitem[{{Nidever} {et~al.}(2010){Nidever}, {Majewski}, {Butler Burton}, \&
  {Nigra}}]{2010..Nidever...MS}
{Nidever}, D.~L., {Majewski}, S.~R., {Butler Burton}, W., \& {Nigra}, L. 2010,
  \apj, 723, 1618, \dodoi{10.1088/0004-637X/723/2/1618}

\bibitem[{{Nidever} {et~al.}(2017){Nidever}, {Olsen}, {Walker}, {Vivas},
  {Blum}, {Kaleida}, {Choi}, {Conn}, {Gruendl}, {Bell}, {Besla}, {Mu{\~n}oz},
  {Gallart}, {Martin}, {Olszewski}, {Saha}, {Monachesi}, {Monelli}, {de Boer},
  {Johnson}, {Zaritsky}, {Stringfellow}, {van der Marel}, {Cioni}, {Jin},
  {Majewski}, {Martinez-Delgado}, {Monteagudo}, {No{\"e}l}, {Bernard},
  {Kunder}, {Chu}, {Bell}, {Santana}, {Frechem}, {Medina}, {Parkash},
  {Navarrete}, \& {Hayes}}]{2017..SMASH..Nidever}
{Nidever}, D.~L., {Olsen}, K., {Walker}, A.~R., {et~al.} 2017, \aj, 154, 199,
  \dodoi{10.3847/1538-3881/aa8d1c}

\bibitem[{{Nidever} {et~al.}(2021){Nidever}, {Olsen}, {Choi}, {Ruiz-Lara},
  {Miller}, {Johnson}, {Bell}, {Blum}, {Cioni}, {Gallart}, {Majewski},
  {Martin}, {Massana}, {Monachesi}, {No{\"e}l}, {Sakowska}, {van der Marel},
  {Walker}, {Zaritsky}, {Bell}, {Conn}, {de Boer}, {Gruendl}, {Monelli},
  {Mu{\~n}oz}, {Saha}, {Vivas}, {Bernard}, {Besla}, {Carballo-Bello}, {Dorta},
  {Martinez-Delgado}, {Goater}, {Rusakov}, \&
  {Stringfellow}}]{2021..Nidever..SMASH..DR2}
{Nidever}, D.~L., {Olsen}, K., {Choi}, Y., {et~al.} 2021, \aj, 161, 74,
  \dodoi{10.3847/1538-3881/abceb7}

\bibitem[{{Niederhofer} {et~al.}(2018){Niederhofer}, {Cioni}, {Rubele},
  {Schmidt}, {Bekki}, {de Grijs}, {Emerson}, {Ivanov}, {Marconi}, {Oliveira},
  {Petr-Gotzens}, {Ripepi}, {van Loon}, \& {Zaggia}}]{2018..Niederhofer..WH}
{Niederhofer}, F., {Cioni}, M. R.~L., {Rubele}, S., {et~al.} 2018, \aap, 613,
  L8, \dodoi{10.1051/0004-6361/201833144}

\bibitem[{{Niederhofer} {et~al.}(2021){Niederhofer}, {Cioni}, {Rubele},
  {Schmidt}, {Diaz}, {Matijev{\u{i}}c}, {Bekki}, {Bell}, {de Grijs}, {El
  Youssoufi}, {Ivanov}, {Oliveira}, {Ripepi}, {Subramanian}, {Sun}, \& {van
  Loon}}]{2021..Niederhofer..MCs..PM}
{Niederhofer}, F., {Cioni}, M.-R.~L., {Rubele}, S., {et~al.} 2021, \mnras, 502,
  2859, \dodoi{10.1093/mnras/stab206}

\bibitem[{{Omkumar} {et~al.}(2021){Omkumar}, {Subramanian}, {Niederhofer},
  {Diaz}, {Cioni}, {El Youssoufi}, {Bekki}, {de Grijs}, \& {van
  Loon}}]{2021..abinaya}
{Omkumar}, A.~O., {Subramanian}, S., {Niederhofer}, F., {et~al.} 2021, \mnras,
  500, 2757, \dodoi{10.1093/mnras/staa3085}

\bibitem[{{Parsons} {et~al.}(2024){Parsons}, {Prinja}, {Bernini-Peron},
  {Fullerton}, {Massa}, {Oskinova}, {Pauli}, {Rickard}, \&
  {Sander}}]{2024..Parsons..B}
{Parsons}, T.~N., {Prinja}, R.~K., {Bernini-Peron}, M., {et~al.} 2024, \mnras,
  527, 11422, \dodoi{10.1093/mnras/stad3966}

\bibitem[{{Patrick} {et~al.}(2022){Patrick}, {Thilker}, {Lennon}, {Bianchi},
  {Schootemeijer}, {Dorda}, {Langer}, \&
  {Negueruela}}]{2022MNRAS..Patrick..RSG..SMC}
{Patrick}, L.~R., {Thilker}, D., {Lennon}, D.~J., {et~al.} 2022, \mnras, 513,
  5847, \dodoi{10.1093/mnras/stac1139}

\bibitem[{{Pieres} {et~al.}(2017){Pieres}, {Santiago}, {Drlica-Wagner},
  {Bechtol}, {Marel}, {Besla}, {Martin}, {Belokurov}, {Gallart},
  {Martinez-Delgado}, {Marshall}, {N{\"o}el}, {Majewski}, {Cioni}, {Li},
  {Hartley}, {Luque}, {Conn}, {Walker}, {Balbinot}, {Stringfellow}, {Olsen},
  {Nidever}, {da Costa}, {Ogando}, {Maia}, {Neto}, {Abbott}, {Abdalla},
  {Allam}, {Annis}, {Benoit-L{\'e}vy}, {Rosell}, {Kind}, {Carretero}, {Cunha},
  {D'Andrea}, {Desai}, {Diehl}, {Doel}, {Flaugher}, {Fosalba},
  {Garc{\'\i}a-Bellido}, {Gruen}, {Gruendl}, {Gschwend}, {Gutierrez},
  {Honscheid}, {James}, {Kuehn}, {Kuropatkin}, {Menanteau}, {Miquel}, {Plazas},
  {Romer}, {Sako}, {Sanchez}, {Scarpine}, {Schubnell}, {Sevilla-Noarbe},
  {Smith}, {Soares-Santos}, {Sobreira}, {Suchyta}, {Swanson}, {Tarle},
  {Tucker}, \& {Wester}}]{2017..Pieres..smc..ss}
{Pieres}, A., {Santiago}, B.~X., {Drlica-Wagner}, A., {et~al.} 2017, \mnras,
  468, 1349, \dodoi{10.1093/mnras/stx507}

\bibitem[{{Pietrzy{\'n}ski} {et~al.}(2019){Pietrzy{\'n}ski}, {Graczyk},
  {Gallenne}, {Gieren}, {Thompson}, {Pilecki}, {Karczmarek}, {G{\'o}rski},
  {Suchomska}, {Taormina}, {Zgirski}, {Wielg{\'o}rski}, {Ko{\l}aczkowski},
  {Konorski}, {Villanova}, {Nardetto}, {Kervella}, {Bresolin}, {Kudritzki},
  {Storm}, {Smolec}, \& {Narloch}}]{2019Pietrzy..LMC..distance}
{Pietrzy{\'n}ski}, G., {Graczyk}, D., {Gallenne}, A., {et~al.} 2019, \nat, 567,
  200, \dodoi{10.1038/s41586-019-0999-4}

\bibitem[{{Podsiadlowski} {et~al.}(1992){Podsiadlowski}, {Joss}, \&
  {Hsu}}]{1992..Podsiadlowski..evolution}
{Podsiadlowski}, P., {Joss}, P.~C., \& {Hsu}, J.~J.~L. 1992, \apj, 391, 246,
  \dodoi{10.1086/171341}

\bibitem[{{Postma} {et~al.}(2011){Postma}, {Hutchings}, \&
  {Leahy}}]{2011PostemaDetectorPC}
{Postma}, J., {Hutchings}, J.~B., \& {Leahy}, D. 2011, \pasp, 123, 833,
  \dodoi{10.1086/661187}

\bibitem[{{Postma} \& {Leahy}(2017)}]{2017Postmaccdlab}
{Postma}, J.~E., \& {Leahy}, D. 2017, \pasp, 129, 115002,
  \dodoi{10.1088/1538-3873/aa8800}

\bibitem[{{Postma} \& {Leahy}(2021)}]{2021Postmaccdlab}
---. 2021, Journal of Astrophysics and Astronomy, 42, 30,
  \dodoi{10.1007/s12036-020-09689-w}

\bibitem[{{Putman} {et~al.}(1998){Putman}, {Gibson}, {Staveley-Smith}, {Banks},
  {Barnes}, {Bhatal}, {Disney}, {Ekers}, {Freeman}, {Haynes}, {Henning},
  {Jerjen}, {Kilborn}, {Koribalski}, {Knezek}, {Malin}, {Mould}, {Oosterloo},
  {Price}, {Ryder}, {Sadler}, {Stewart}, {Stootman}, {Vaile}, {Webster}, \&
  {Wright}}]{1998..mw-lmc-smc-inter}
{Putman}, M.~E., {Gibson}, B.~K., {Staveley-Smith}, L., {et~al.} 1998, \nat,
  394, 752, \dodoi{10.1038/29466}

\bibitem[{{Ripepi} {et~al.}(2017){Ripepi}, {Cioni}, {Moretti}, {Marconi},
  {Bekki}, {Clementini}, {de Grijs}, {Emerson}, {Groenewegen}, {Ivanov},
  {Molinaro}, {Muraveva}, {Oliveira}, {Piatti}, {Subramanian}, \& {van
  Loon}}]{2017..Ripepi..NE-SW..SMC}
{Ripepi}, V., {Cioni}, M.-R.~L., {Moretti}, M.~I., {et~al.} 2017, \mnras, 472,
  808, \dodoi{10.1093/mnras/stx2096}

\bibitem[{{Romaniello} {et~al.}(2008){Romaniello}, {Primas}, {Mottini},
  {Pedicelli}, {Lemasle}, {Bono}, {Fran{\c{c}}ois}, {Groenewegen}, \&
  {Laney}}]{2008A&A...488..731R}
{Romaniello}, M., {Primas}, F., {Mottini}, M., {et~al.} 2008, \aap, 488, 731,
  \dodoi{10.1051/0004-6361:20065661}

\bibitem[{{Rubele} {et~al.}(2015){Rubele}, {Girardi}, {Kerber}, {Cioni},
  {Piatti}, {Zaggia}, {Bekki}, {Bressan}, {Clementini}, {de Grijs}, {Emerson},
  {Groenewegen}, {Ivanov}, {Marconi}, {Marigo}, {Moretti}, {Ripepi},
  {Subramanian}, {Tatton}, \& {van Loon}}]{2015MNRAS.449..639Rubele..SFH}
{Rubele}, S., {Girardi}, L., {Kerber}, L., {et~al.} 2015, \mnras, 449, 639,
  \dodoi{10.1093/mnras/stv141}

\bibitem[{{Rubele} {et~al.}(2018){Rubele}, {Pastorelli}, {Girardi}, {Cioni},
  {Zaggia}, {Marigo}, {Bekki}, {Bressan}, {Clementini}, {de Grijs}, {Emerson},
  {Groenewegen}, {Ivanov}, {Muraveva}, {Nanni}, {Oliveira}, {Ripepi}, {Sun}, \&
  {van Loon}}]{2018..Rubele..MCs..int}
{Rubele}, S., {Pastorelli}, G., {Girardi}, L., {et~al.} 2018, \mnras, 478,
  5017, \dodoi{10.1093/mnras/sty1279}

\bibitem[{{Russell} \& {Dopita}(1992)}]{1992ApJ...384..508R}
{Russell}, S.~C., \& {Dopita}, M.~A. 1992, \apj, 384, 508,
  \dodoi{10.1086/170893}

\bibitem[{{Sagar} \& {Richtler}(1991)}]{1991..Ram..Sagar..completeness}
{Sagar}, R., \& {Richtler}, T. 1991, \aap, 250, 324

\bibitem[{{Sahu} {et~al.}(2022){Sahu}, {Subramaniam}, {Singh}, {Yadav},
  {Valcarce}, {Choudhury}, {Rani}, {Prabhu}, {Chung}, {C{\^o}t{\'e}}, {Leigh},
  {Geller}, {Chatterjee}, {Kameswara Rao}, {Bandyopadhyay}, {Shara},
  {Dalessandro}, {Pandey}, {Postma}, {Hutchings}, {Simunovic}, {Stetson},
  {Thirupathi}, {Puzia}, \& {Sohn}}]{2022..sahu..AST..PSF}
{Sahu}, S., {Subramaniam}, A., {Singh}, G., {et~al.} 2022, \mnras, 514, 1122,
  \dodoi{10.1093/mnras/stac1209}

\bibitem[{{Sakowska} {et~al.}(2024){Sakowska}, {No{\"e}l}, {Ruiz-Lara},
  {Gallart}, {Massana}, {Nidever}, {Cassisi}, {Correa-Amaro}, {Choi}, {Besla},
  {Erkal}, {Mart{\'\i}nez-Delgado}, {Monelli}, {Olsen}, \&
  {Stringfellow}}]{2024..Sakowska..shell}
{Sakowska}, J.~D., {No{\"e}l}, N. E.~D., {Ruiz-Lara}, T., {et~al.} 2024,
  \mnras, 532, 4272, \dodoi{10.1093/mnras/stae1766}

\bibitem[{{Sana} {et~al.}(2012){Sana}, {de Mink}, {de Koter}, {Langer},
  {Evans}, {Gieles}, {Gosset}, {Izzard}, {Le Bouquin}, \&
  {Schneider}}]{2012..Sana..Binary..70}
{Sana}, H., {de Mink}, S.~E., {de Koter}, A., {et~al.} 2012, Science, 337, 444,
  \dodoi{10.1126/science.1223344}

\bibitem[{{Sana} {et~al.}(2014){Sana}, {Le Bouquin}, {Lacour}, {Berger},
  {Duvert}, {Gauchet}, {Norris}, {Olofsson}, {Pickel}, {Zins}, {Absil}, {de
  Koter}, {Kratter}, {Schnurr}, \& {Zinnecker}}]{2014..Sana..Binary}
{Sana}, H., {Le Bouquin}, J.~B., {Lacour}, S., {et~al.} 2014, \apjs, 215, 15,
  \dodoi{10.1088/0067-0049/215/1/15}

\bibitem[{{Scowcroft} {et~al.}(2016){Scowcroft}, {Freedman}, {Madore},
  {Monson}, {Persson}, {Rich}, {Seibert}, \&
  {Rigby}}]{2016..Scowcroft..NE-SW..SMC}
{Scowcroft}, V., {Freedman}, W.~L., {Madore}, B.~F., {et~al.} 2016, \apj, 816,
  49, \dodoi{10.3847/0004-637X/816/2/49}

\bibitem[{{Shenar} {et~al.}(2024){Shenar}, {Bodensteiner}, {Sana}, {Crowther},
  {Lennon}, {Abdul-Masih}, {Almeida}, {Backs}, {Berlanas}, {Bernini-Peron},
  {Bestenlehner}, {Bowman}, {Bronner}, {Britavskiy}, {de Koter}, {de Mink},
  {Deshmukh}, {Evans}, {Fabry}, {Gieles}, {Gilkis}, {Gonz{\'a}lez-Tor{\`a}},
  {Gr{\"a}fener}, {G{\"o}tberg}, {Hawcroft}, {H{\'e}nault-Brunet}, {Herrero},
  {Holgado}, {Janssens}, {Johnston}, {Josiek}, {Justham}, {Kalari}, {Katabi},
  {Keszthelyi}, {Klencki}, {Kub{\'a}t}, {Kub{\'a}tov{\'a}}, {Langer},
  {Lefever}, {Ludwig}, {Mackey}, {Mahy}, {Ma{\'\i}z Apell{\'a}niz}, {Mandel},
  {Maravelias}, {Marchant}, {Menon}, {Najarro}, {Oskinova}, {Ovadia},
  {Patrick}, {Pauli}, {Pawlak}, {Ramachandran}, {Renzo}, {Rocha}, {Sander},
  {Sayada}, {Schneider}, {Schootemeijer}, {Sch{\"o}sser}, {Sch{\"u}rmann},
  {Sen}, {Shahaf}, {Sim{\'o}n-D{\'\i}az}, {Stoop}, {van Loon}, {Toonen},
  {Tramper}, {Valli}, {van Son}, {Vigna-G{\'o}mez}, {Villase{\~n}or}, {Vink},
  {Wang}, \& {Willcox}}]{2024..Shenar..BLOeM}
{Shenar}, T., {Bodensteiner}, J., {Sana}, H., {et~al.} 2024, arXiv e-prints,
  arXiv:2407.14593, \dodoi{10.48550/arXiv.2407.14593}

\bibitem[{Simons {et~al.}(2014)Simons, Thilker, Bianchi, \& Wyder}]{SIMGALEX}
Simons, R., Thilker, D., Bianchi, L., \& Wyder, T. 2014, Advances in Space
  Research, 53, 939, \dodoi{https://doi.org/10.1016/j.asr.2013.07.016}

\bibitem[{{Sriram} {et~al.}(2023){Sriram}, {Valsan}, {Subramaniam}, {Unni},
  {Maheswar}, \& {Chand}}]{2023..INSIST}
{Sriram}, S., {Valsan}, V., {Subramaniam}, A., {et~al.} 2023, Journal of
  Astrophysics and Astronomy, 44, 55, \dodoi{10.1007/s12036-023-09934-y}

\bibitem[{{Stanimirovi{\'c}} {et~al.}(2004){Stanimirovi{\'c}},
  {Staveley-Smith}, \& {Jones}}]{2004..stanimirovi..SMC..mass}
{Stanimirovi{\'c}}, S., {Staveley-Smith}, L., \& {Jones}, P.~A. 2004, \apj,
  604, 176, \dodoi{10.1086/381869}

\bibitem[{{Stetson}(1987)}]{1987..Stetson..IRAF}
{Stetson}, P.~B. 1987, \pasp, 99, 191, \dodoi{10.1086/131977}

\bibitem[{{Subramaniam}(2022)}]{2022JApA..INSIST}
{Subramaniam}, A. 2022, Journal of Astrophysics and Astronomy, 43, 80,
  \dodoi{10.1007/s12036-022-09870-3}

\bibitem[{Subramaniam {et~al.}(2016{\natexlab{a}})Subramaniam, Tandon,
  Hutchings, Ghosh, George, Girish, Kamath, Kathiravan, Kumar, Lancelot,
  Mahesh, Mohan, Murthy, Nagabhushana, Pati, Postma, Rao, Sankarasubramanian,
  Sreekumar, Sriram, Stalin, Sutaria, Sreedhar, Barve, Mondal, \&
  Sahu}]{10.1117/12.2235271}
Subramaniam, A., Tandon, S.~N., Hutchings, J., {et~al.} 2016{\natexlab{a}}, in
  Space Telescopes and Instrumentation 2016: Ultraviolet to Gamma Ray, ed.
  J.-W.~A. den Herder, T.~Takahashi, \& M.~Bautz, Vol. 9905, International
  Society for Optics and Photonics (SPIE), 99051F, \dodoi{10.1117/12.2235271}

\bibitem[{Subramaniam {et~al.}(2016{\natexlab{b}})Subramaniam, Sindhu, Tandon,
  Rao, Postma, Côté, Hutchings, Ghosh, George, Girish, Mohan, Murthy,
  Sankarasubramanian, Stalin, Sutaria, Mondal, \& Sahu}]{Subramaniam_2016}
Subramaniam, A., Sindhu, N., Tandon, S.~N., {et~al.} 2016{\natexlab{b}}, The
  Astrophysical Journal Letters, 833, L27, \dodoi{10.3847/2041-8213/833/2/L27}

\bibitem[{{Subramanian} \& {Subramaniam}(2012)}]{2012..Smitha..NE-SW..SMC}
{Subramanian}, S., \& {Subramaniam}, A. 2012, \apj, 744, 128,
  \dodoi{10.1088/0004-637X/744/2/128}

\bibitem[{{Tandon} {et~al.}(2017{\natexlab{a}}){Tandon}, {Hutchings}, {Ghosh},
  {Subramaniam}, {Koshy}, {Girish}, {Kamath}, {Kathiravan}, {Kumar},
  {Lancelot}, {Mahesh}, {Mohan}, {Murthy}, {Nagabhushana}, {Pati}, {Postma},
  {Rao}, {Sankarasubramanian}, {Sreekumar}, {Sriram}, {Stalin}, {Sutaria},
  {Sreedhar}, {Barve}, {Mondal}, \& {Sahu}}]{2017June..TondonUVIT}
{Tandon}, S.~N., {Hutchings}, J.~B., {Ghosh}, S.~K., {et~al.}
  2017{\natexlab{a}}, Journal of Astrophysics and Astronomy, 38, 28,
  \dodoi{10.1007/s12036-017-9445-x}

\bibitem[{{Tandon} {et~al.}(2017{\natexlab{b}}){Tandon}, {Subramaniam},
  {Girish}, {Postma}, {Sankarasubramanian}, {Sriram}, {Stalin}, {Mondal},
  {Sahu}, {Joseph}, {Hutchings}, {Ghosh}, {Barve}, {George}, {Kamath},
  {Kathiravan}, {Kumar}, {Lancelot}, {Leahy}, {Mahesh}, {Mohan},
  {Nagabhushana}, {Pati}, {Kameswara Rao}, {Sreedhar}, \&
  {Sreekumar}}]{2017Sep..TondonUVIT}
{Tandon}, S.~N., {Subramaniam}, A., {Girish}, V., {et~al.} 2017{\natexlab{b}},
  \aj, 154, 128, \dodoi{10.3847/1538-3881/aa8451}

\bibitem[{{Tandon} {et~al.}(2020){Tandon}, {Postma}, {Joseph}, {Devaraj},
  {Subramaniam}, {Barve}, {George}, {Ghosh}, {Girish}, {Hutchings}, {Kamath},
  {Kathiravan}, {Kumar}, {Lancelot}, {Leahy}, {Mahesh}, {Mohan},
  {Nagabhushana}, {Pati}, {Rao}, {Sankarasubramanian}, {Sriram}, \&
  {Stalin}}]{2020TondonUVITfilters}
{Tandon}, S.~N., {Postma}, J., {Joseph}, P., {et~al.} 2020, \aj, 159, 158,
  \dodoi{10.3847/1538-3881/ab72a3}

\bibitem[{Tatton {et~al.}(2021)Tatton, van Loon, Cioni, Bekki, Bell,
  Choudhury, de Grijs, Groenewegen, Ivanov, Marconi, Oliveira, Ripepi, Rubele,
  Subramanian, \& Sun}]{Tatton..2021..inter}
Tatton, B.~L., van Loon, J.~T., Cioni, M.-R.~L., {et~al.} 2021, Monthly
  Notices of the Royal Astronomical Society, 504, 2983,
  \dodoi{10.1093/mnras/staa3857}

\bibitem[{{van der Marel} \&
  {Cioni}(2001)}]{2001AJ..van..der..Marel..&..Cioni..SMC..projection}
{van der Marel}, R.~P., \& {Cioni}, M.-R.~L. 2001, \aj, 122, 1807,
  \dodoi{10.1086/323099}

\bibitem[{{van der Marel} \&
  {Kallivayalil}(2014)}]{2014..van..der..Marel...LMC..mass}
{van der Marel}, R.~P., \& {Kallivayalil}, N. 2014, \apj, 781, 121,
  \dodoi{10.1088/0004-637X/781/2/121}

\bibitem[{van~der Marel {et~al.}(2008)van~der Marel, Kallivayalil, \&
  Besla}]{van_der_Marel_2008}
van~der Marel, R.~P., Kallivayalil, N., \& Besla, G. 2008, Proceedings of the
  International Astronomical Union, 4, 81, \dodoi{10.1017/s1743921308028299}

\bibitem[{{van der Marel} \& {Sahlmann}(2016)}]{2016van..der..Marel.MCs..PM}
{van der Marel}, R.~P., \& {Sahlmann}, J. 2016, \apjl, 832, L23,
  \dodoi{10.3847/2041-8205/832/2/L23}

\bibitem[{Vasiliev(2023)}]{2023..Vasiliev}
Vasiliev, E. 2023, Monthly Notices of the Royal Astronomical Society, 527, 437,
  \dodoi{10.1093/mnras/stad2612}

\bibitem[{{Virtanen} {et~al.}(2020){Virtanen}, {Gommers}, {Oliphant},
  {Haberland}, {Reddy}, {Cournapeau}, {Burovski}, {Peterson}, {Weckesser},
  {Bright}, {van der Walt}, {Brett}, {Wilson}, {Millman}, {Mayorov}, {Nelson},
  {Jones}, {Kern}, {Larson}, {Carey}, {Polat}, {Feng}, {Moore}, {VanderPlas},
  {Laxalde}, {Perktold}, {Cimrman}, {Henriksen}, {Quintero}, {Harris},
  {Archibald}, {Ribeiro}, {Pedregosa}, {van Mulbregt}, \& {SciPy 1. 0
  Contributors}}]{2020NatMe..17..261V}
{Virtanen}, P., {Gommers}, R., {Oliphant}, T.~E., {et~al.} 2020, Nature
  Methods, 17, 261, \dodoi{10.1038/s41592-019-0686-2}

\bibitem[{{Zaritsky} {et~al.}(2000){Zaritsky}, {Harris}, {Grebel}, \&
  {Thompson}}]{2000..Zaritsky..SMC..morphology}
{Zaritsky}, D., {Harris}, J., {Grebel}, E.~K., \& {Thompson}, I.~B. 2000,
  \apjl, 534, L53, \dodoi{10.1086/312649}

\bibitem[{{Zivick} {et~al.}(2018){Zivick}, {Kallivayalil}, {van der Marel},
  {Besla}, {Linden}, {Koz{\l}owski}, {Fritz}, {Kochanek}, {Anderson}, {Sohn},
  {Geha}, \& {Alcock}}]{2018..Zivic..MCs..PM}
{Zivick}, P., {Kallivayalil}, N., {van der Marel}, R.~P., {et~al.} 2018, \apj,
  864, 55, \dodoi{10.3847/1538-4357/aad4b0}

\bibitem[{{Zivick} {et~al.}(2019){Zivick}, {Kallivayalil}, {Besla}, {Sohn},
  {van der Marel}, {del Pino}, {Linden}, {Fritz}, \&
  {Anderson}}]{2019Zevick..MB}
{Zivick}, P., {Kallivayalil}, N., {Besla}, G., {et~al.} 2019, \apj, 874, 78,
  \dodoi{10.3847/1538-4357/ab0554}

\end{thebibliography}
%\bibliographystyle{aasjournal}

\end{document}